\makeatletter
\def\input@path{{./packages/}}
\makeatother
\PassOptionsToPackage{dvipsnames}{xcolor}
\documentclass[sigconf, nonacm]{acmart}

\AtBeginDocument{%
  }

\setcopyright{acmlicensed}
\copyrightyear{2024}
\acmYear{2024}
\usepackage{graphicx}
\usepackage{xcolor}
\definecolor{PURPLE}{rgb}{0.5,0,0.5}
\usepackage{tablefootnote}
\usepackage{subcaption}
\newcommand{\eg}{\emph{e.g.,} }
\newcommand{\ie}{\emph{i.e.,} }
\usepackage[most]{tcolorbox}
\usepackage{enumitem}
\usepackage{chngcntr}
\usepackage{appendix}
\usepackage{placeins}
\usepackage{pdflscape}
\usepackage{stfloats}

\counterwithin{table}{section}

\definecolor{forestgreen}{rgb}{0.0, 0.27, 0.13}
\definecolor{tropicalrainforest}{rgb}{0.0, 0.46, 0.37}

\AtBeginDocument{\thanks{\llap{\hspace*{-35pt}}${\star}$Both authors contributed equally to this research.}}

\begin{document}
 \title{SafarDB: FPGA-Accelerated Distributed Transactions via Replicated Data Types \\
% \sadoghi{SafarDB or SafarDBDB: FPGA-Accelerated Distributed Transactions through Replicated Data Types}
}


\author{Javad Saberlatibari$^{\star\star}$, Prithviraj Yuvaraj$^{\star\star}$, Philip Brisk$^{\star}$, Mohammad Sadoghi$^{\ddagger}$, Mohsen Lesani$^{\dagger}$}
\affiliation{%
  \institution{University of California, $^{\star}$Riverside, $^{\ddagger}$Davis, $^{\dagger}$Santa Cruz}
  \country{USA}
}
% \affiliation{%
%   \state{Ireland}

% \orcid{0000-0002-1825-0097}
% \affiliation{%
%   \country{Iceland}

% \orcid{0000-0001-5109-3700}
% \affiliation{%
% \email{vb@rocquencourt.com}

% \author{J\"org von \"Arbach}
% \affiliation{%
%   \institution{University of T\"ubingen}
%   \city{T\"ubingen}
%   \country{Germany}
% }
% \email{jaerbach@uni-tuebingen.edu}
% \email{myprivate@email.com}
% \email{second@affiliation.mail}

% \author{Wang Xiu Ying}
% \author{Zhe Zuo}
% \affiliation{%
%   \institution{East China Normal University}
%   \city{Shanghai}
%   \country{China}
% }
% \email{firstname.lastname@ecnu.edu.cn}

\renewcommand{\shortauthors}{-}


\begin{abstract}
Data replication is a critical aspect of data center design, as it ensures high availability, scalability, and fault tolerance. However, replicas need to be coordinated to maintain convergence and database integrity constraints under transactional workloads.
Commutative Replicated Data Types (RDTs) provide convergence for conflict-free objects
using relaxed consistency,
and
Well-coordinated Replicated Data Types (WRDTs) provide convergence and integrity for general objects using a hybrid model, relaxed when possible and strong when necessary.
While state-of-the-art RDT hardware acceleration uses Remote Direct Memory Access (RDMA), we observe that trends towards lower latency and higher throughput have driven recent data center architectures to leverage FPGAs as application accelerators.
% 
% \sadoghi{I have improved the above. We have a gap here, jumping into NIC before even describing what we are aiming to do. Better to first build the story RDMA and FPGA, then perhaps bringforth the implementation details, which is to further leverage the NIC.}
In contrast to deploying an FPGA-based SmartNIC, this paper connects an FPGA accelerator card directly to the network, which allows a complete redesign of the NIC to match the needs of the FPGA-hosted application.
% 
% \sadoghi{The terminology verb is extremely low-level to be avoided when we are describing the higher-level motivation, gap, problem statement, and the contributions.}
% We design specialized RDMA verbs that simultaneously transmit data to FPGA internal storage while invoking FPGA-resident accelerators as Remote Procedure Calls (RPCs), as well as mechanisms to allow accelerators to access the NIC's internal state. 
We co-design a network-attached FPGA replication engine with an FPGA-resident network interface, enabling near-network execution of replicated transactions and direct invocation of FPGA-resident operators.
% \sadoghi{it would be important to organize the flow (1) what is the motivation/trends, what is our aim, what is the gap we are addressing, then what is our problem statement, and then what are the technical contributions, again, we want to stay at higher level without too much low level hardware details, what is the high level picture, which we can refine and go deeper in the intro and the body.}
% 
Following this approach, we introduce SafarDB, an FPGA-accelerated system for Conflict-Free Replicated Data Types (CRDTs) and WRDTs.
% 
% SafarDB accelerates both relaxed and strong consistency,
% in particular, the leader election for consensus. 
SafarDB accelerates both relaxed and strongly ordered replication paths; when strong ordering is required, SafarDB accelerates the underlying consensus control path.
% \sadoghi{See now from NIC, verb, simultaneous transmit, we are now bringing consensus and leader election. This is the high-level problem that needs to be introduced first before going into all the details.}
\textcolor{black}{In FPGA-only CRDT and WRDT experiments where the complete application state fits in FPGA-resident memory, SafarDB provides lower average response time and higher average throughput than a state-of-the-art RDMA-based implementation.}
Further, experiments demonstrate that 
SafarDB is more resilient to crash failures than existing CPU/RDMA-based CRDT and WRDT implementations, and
SafarDB can detect leader failures and elect new leaders much faster than
% previous work.
% was 
previously possible.

% \sadoghi{The relationship between replicated data types and consensus and leader election is not clear; we cannot assume the reader is aware. So you could motivate first the replication problem, then bring consensus (and leader election if it is relevant at this stage or leave for later), then say well here are the hardware trends RDMAs and FPGAs, then say here are the replciation advaccement such as RDTs, etc... then what is the gap and shortcoming, what is the our precise problem statement, what are contributions and results.}
% 
\end{abstract}

\maketitle

\section{Introduction}
\label{sec:introduction}
% \sadoghi{We have tension between consensus and RDTs, it is better framed with consensus than internally as RDTs. A special structure/scheme for achieving replication. Folks in DBs are familiar with consensus, replication, CRDTs, but RDTs are not a common term, so we need to introduce them. The goal is replication and consensus, and RDTs provide a unique abstract or efficient abstraction to solve the consensus problem needed for replication. Also, maybe we can say that RDTs are an easier abstraction, at the object/data types level, than building distributed data types. We need a motivation as to why we should care about RDTs, also for the DB community, they care more about transactions as an abstraction, not so much as a programming language construct. We can conceive our application or transaction consists of RDTs abstraction, for DB, better to reframe our transactions as modeled RDTs, since our evaluation is also about distributed transaction not distributed object programming.}
Replicated Data Types (RDTs) are an object-level abstraction used inside replicated databases to implement distributed transactions that run on data center networks \cite{shapiro11, BurckhardtGotsman14, calm}. 
RDTs provide common abstractions, \eg registers, sets, and lists, using coordination strategies that guarantee convergence, availability, and integrity properties.
% in a manner that is transparent to the programmer. 
Database systems use RDTs as building blocks for distributed transactions %\sadoghi{building blocks of distributed transactions, for example} 
\cite{calm,dynamo,flighttracker,peritext} without requiring application developers to directly implement or interface with consistency models and coordination algorithms, or handle node and network failures.

We use RDT as a general category for replicated objects in transactional systems. Conflict-Free Replicated Data Types (CRDTs) are a subclass of RDTs whose operations commute, enabling relaxed consistency without coordination. Well-coordinated Replicated Data Types (WRDTs) extend this idea to support integrity constraints by using relaxed execution when safe and invoking strongly ordered replication (consensus) only for operations that can violate integrity if reordered.

% The objective of this work is to improve RDT performance \sadoghi{in the context of distributed transactions. It would be important to expand our title to include FPGA-Accelerated Distributed Transactions using Replicated Data Types.} 
The objective of this work is to improve the performance of distributed transactions by accelerating RDT execution on network-attached FPGAs.
State-of-the-art RDT systems are specialized for data centers equipped with RDMA networks \cite{hamband, damon}, which enable microsecond-scale remote memory access latencies.
% The current state-of-the-art RDTs are specialized for data center which are enabled with RDMA networks \cite{hamband}, which provide $\mu$s-scale remote memory access latencies.
One study by Microsoft reports that RDMA traffic accounts for up to 70\% of total network traffic in the Azure public cloud \cite{azure_nsdi23}. We observe that RDMA performance is limited by several sources of latency at each network node: PCIe transaction latency, and the use of host memory as an intermediary between the CPU and NIC; this complements recent research that has identified microarchitectural bottlenecks within RDMA NICs \cite{srnic, smart_rdma}. Existing RDT/WRDT systems, including RDMA-based implementations, largely assume CPU-hosted execution \cite{hamsaz,hamband}. Prior network-attached FPGA, SmartNIC, and programmable-network systems show that storage, protocol/verb processing, or ordering logic can be moved into hardware \cite{caribou,waverunner,cinbox,p4xos,netpaxos,StRoM}. However, these systems target key-value storage, verb/protocol offload, or strongly ordered consensus rather than RDMA-based RDT execution. As a result, prior systems do not simultaneously provide (i) FPGA-resident RDT operation handlers, (ii) a hybrid FPGA/host storage and execution model under one replication interface, (iii) both relaxed and strongly ordered replication paths, and (iv) low-overhead failure-control paths such as leader changes and permission updates \cite{mu,hamband}.

% \sadoghi{we need to identify a few more shortcomings and gaps, at least cover 4 shortcomings, in say 5-10 papers in this area (at least a few from SIGMOD, VLDB, ICDE, EDBT). Of course, one shortcoming could be overlooking a hardware opportunity, so we can list a few related papers and say none of them consider FPGAs and RDMAs. Also, we need to differentiate between CRDTs and RDTs, and again, provide the reference. Also, citing related CRDTs in DB.}

To alleviate these bottlenecks, we propose SafarDB, an RDT design and implementation on network-attached FPGAs. SafarDB co-locates the RDT and NIC within the FPGA, using onboard memory sparingly. 
SafarDB specializes the NIC for communication between local and remote FPGA-resident accelerators, rather than supporting tens of thousands of concurrent threads running on the local CPU host. 
In doing so, we make a number of contributions that are fundamental to RDMA implementations for network-attached FPGAs and FPGA-based implementations of distributed systems:
\begin{itemize}[leftmargin=*,topsep=3pt]
\item 
SafarDB augments the NIC with novel one-sided RDMA verbs that access integrated FPGA storage resources directly, and repurposes them to implement efficient Remote Procedure Calls (RPCs) to FPGA-resident accelerators. 
\item 
SafarDB provides FPGA-resident accelerators with direct access to the NIC's microarchitectural state.
SafarDB leverages this access to accelerate RDTs,
which can use both relaxed and strong consistency models.
% \sadoghi{This is the first mention of consistency models; this needs to be included in the shortcomings, say X and Y have limited consistency models, while SafarDB, etc.}
In particular, it eliminates inefficiencies associated with leader election, a critical component of consensus.
% In particular, it removes inefficiencies for leader election, a necessary ingredient for consensus.
% under a crash-failure model.
% \item Compared to state-of-the-art RDMA-based RDTs \cite{hamband}:
%
\item 
% SafarDB provides a hybrid mode that utilizes the host CPU and its memory, enabling capacity beyond the FPGA's available memory.
SafarDB provides a hybrid mode that utilizes the host CPU and its memory to scale capacity beyond the FPGA's available memory, while maintaining a single replication/consistency interface across FPGA- and host-resident data, a challenge not addressed by prior FPGA-only storage or replication designs \cite{caribou,cinbox}.
% 
% SafarDB provides a hybrid mode that involves the host CPU and CPU memory to expand beyond the FPGAs available memory. 
% 
% \item 
% SafarDB reduces latency and increases throughput of Conflict-Free Replicated Data Types (CRDTs) \cite{shapiro11, calm}, 
% % which do not require synchronization, 
% which uses relaxed consistency. 
% % \sadoghi{first mention of CRDTs and its link, how we differentiate between RDTs, and CRDT remains unclear.}
% % 
% \item 
% SafarDB reduces latency and increases throughput of Well-coordinated Replicated Data Types (WRDTs) \cite{hamsaz, hampa, hamband}, 
% which uses relaxed consistency when possible, and strong consistency when necessary.
% % 
% \item 
%  SafarDB reduces latency and increases throughput of YCSB \cite{ycsb} and SmallBank \cite{oltp}.
% %
\item 
\textcolor{black}{In FPGA-only CRDT and WRDT experiments where the complete application state fits in FPGA-resident memory, SafarDB provides lower average response time and higher average throughput than the CPU/RDMA baseline. Evaluations with YCSB \cite{ycsb} and SmallBank \cite{oltp} further characterize FPGA-resident and hybrid execution. The configured application-state and dataset sizes are reported in Appendix Table~\ref{tab:application-state-datasets}.\footnote{Extended appendix: \url{https://tinyurl.com/78d9dzmj}.}}
\end{itemize}

We anticipate that this work will foment further investigation into distributed system deployment on network-attached FPGAs, especially for replicated transactional systems and database replication under a hybrid (FPGA+host) design model. FPGAs tend to obviate the distinction between application, operating system, and (micro)architecture: application and kernel functions can easily be co-located on a programmable FPGA fabric, with system calls reduced to wires between modules. SafarDB embodies this unification as a hardware-accelerated replication engine for transactional workloads, achieving performance that would otherwise be unattainable. Similarly, SafarDB demonstrates the benefits of customizing an FPGA-resident NIC to meet the needs of a transactional workload. We start with a NIC that provides one-sided RDMA verbs, which read from and write to onboard high-bandwidth memory. We then design and evaluate RDT-specific RPC mechanisms that eliminate otherwise spurious memory accesses. A similar approach can be applied to a wide variety of replicated databases and transactional workloads.

% \sadoghi{I suggest, let's decide what SafarDB is, and if its replicated DB, then let's focus on transactions, distributed transactions, etc.}
% and services.
% 
% 
% A similar approach could be applied to a wide variety of distributed applications. 

%which can generalize to principally similar distributed applications and services, to achieve high performance. 

% \newpage
%SafarDB augments the NIC with novel one-sided RDMA verbs that access integrated FPGA storage resources directly, instead of or in addition to onboard memory, and then repurposes these new verbs to provide low-overhead Remote Procedure Calls (RPCs) that invoke FPGA-resident accelerators. Additionally, SafarDB provides accelerators with direct access to NIC microarchitectural state, which play an important, albeit limited, role in an RDMA-based consensus protocol \cite{mu}, which provides coordination for conflicting \cite{hamband} RDT method calls. These ingredients enable SafarDB to outperform the current state-of-the-art RDT implementation for RDMA networks.

%\input{architecture} % New
% \section{SafarDB Implementation}
\section{SafarDB Overview}
% SafarDB
\label{sec:rdt_implementation}

% \sadoghi{This is the first time that we have a definition of what SafarDB is. It is a bit long, so maybe SafarDB is a distributed DB connected through RDMA with a network-attached FPGA that supports a hybrid storage hierarchy including both host memory and device memory, etc.}

We first formalize the system settings and assumptions under which SafarDB operates. 

\textbf{System Model}. We consider a distributed database deployed across $N$ replicas. Each replica consists of a server equipped with a CPU and network-attached FPGA. Replicas communicate transactions\footnote{Transactions are assumed to be single-statement transactions.} through RDMA over a reliable network.

\textbf{Network Model}. Replicas communicate via RDMA over Converged Ethernet (RoCEv2). We assume reliable, in-order delivery of RDMA messages.
\textcolor{black}{Detailed background on traditional and FPGA-based RDMA, including SafarDB's FPGA-specific RPC path, is provided in \autoref{sec:rdma}.}

\textbf{Consistency Model}. SafarDB implements a hybrid consistency model based on Hamband \cite{hamband}. Transactions are categorized into reducible, irreducible, and conflicting operations. Conflicting transactions require total ordering and are serialized using a State Machine Replication (SMR) hardware module that accelerates Mu \cite{mu}, a primary-backup, RDMA-based consensus protocol in which one replica is designated as the leader and the rest are followers. The leader performs coordination and commits transactions by writing into the followers' \textit{replication logs}, which are circular buffers in memory.

\textbf{Fault Model}. We assume a crash fault model. A faulty replica may halt at any time and remain non-responsive thereafter, or return to functionality. Replicas are assumed not to exhibit Byzantine behavior. Both the leader replica and follower replicas can crash. When a follower crashes and comes back up, the leader can use the log for recovery by reissuing the previously committed transactions to the newly returned follower. When a leader crashes, the remaining followers elect a new leader. Replica crashes are detected by monitoring each replica's \textit{liveness}, which measures its activity and responsiveness.

\textbf{System Scope and Limitations.}\label{sec:system-scope-limitations}
SafarDB is a hardware-accelerated RDT/WRDT replication design rather than a complete general-purpose database engine. It currently assumes single-statement transactions within a synchronization group and does not provide durable commits, general multi-statement transactions across RDT objects, or a query/indexing layer. SafarDB provides convergence, integrity checks, and SMR ordering for conflicting operations, but not full cross-object ACID transactions.

\textbf{Durability.} SafarDB provides crash fault tolerance through replicated logs, leader election, and remote recovery after a node failure: a returning follower can recover by replaying committed transactions from the leader's log, and a failed leader triggers a new election among the surviving replicas. This recovery model is not ACID durability because the replication log and application state currently reside in volatile FPGA HBM and host memory. A simultaneous power loss across replicas would result in data loss. Providing durable commits would require flushing the replication log to NVMe or persistent memory before acknowledging a commit. This is architecturally compatible with SafarDB because the replication log is already a structured circular buffer, and a write-through path to persistent storage could be added as an additional FPGA kernel or handled by the host CPU in hybrid mode.

\textbf{Multi-statement transactions.} SafarDB assumes single-statement transactions. Transactions that span multiple RDT objects would require coordination across synchronization groups, a distributed transaction manager, and abort/rollback logic. These are non-trivial additions beyond the current system scope. The RDT/WRDT model partially mitigates this limitation: many operations that would require multi-statement transactions in a traditional database, such as an update followed by an integrity check, can be expressed as a single WRDT method with an integrity predicate.

\textbf{Indexing.} SafarDB does not provide a query layer or indexing structures. Application state is accessed directly by the FPGA-resident accelerator via on-chip BRAM or HBM. Adding indexing would require implementing index structures, such as B-trees or hash indexes, as FPGA kernels or on the host CPU side in hybrid mode. KV-Direct~\cite{kv_direct} demonstrates that key-value indexing on FPGAs is feasible; integrating such structures with SafarDB's replication interface is a natural extension.

\textbf{ACID properties.} SafarDB provides convergence and integrity guarantees through the RDT/WRDT model rather than full ACID semantics. Atomicity holds for single-statement transactions within a synchronization group; consistency is maintained through permissibility checks and integrity predicates; and conflicting operations are isolated through SMR serialization. Durability is not provided, as discussed above. Applications requiring full ACID semantics across multiple objects would need multi-statement transaction support and durable commit mechanisms.

% \subsection{Architecture}
\textbf{Architecture. \ }
SafarDB is a distributed database connected through RDMA utilizing a network-attached FPGA accelerator that co-locates application logic, replication protocols, and RDMA networking on a single device to eliminate host-side RDMA overheads,
and supports a hybrid execution model with a storage hierarchy that includes both host memory and device memory.
It enables efficient implementations of CRDTs, WRDTs, and larger distributed applications and benchmarks, including YCSB and SmallBank. 
SafarDB accelerates replication for the three categories of transactions: reducible, irreducible, and conflicting, explained in \autoref{sec:rdts}.
In addition to RDMA Read and RDMA Write verbs, 
we introduce and use FPGA-specific RDMA RPC verbs to further improve performance. 
The accelerator consists of multiple modules and memories on the FPGA, along with a CPU-resident application and CPU memory to expand the supported workload size.

\begin{figure}[t]
\vspace*{-8mm}
\centering
\includegraphics[width=0.8\linewidth]{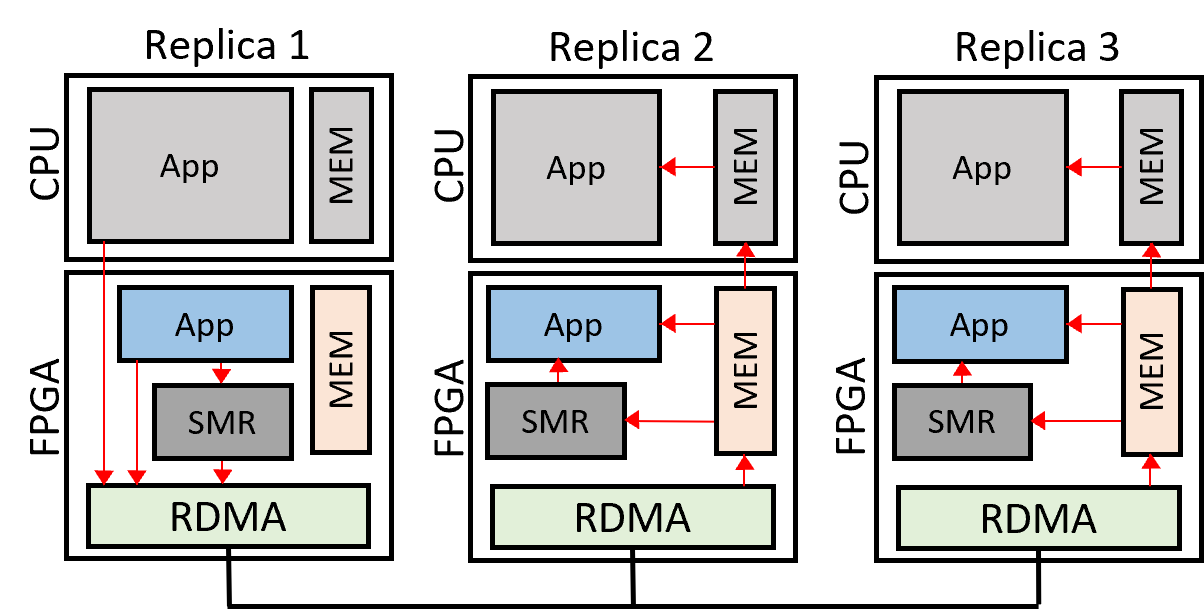}
\vspace*{-2mm}
\caption{SafarDB's architecture stack: The red arrows depict the different paths transactions can take depending on the category of transaction.}
\label{fig:arch:software-stack}
\vspace*{-6mm}
\end{figure}

% \todo{Citation missing?}
\autoref{fig:arch:software-stack} provides the holistic picture of SafarDB. 
Transactions can originate from either the host CPU or FPGA, invoking the SMR module if total ordering is required \ie conflicting transactions. RDMA is used to propagate these transactions to other replicas. Receiving replicas can update FPGA and CPU data by reading from memory. For totally ordered transactions, SMR will read from memory and update the data.

% \sadoghi{The section heading says design of RDTs, but we really provide the architecture of SafarDB. So this should be SafarDB Architecture. SafarDB is more than just RDTs because we now have hybrid memory, RDMA, and FPGA, all of which go beyond just RDTs. So we need to expand the scope, as mentioned earlier, and reflected in the title. In fact, I suggest SafarDBDB.}

\begin{figure}[t]
\vspace{-10mm}
\centering
\includegraphics[width=\linewidth]{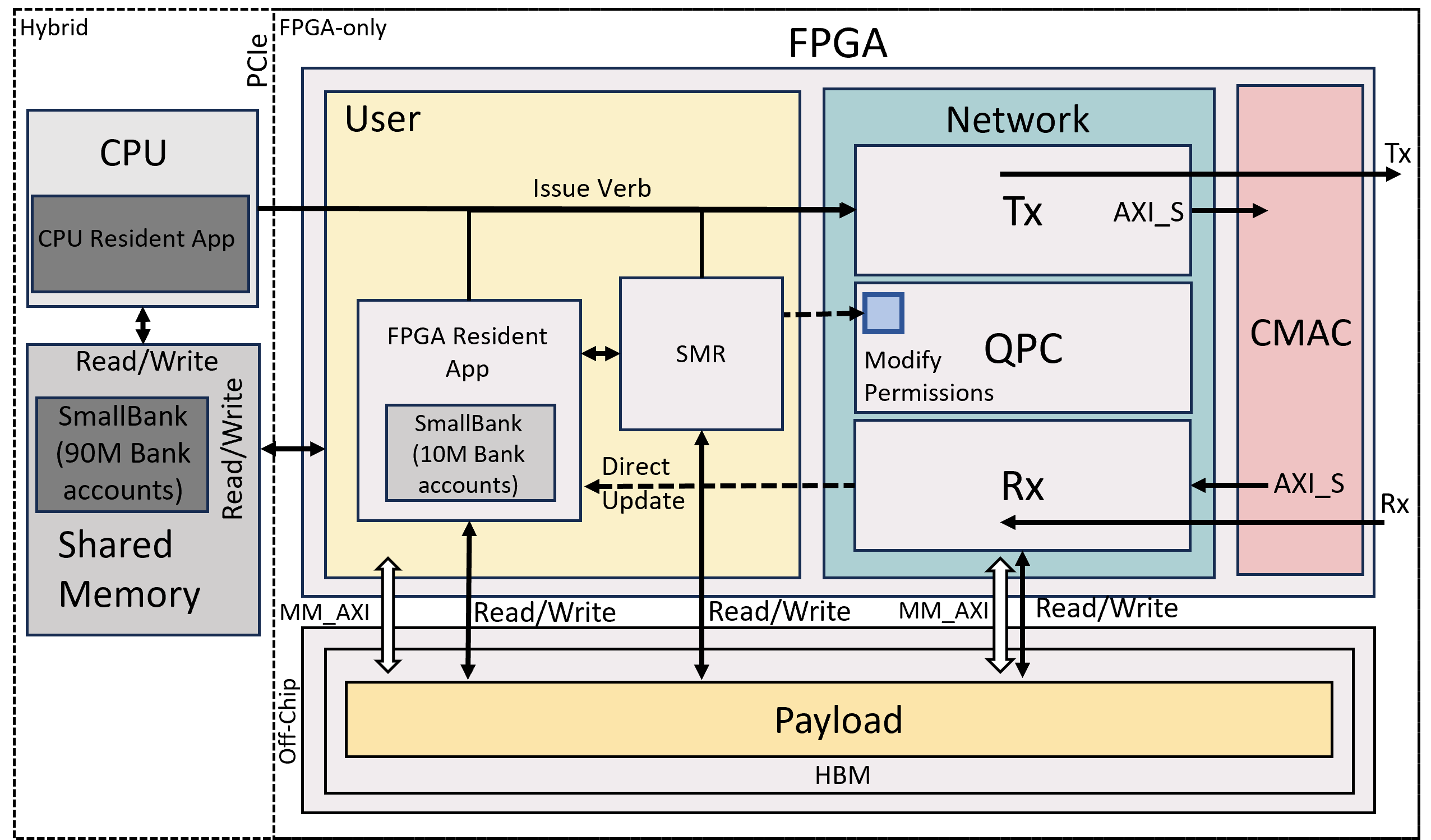}
\vspace{-6mm}
\caption{SafarDB architecture: SafarDB can operate in an FPGA-only mode, in which the data is stored entirely on the FPGA. It can further operate in hybrid mode, in which the CPU memory is used to store additional data.
%\sadoghi{No details of benchmark should appear in the figure caption. The only thing that can be said here is that data is stored in X and Y, with no reference to any benchmark.}
}
\label{fig:arch:overview}
\vspace{-6mm}
\end{figure}

\autoref{fig:arch:overview} depicts SafarDB’s accelerator in both FPGA-only and hybrid modes. The hardware modules on the FPGA consist of a CMAC kernel, which houses the 100G Ethernet Subsystem Xilinx IP used for Ethernet communication; a network stack that performs RoCEv2 RDMA operations; and a user kernel. The user kernel consists of a collection of user-specified operations for the FPGA-resident application, logic for the transactions, an SMR module, and the application’s data stored in on-chip BRAM. The FPGA also includes off-chip memory in the form of an 8 GB HBM buffer, which can be read from and written to via RDMA. On the CPU side, a CPU-resident application provides equivalent functionality to the FPGA-resident application in software, along with a shared memory buffer in main memory.

% \sadoghi{We need a software-stack architecture of SafarDB that demonstrates the holistic picture of RMDA, FPGA, several replicas, the hybrid memory design, host cpu, and perhaps a high-level image of transactions are routed among replicas and within the FPGA, CPU, and memory.}

SafarDB can execute operations in an FPGA-only mode, where the entirety of the application state resides on the FPGA, or in a hybrid mode that allows the application state to be distributed between FPGA and CPU memory. Operations can be replicated to remote replicas from either the FPGA- or CPU-resident application using the three different transaction categories (see \autoref{sec:rdts}).
%\sadoghi{It would be good to mention these earlier and their relations to various consistency models, and perhaps their relation to RDTs.} 
Conflicting transactions are totally ordered and use a dedicated SMR module, explained in \autoref{sec:Mu}.

In FPGA-only mode, SafarDB is limited by the memory available on the FPGA. However, because the application resides on the same fabric as the RDMA network stack, RDMA operations are extremely fast. This is achieved by replacing PCIe communication with the Advanced eXtensible Interface (AXI) protocol, a lightweight synchronous on-chip communication protocol. 
%An example method call is a deposit or withdrawal in the SmallBank benchmark. 
% \sadoghi{We should not mention benchmark details here; all descriptions should be generic, unless we have an explicit running example, which is not the case here.} 
In order to execute a reducible transaction, the FPGA-resident application issues a local update to its on-chip BRAM, and then issues an RDMA Write, which travels through the network module on the transmit path, exits through the CMAC kernel, and reaches the receiving replicas’ memory. 
The replicas then read from HBM memory to update their local data. 
In order to execute a conflicting transaction, the FPGA-resident application forwards the transaction to SMR for total ordering. SMR performs a round of consensus, and commits the totally ordered operation into the replicas’ logs in HBM. The replicas, similar to the reducible case, are responsible for retrieving the value from memory.

% \bigskip
% \ \\
\vspace{1ex}
\noindent
\fbox{%
    \parbox{\linewidth}{
%     \textbf{Design Principle \#\todo{NUMBER}}:
    \textbf{Design Principle \# 1}:
    In traditional RDMA systems, the CPU running the host application issues RDMA verbs to the RNIC via PCIe, incurring communication overhead. 
    By collocating the application and RNIC hardware on a single FPGA chip, PCIe communication overhead is replaced by lightweight on-chip communication, and near-network processing is enabled.
    }
}
\vspace{1ex}
% \ \\

SafarDB provides the hybrid mode for cases where the entire dataset cannot fit within the FPGA memory. 
%\sadoghi{We use 'application state' as programming terminology; in DB, we have the entire dataset fit in X and Y. State implies intermediate transaction state, but when you say 'state,' you mean the data.}
In this mode, the FPGA performs transactions on FPGA-resident data. 
% \sadoghi{it should be FPGA-resident data (not application), also we do not have method calls, we have transactions and operations (or we can say transactions consist of a set of reads or writes) within transactions. Method call is a programming language.} 
Additionally, the CPU can perform transactions by either issuing RDMA operations directly to the network module through PCIe or by triggering SMR to perform consensus for conflicting transactions. On the receiving side, after reading from the HBM memory, a dedicated module writes to host memory to update the CPU-resident data.

\section{RDT Acceleration in SafarDB}
In this section, we present in-depth explanations of how SafarDB implements the three categories of transactions, their optimizations, and a detailed description of the hardware implementing SMR. 

% accelerates 
% different types of transactions
% for WRDTs (Well-coordinated Replicated Data Types).

% The remainder of this section presents in-depth explanations of the three categories of transactions, their optimizations, and a detailed description of the SMR protocol.

% \todo{Some introduction text maybe from the previous section.}

\subsection{Reducible Transactions}
\vspace*{-1mm}
\label{ssec:reducible-methods}

% \autoref{fig:arch:reducible} depicts SafarDB's three configurations of the reducible category of transactions.

SafarDB implements three configurations for reducible transactions.
% \todo{Check this for correctness:}
% in the order of increasing efficiency.
They use the RDMA Write and SafarDB's custom RDMA RPC verbs.
For example, let the system comprise $N$ replicas, and the replicated data comprise one scalar $B$. 
%\sadoghi{Up to this point, we have not formally defined our setting. We need a proper section that defines the setting, meaning that we have n number of replicas, what is our fault model (e.g., crash tolerant), what is the network model (partial synch, synch, or asynch), reliable or unreliable network (whether we have message loss or not). All of these need to go into the preliminary. Talking about a replica is unexpected and breaks the flow.}
% Each call to reducible method $\mathit{deposit}$ adds the positive deposited value to $B$. 

% \bigskip
% \noindent
(1) $\mathsf{RDMA\ Write}$ (no buffer):
%In \autoref{fig:arch:reducible} ($\mathit{a}$, $\mathit{b}$) , 
Each replica allocates an $N$-element array, $A$, in the HBM memory. The element $A[i]$ is written exclusively by the $i^{th}$ replica, and is read locally. 
The $i^{th}$ replica summarizes its scalar into a single value, and RDMA Writes that summary to $A[i]$ at all remote replicas 
%(through $\mathit{a}$). 
At each remote replica, the next data access updates $B$. 
%(through $\mathit{b}$).

%At each remote replica, the next call to $\mathit{query}$, or the next permissibility check triggers the update (through $\mathit{b}$).% $B = B + \Sigma_{i=1}^N A[i]$. 

%Strong eventual consistency requires each replica to propagate remotely-written values from $A$ to $B$ at some indeterminate point in the future. The next call to $\mathit{query}$ or the next permissibility check triggers the update $B = B + \Sigma_{i=1}^N A[i]$. 

% \bigskip
% \noindent
(2) $\mathsf{RDMA\ Write}$ (with buffer):
%In \autoref{fig:arch:reducible} ($\mathit{a}$, $\mathit{c}$) , 
A copy of $A$ is allocated within the FPGA in addition to HBM memory. Replicas RDMA Write summaries to $A$ in memory. 
%(through $\mathit{a}$). 
In the background, the replicas periodically poll memory to update the FPGA-resident copy of $A$. 
%(through $\mathit{c}$). 
The next data access updates $B$ from the FPGA-resident copy of $A$ without accessing memory. %removing HBM reads from the critical path. 

%\sadoghi{better to simplify as read and write operations, and transactions that may consist of a set of read and write operations, namely, multi-statement transactions. We now have several terms: query, method calls, deposit, update, etc. Also, it would be good to define and lay out this terminology in the preliminary sections. In fact, it is better to reformulate as \textbf{Get and Set operations} to avoid mixing with our RDMA reads/writes. Essentially, focusing on NoSQL DB that supports gets and sets operations, but also supports multi-statement transactions of gets/sets, and adjust as needed to ensure we formulate what we can support because, as we discussed in the past, we do have some limitations and restrictions, so we cannot support general transactions.}

% \bigskip
% \noindent
(3) $\mathsf{RDMA\ RPC}$: 
%In \autoref{fig:arch:reducible} ($\mathit{d}$), 
Reducible transactions are propagated as RPCs to remote replicas (\autoref{fig:bg:rpc}), eliminating array $A$. 
%(through $\mathit{d}$). %After calling $deposit$ at least once, the local replica propagates the opcode for $deposit$ and the aggregate deposit amount to all remote replicas, each of which invokes $deposit$ immediately. 
% The $query$ calls and permissibility checks do not trigger remote updates.

\vspace{-2.7mm}
\subsection{Irreducible Conflict-free Transactions}
\label{ssec:irreducible-methods}

%\autoref{fig:arch:irreducible} depicts how 
SafarDB implements irreducible conflict-free transactions using RDMA Write and SafarDB's custom RDMA RPC verbs. 
For example, let the system comprise $N$ replicas, and let each replica's data comprise a set $S$.

% \bigskip
% \noindent
(1) $\mathsf{RDMA\ Write}$: 
In \autoref{fig:arch:irreducible}, each replica allocates $N$ queues in memory, where $Q_i$ is written exclusively by the $i^{th}$ replica and is read locally. 
The $i^{th}$ replica locally updates $S$, and propagates the transaction to remote replicas by RDMA Writing the opcode to $Q_i$ (through $\mathit{a}$). 
The remote replica periodically polls the queues by issuing memory reads (through $\mathit{b}$), and updates $S$. 
%The $\mathit{query}$ calls and permissibility checks do not trigger remote updates under this scheme. 
%\sadoghi{Courseware has not been defined, unclear what it is, and we should not be referencing benchmark. The description should be abstract.}

% \bigskip
% \noindent
(2) $\mathsf{RDMA\ RPC}$: 
In \autoref{fig:arch:irreducible} ($\mathit{c}$), irreducible conflict-free transactions are propagated as RPCs to remote replicas (\autoref{fig:bg:rpc}), eliminating queues and memory reads (through $\mathit{c}$).

\subsection{Conflicting Transactions}
\label{ssec:conflicting-methods}
\label{sec:conflicting}

\begin{figure*}[t]
\centering
\vspace*{-9mm}
\includegraphics[width=0.9\linewidth]{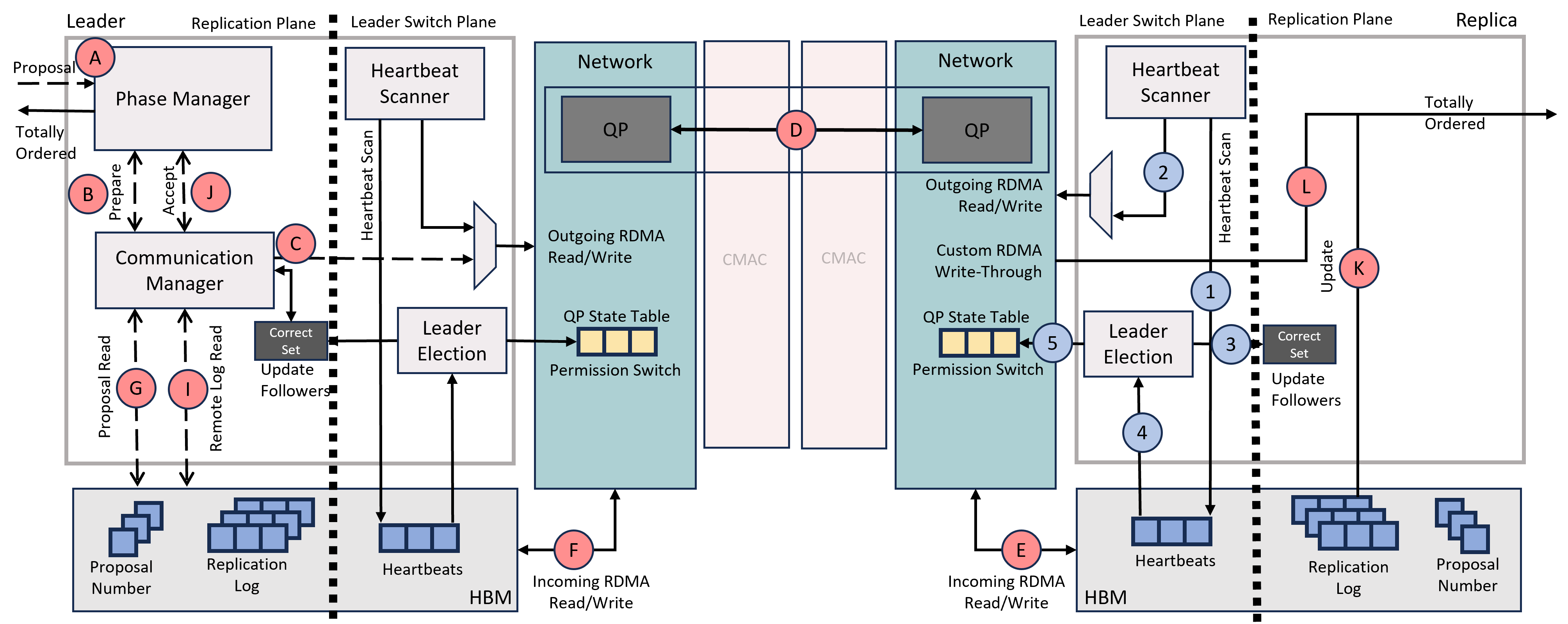}
% \vspace*{-5mm}
\vspace*{-4mm}
\caption{SMR in SafarDB: The solid lines represent the portions of the SMR that operate regardless of leader status and the dashed lines represent the leader-specific operations.}
\label{fig:arch:smr}
\vspace*{-4mm}
\end{figure*}

%\autoref{fig:arch:conflicting} depicts how 
SafarDB implements conflicting transactions using (a) RDMA Write verbs and (b) a new RDMA RPC Write-Through verb.
For example, consider a system with $N$ replicas, and let each replica's data comprise a set $S$.
%\sadoghi{no Movie, describe in an abstract sense, this will be criticized and viewed very negatively.}
% The state is a pair of sets: customers $C$ and movies $M$; Movie has four conflicting methods that form two synchronization groups. 
%\sadoghi{Unless this is supposed to be an example, which should be placed in the Example block, otherwise the algorithm and design should be narrated using non-abstract language.}
% 
Conflicting transactions require synchronization by an SMR protocol (see \autoref{fig:arch:smr}).
Each synchronization group maintains an instance of SMR,
and
each conflicting transaction invokes replication on the SMR of its synchronization group. 
We subsequently describe SMR in \autoref{sec:Mu}.

Each replica allocates a \textit{replication log} in memory. Each replication log entry is a pair of values: a proposal number and an operation. The proposal number ensures total ordering, and an operation consists of an opcode, which identifies the conflicting transaction to execute, and input parameters for the transaction. The replication log is used to buffer committed transactions as well as for recovery.
We consider that the replication log can grow beyond the capacity of FPGA-integrated storage; therefore, it is allocated in memory.

% % Many SMR protocols, including Mu \cite{mu} (see \autoref{sec:Mu}), 
% One replica is designated as the \textit{leader},
% and when the leader fails, a new leader is elected.
% Different synchronization groups may have different elected leaders. 
% % at the same time
% In order to help other replicas when they fail,
% log entries are not discarded when their method calls are invoked.
% % 
% When a failed replica comes back online,
% the leader brings it up-to-date:
% the leader retrieves from the replication log method calls that the replica missed, and (re-)issues them. %\sadoghi{This notion of leader is also first mentioned, again, and we have also not said what consensus protocols we are using; none of these assumptions have explained. Again, the best place to ensure these are included in the preliminary section. For example, are we assuming Paxos, Raft, PBFT, etc? Also, further mentioning whether we can be consensus-agnostic, meaning that any consensus protocols can be employed. But this notion of leader in agnostic model requires more description, so maybe easier to just assume one protocol, and say that it can be extended to consensus protocols, but without loss of generality to simplify the exposition, we rely on a single protocol X.}

% \bigskip
% \noindent
(1) $\mathsf{RDMA\ Write}$: 
In \autoref{fig:arch:conflicting}, RDMA Writes propagate conflicting transactions to remote replicas by appending them to the replication log. 
% (rather than periodically)
% consistently and 
The replica repeatedly polls the log to identify newly-appended transactions as soon as they are written. New transactions are read and invoked via the Dispatcher. 
Accesses to $S$ read the log and execute any conflicting transactions that are present, ensuring that the state is up-to-date. 
%\sadoghi{In DB, log means that log is used for recovery, but here the logs imply buffer or queue, so we need to avoid using log unless we mean recovery log.}

% \bigskip
% \noindent
(2) $\mathsf{RDMA\ RPC\ Write\text{-}Through}$:
In \autoref{fig:arch:conflicting}, conflicting transactions are propagated as RPCs to remote replicas (through $\mathit{c}$) and are simultaneously appended to the replication log (through $\mathit{a}$). 
The RPC does not access the replication log; however, 
the log is updated to support failure recovery.

% \sadoghi{Log is used to maintain the operations that have already been applied or will be applied, but log will not be used as a queue/buffer to determine what the outstanding operations are. So I am not sure how you are using the notion of log, which contradicts what was mentioned earlier.}

% \begin{figure}[t]
% \begin{subfigure}{.47\textwidth}
%   \centering
%   % include first image
%   \includegraphics[width=\linewidth]{figures/arch-new-colors/conflicting-method.png}  
%   \caption{RDMA\_Writes append conflicting calls to the replication log in memory. Like Hamband \cite{hamband}, a polling process consistently and repeatedly accesses the logs to call conflicting methods. $\mathit{query}$ calls and permissibility checks also access the logs to call conflicting methods to ensure that the state is up-to-date before they are invoked.}
%   \label{fig:movie-rdma-write}
% \end{subfigure}
% \begin{subfigure}{.47\textwidth}
%   \centering
%   % include second image
%   \includegraphics[width=\linewidth]{figures/arch-new-colors/conflicting-method-custom.png}  
%   \caption{An RDMA\_RPC-Write-Through verb calls conflicting methods as RPCs and concurrently updates the replication log in memory. The replication log is used exclusively for failure recovery.}
%   \label{fig:movie-rdma-writethrough}
% \end{subfigure}
% \caption{Two network-attached FPGA implementations of conflicting method calls for Movie.}
% \label{fig:movie}
% \Description[]{}
% \end{figure}

% \bigskip
\vspace{1.5ex}
\noindent
\fbox{%
    \parbox{\linewidth}{
%         \textbf{Design Principle \#\todo{NUMBER}}:
        \textbf{Design Principle \# 2}:
        RDMA-based communication requires receiving replicas to access memory, incurring memory access latency. Although buffering and background polling can mitigate this overhead, direct updates enabled by custom RDMA RPC verbs can eliminate the access entirely.
    }
}

\vspace{-2mm}
\subsection{Accelerated Consensus}
\label{sec:Mu}

% \begin{figure}[th]
% \centering
% \includegraphics[width=\linewidth]{figures/arch-new-colors/smr-custom-verb.png}
% \caption{RDMA Write-Through modifications to SMR receiving logic: The red line shows the path of the Totally Ordered data which directly comes from the network, rather than going to memory before being read.} 
% \label{fig:arch:mod-smr}
% \Description[]{}
% \end{figure}

SafarDB provides strong consistency for conflicting transactions through an FPGA-accelerated implementation of Mu \cite{mu} (\autoref{fig:arch:smr}), an RDMA-based SMR protocol for
% We consider a 
the crash-failure model:
a replica either crashes and stops, or executes correctly.
The SMR picks a replica as the leader, with the remaining replicas designated as followers. Functionality is partitioned between two planes, 
a \textit{Replication Plane} and a \textit{Leader Switch Plane}, 
described in the following subsections.

\textbf{Replication Plane. \ }
% \subsubsection{Replication Plane}
% \hfill \break
In the $\mathit{Propose}$ phase, a new leader confirms the follower list by obtaining write permissions from a quorum (majority) of replicas, and then proposes a new transaction (at A). 
After transitioning into the $\mathit{Prepare}$ phase (at B), the leader $\mathsf{RDMA\ reads}$ the latest proposal numbers from followers (through C, D, E, F and G). The leader $\mathsf{RDMA\ writes}$ the next highest proposal number to the followers (through C, D, E).
It then $\mathsf{RDMA\ reads}$ each follower's log slot that it intends to write (through C, D, E, F and I). If all followers' log slots are empty, the leader proceeds to the $\mathit{Accept}$ phase with its own transaction. If the leader reads a non-empty log slot, then that follower previously committed a different conflicting transaction; if so, the leader adopts that transaction instead. If multiple follower slots are non-empty, the leader adopts the transaction from the slot with the highest proposal number.

% \sadoghi{
% %The notion of leader and follower, again, is an essential concept that has not been discussed or defined before; it is assumed that readers know what we mean. Again, this is introduced abruptly; all of these formulations should have been defined clearly in the preliminary and problem setting. Also unclear what the difference is between follower and replica. 
% Are we relying on a primary-backup model?}
% \py{Not sure how to address this comment.}

% \sadoghi{We can say RDMA writes, not need to say Writes.}

In the $\mathit{Accept}$ phase (at J), the leader executes the conflicting transaction and RDMA writes it to the followers' replication logs (through C, D, E). Followers read their logs (at K) and execute the transaction. If the originally proposed transaction was executed, the leader restarts from the first phase to prepare for the next conflicting transaction; if not, the leader repeats the $\mathit{Prepare}$ phase for the originally proposed transaction. 
% If a leader commits conflicting methods consecutively without encountering a non-empty log slot, a \textit{fast path} is triggered which skips checking proposal numbers or log slots.
% \ml{That's not true. The remote operations are the same in the two cases. We don't have a fast-path in the mu paper either.}

With custom RDMA verbs, the final RDMA write in the $\mathit{Accept}$ phase is replaced with RDMA RPC write-through.
Thus, the application states for the followers are directly updated from the network (at L).
The followers no longer need to read their logs (at K),
although the logs are updated (through C, D, E), similarly to a traditional RDMA write.

% \autoref{fig:arch:mod-smr} depict the changes to the replication plane that occur when utilizing \textbf{RDMA\_RPC\_Write-Through}. The receiving side replication plane no longer needs to read the log from memory and the network can directly pass the totally ordered operation to the application. 

% \noindent\fbox{%
%     \parbox{\linewidth}{
%         \textbf{Design Principle \#\todo{NUMBER}}:
%         By developing the SMR protocol on the FPGA with dedicated hardware as well as near-network processing consensus can be achieved much faster.
%     }
% }

\textbf{Leader Switch Plane. \ }
% \subsubsection{Leader Switch Plane}
% \label{sec:leader-switch}
% \hfill \break
The leader of SMR may fail. 
In order to maintain liveness, replicas need to detect leader failure when it occurs and elect a new leader. Each replica stores an array of \textit{heartbeats} in memory, one for each other replica.
% one for each other process. 
Each heartbeat is an RDMA-exposed incrementing counter
that allows the replica to express its liveness to a remote replica.

% \sadoghi{The notion of liveness and safety, first mentioned, needs to be defined clearly in the preliminary problem setting section.}

Each replica's \textit{Heartbeat Scanner} periodically increments its heartbeats
(\autoref{fig:arch:smr} at 1), and 
RDMA Reads heartbeats of remote replicas (through 2, D, F, E, 4). 
If a remote replica's heartbeat is constant after several reads, 
then the replica is assumed to have failed,
and it is removed from the set of correct replicas (at 3). 

% The \textit{Leader Election} module regularly reads each replica's heartbeats; 

If a leader is thought to have failed, then a new leader is elected:
% then deterministically selects a new leader
% deterministically 
the live replica with the smallest ID.
% an election is held. 
% (see Ref. \cite{mu} for details)
Leader operation relies on RDMA QP connections (\autoref{sec:cpu-rdma}). The QP state is \textit{open} or \textit{closed}; RDMA Writes to a closed QP fail. Each follower has one open QP that grants write permission to a leader. If a follower determines that a leader has failed, it closes its QP with the leader, and then opens a new QP with the new leader (at 5). 

Changing leaders in this manner is called a \textit{permission switch}. The fastest way to switch permissions is to change QP access flags; however, doing so while RDMA verbs are in flight may induce a QP error. Software Mu \cite{mu} implements a fast path when no error occurs and a slow path that detects and recovers from errors; only the slow path modifies the QP state. In a traditional RDMA network (\autoref{sec:cpu-rdma}), modifying RDMA write permissions takes hundreds of microseconds, accounting for \textasciitilde$30\%$ of failover time. The developers of Mu attribute this overhead to the lack of hardware optimization in the RNIC and its interface \cite{mu}, which points to PCIe round-trip latencies. In our network-attached FPGA (\autoref{sec:rdma-fpga}), the SMR kernel can access the QP state directly.

% \sadoghi{We are now going deep into the consensus protocol. The moment we get into leader election, it is essential to clarify whether we are proposing a new protocol or we are providing a hardware-assisted method to implement leader selection in the existing protocol X.}

% \bigskip
\vspace{1.5ex}
\noindent
\fbox{%
    \parbox{\linewidth}{
%         \textbf{Design Principle \#\todo{NUMBER}}:
        \textbf{Design Principle \# 3}:
        Changing RDMA permissions can take hundreds of microseconds. By colocating Mu and RNIC hardware on a single FPGA chip and exposing internal RNIC registers to the SMR protocol, permission-switch latency can be reduced to the order of nanoseconds.
    }
}

%In Mu, the CPU communicates these adjustments to the RNIC over PCIe. Since SafarDB adheres to \textbf{Architectural Principle 1}, the application avoids PCIe communication overhead. By using a soft-RNIC, internal signals are exposed allowing for optimization and customization to RDMA functionality. SafarDB adds a direct connection from the RDT application kernel directly into the RNIC to modify the QP state table, enabling fast permission switching. 

%\section{Implementation of Mu}

% \subsection{hybrid \todo{NEW}}

% \noindent\fbox{%
%     \parbox{\linewidth}{
%         \textbf{Design Principle \#\todo{NUMBER}}:
%         By involving CPU host memory, SafarDB can run larger applications where the state doesn't entirely fit in FPGA memory, while maintaining the accelerated replication provided by the FPGA.
%     }
% }
 % Old

\section{Evaluation}
\label{sec:eval}
%In this section, we evaluate the CRDTs and WRDTs of SafarDB in comparison with Hamband~\cite{hamband} and Waverunner~\cite{waverunner} in FPGA-only mode. We then evaluate SafarDB in hybrid mode using the YCSB and SmallBank applications~\cite{PolyphenySimpleClient}. In this section, we answer the following questions:

Our evaluation aims to answer the following:

\begin{enumerate}[label=Q\arabic*, ref=Q\arabic*,leftmargin=*]
    \item\label{q:pci-elimination}
    \textcolor{black}{(PCIe): How does eliminating PCIe communication affect SafarDB’s performance?}

    \item\label{q:memory-access}
    (Memory): How does mitigating memory accesses affect SafarDB’s performance?

    \item\label{q:consistency-selection}
    \textcolor{black}{(Consistency): How does consistency selection affect SafarDB’s performance?}

    \item\label{q:fpga-scaling}
    (Scaling): How does SafarDB (in FPGA-only mode) scale compared to existing systems?

    \item\label{q:update-percentage}
    (Updates): How does SafarDB’s performance change with varying update percentages?

    \item\label{q:crash-faults}
    (Fault): What are the effects of crash faults on SafarDB’s performance compared to existing systems?

    \item\label{q:permission-switch}
    (Permission Switch): Does SafarDB benefit from an improved permission switch during crash-fault recovery?

    \item\label{q:hybrid-mode}
    (Hybrid): What are the trade-offs of involving the host CPU in SafarDB’s hybrid mode?

    \item\label{q:energy}
    (Power): What are the power implications of SafarDB?
\end{enumerate}

\textbf{Setup.}
SafarDB was deployed on eight AMD Xilinx Alveo U280s connected via 100GbE switches, with host Intel(R) Xeon(R) Gold 6226R CPUs @ 2.90GHz at the Open Cloud Testbed (OCT) \cite{oct}. Each experiment runs for a total of 4 million operations with varying update percentages, numbers of nodes, crash faults, and workload distributions, depending on the experiment.

% \noindent
\textbf{Implementation.}
We generate FPGA bitstreams for SafarDB use cases using Vivado 2023.2 and develop them with Vitis HLS 2023.2, with intermediate IPs for the RDMA network stack generated using Vitis HLS 2020.2 and SystemVerilog. The RDMA stack is based on StRoM \cite{StRoM}. The host CPU application is developed in C++ (version 11.4.0).

% \noindent
\textbf{Baselines.}
We compare SafarDB against the following systems:
\begin{enumerate}[leftmargin=*]
    \item \textcolor{black}{Hamband \cite{hamband}: Software RDT implementation utilizing Mu \cite{mu} for consensus. We evaluate two deployments. For the controlled comparison, our port uses 100 Gbps hardware RoCEv2 RNICs in three- and four-replica configurations on the same OCT hosts used by SafarDB. For the scalability comparison, the original InfiniBand version runs on up to 8 Sapphire Rapids nodes at Texas A\&M's HPRC ACES cluster containing dual Intel Xeon Platinum 8468 CPUs with 512 GB DDR5 memory, an NVIDIA Mellanox NDR200 InfiniBand RNIC (200Gbps), and 400Gbps NDR400 InfiniBand network switches. \textcolor{black}{Each Hamband replica is pinned to four physical CPU cores in both deployments. Both Hamband deployments are compiled with GCC/G++ 8.3.0.}}
    \item Waverunner \cite{waverunner}: Network-attached FPGA implementation of RAFT \cite{raft}; deployed on OCT with the same configuration as SafarDB, 
%     limited to 
    with 3 nodes due to Waverunner's implementation.
    \item Caribou \cite{caribou}: Network-attached FPGA replicated key-value storage system. We ported the storage path to the OCT U280 setup using EasyNet TCP/IP \cite{easynet} and U280 HBM, and configured three replicas with one leader and two followers. PUTs are routed to the leader for ordering and replication, while GETs can be served by any replica after it has received and applied the replicated update.
    \item P4xos \cite{p4xos}: Programmable-network Paxos service. We ported the packet-oriented Paxos datapath to the OCT U280 setup using UDP/EasyNet \cite{easynet}, with one proposer/coordinator FPGA and two acceptor/learner FPGAs. In this P4xos/LevelDB-style baseline, client GET and PUT requests are submitted through the proposer/coordinator as Paxos values; the key-value state machine replies after a learner observes a quorum of acceptor votes and delivers the chosen request.
    % \todo{Is majority-delivery path a term in that paper? Also used in experiments section}
\end{enumerate}

% \noindent
\textbf{Workloads.} Our evaluation uses a microbenchmark suite consisting of CRDTs (\autoref{tab:crdt-methods}) \cite{shapiro11} and WRDTs (\autoref{tab:wrdt-methods}) \cite{hamsaz,hamband}, 
% along with 
and
two widely used benchmarks: YCSB \cite{ycsb} and SmallBank \cite{oltp}.

\textbf{Number of clients.} Unless otherwise noted, an experiment with $N$ replicas uses $N$ logical clients, with one client assigned to each replica; e.g., a three-replica experiment uses three clients.

%\todo{CRDTs}
% \noindent
CRDTs:
% \begin{enumerate}[leftmargin=*]
% \item 
(1) PN-Counter: A positive-negative counter that supports increment and decrement operations. A PN-Counter comprises two G-Counters: one for increments and one for decrements.
% \item 
(2) LWW-Register: A last-writer-wins register that supports assigning values to the register. Unique timestamps are associated with each assignment, ensuring a total order of operations. The register always retains the most recently written value.
% \item 
(3) G-Set: A grow-only set that supports addition, but not removal, of elements.
% \item 
(4) PN-Set: A set where a counter is associated with each element: addition increments the counter, and removal decrements the counter. An element is present if its counter is positive. 
% \item
(5) 2P-Set: A two-phase set that supports addition and removal of elements: once removed, an element cannot be reinserted into the set. A 2P-Set is typically implemented using two G-Sets.
% \end{enumerate}

% \noindent
WRDTs:
% \begin{enumerate}[leftmargin=*]
% \item 
(1) Account: A small-scale banking system which allows for depositing, withdrawing, and querying a single account.
% \item 
(2) Courseware: A school management system that allows courses and students to be added and deleted, and students to enroll in courses.
% \item 
(3) Project: A project management system that allows for projects and employees to be added, deleted, and employees to be assigned to projects.
% \item 
(4) Movie: A movie theater ticketing system that manages a set of customers and movies.
% \item
(5) Auction: An online auction system based on RUBis \cite{rubis}.
% \end{enumerate}
\begin{figure}[t]
\vspace{-10mm}
\centering
\includegraphics[width=\linewidth]{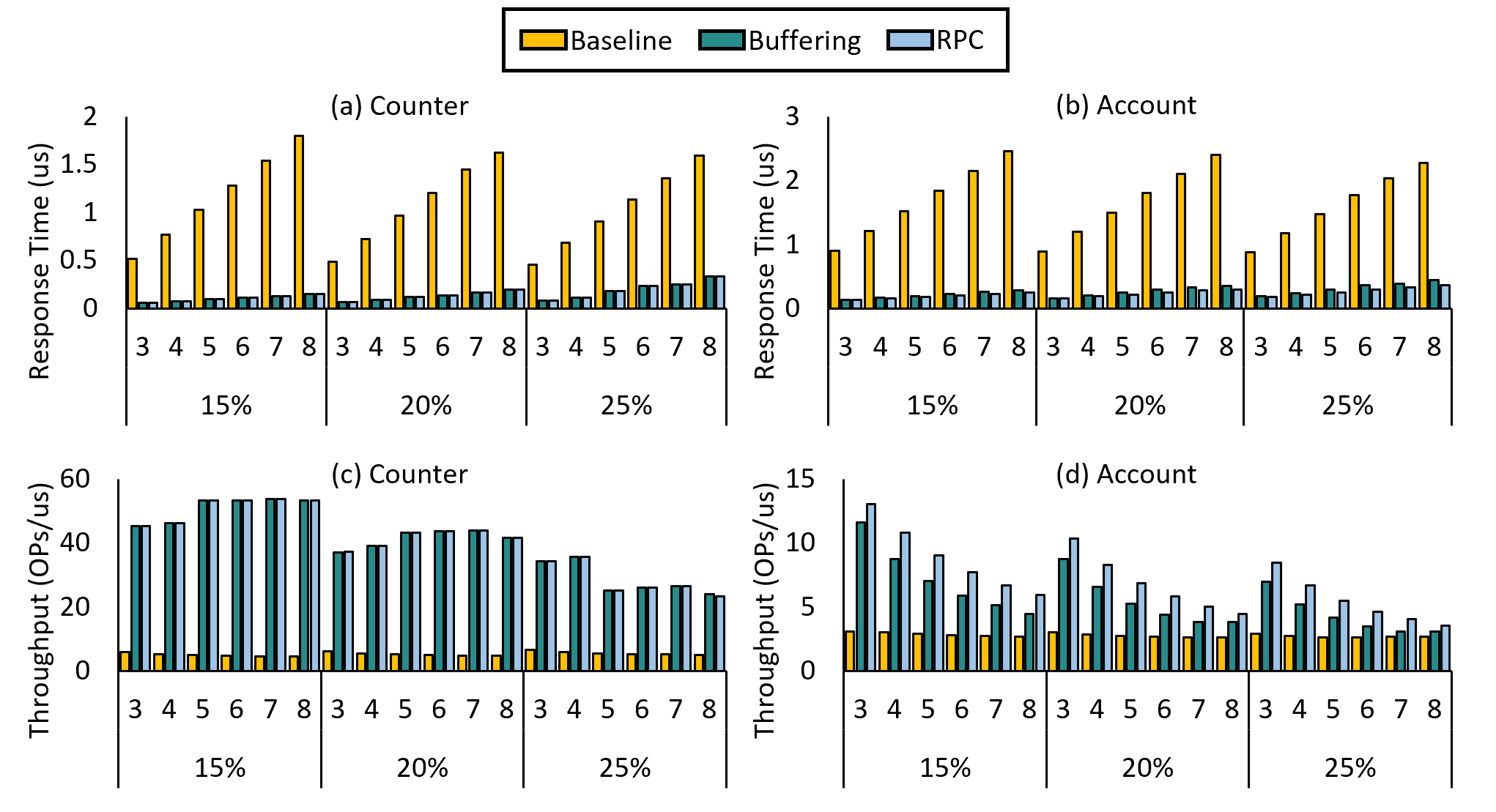}
\vspace*{-6mm}
\caption{Response time and throughput for one CRDT and one WRDT for three reducible transaction implementations.} 
\label{fig:results:custom-verbs:reducible}
\vspace*{-5mm}
\end{figure}

% \noindent
\textbf{Metrics.} We focus on three metrics:
% \begin{enumerate}[leftmargin=*]
% 
% \item 
(1) \textit{Response Time ($\mu s$)}: The average time from when a client issues an operation until it receives the corresponding response.
% \item 
(2) \textit{Throughput ($OPs/\mu s$)}: The maximum number of operations per microsecond which the system completes.
We report aggregate system throughput: the total number of client operations completed across all logical clients during the experiment divided by the total execution time. Thus, the reported value is a system-level throughput, not a per-client rate.
\textcolor{black}{Aggregate CRDT/WRDT results average per-configuration ratios equally across five applications, update percentages of 15\%, 20\%, and 25\%, and 3--8 replicas (90 configurations). Total logical application state per replica ranges from 4~B to 800~KB for CRDTs and from 12~B to approximately 1~MB for WRDTs; Appendix Table~\ref{tab:application-state-datasets} reports each configuration.}
% \item 
(3) \textit{Power Consumption ($W$)}: FPGA power consumption is measured following AMD Xilinx Documentation \cite{xilinx-docs}. CPU and I/O power consumption was measured with a combination of PowerTop \cite{powertop} and Scaphandre \cite{scaphandre}. Detailed power results are reported in \autoref{fig:app:power-consumption} in the appendix.
% \end{enumerate}

\textcolor{black}{\textbf{Controlled and cross-cluster comparisons.} Four OCT hosts provide both Alveo U280 cards for SafarDB and CPU-accessible 100 Gbps RoCEv2 RNICs for Hamband. We therefore ported Hamband to RoCEv2 and ran both systems in the matched OCT configurations. The port preserves Hamband's protocol and workload semantics; its changes are confined to transport-specific addressing, queue-completion handling, and device selection. On OCT, Hamband uses CPU-hosted execution over RoCEv2, whereas SafarDB uses its FPGA-resident RoCEv2 path. Because only four compatible hosts were available, the controlled experiment is limited to three and four replicas. To study scaling through eight replicas, we additionally report Hamband's original InfiniBand deployment on ACES. \autoref{tab:cluster-comparison} in the appendix summarizes the two clusters.}

\begin{figure}[t]
\vspace{-10mm}
\centering
\vspace*{-3mm}
\includegraphics[width=\linewidth]{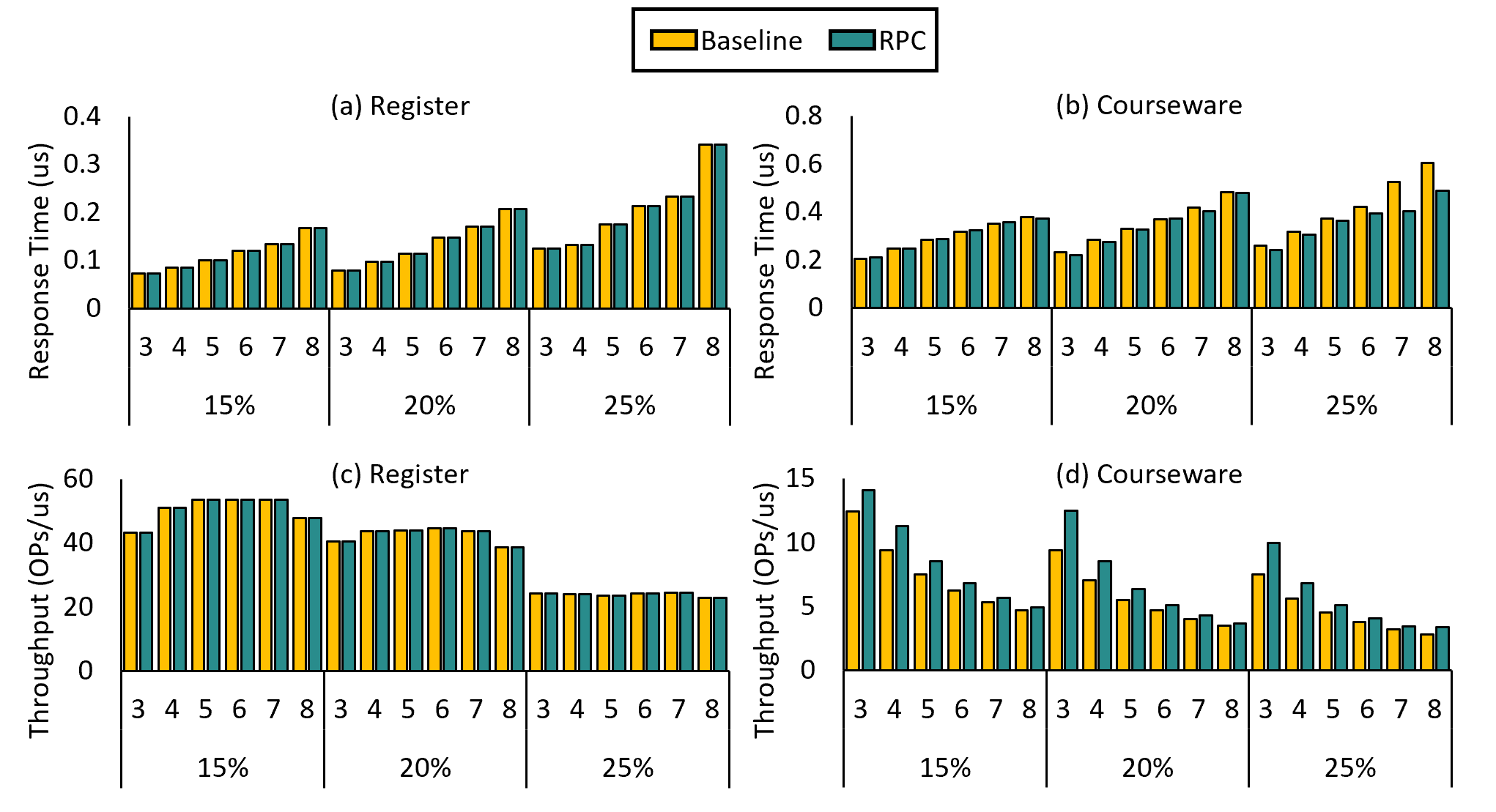}
\vspace*{-5mm}
\caption{Response time and throughput for one CRDT and one WRDT for two irreducible transaction implementations.}
\label{fig:results:custom-verbs:irreducible}
% \vspace{-8mm}
\end{figure}

\subsection{Ablation Study}

\textcolor{black}{To explore and justify SafarDB's design choices, we provide a detailed ablation study exploring the different design areas: memory access, PCIe elimination, FPGA-resident execution and consistency selection. }

\begin{figure}[t]
% \vspace{-9mm}
\centering
\vspace*{-3mm}
\includegraphics[width=\linewidth]{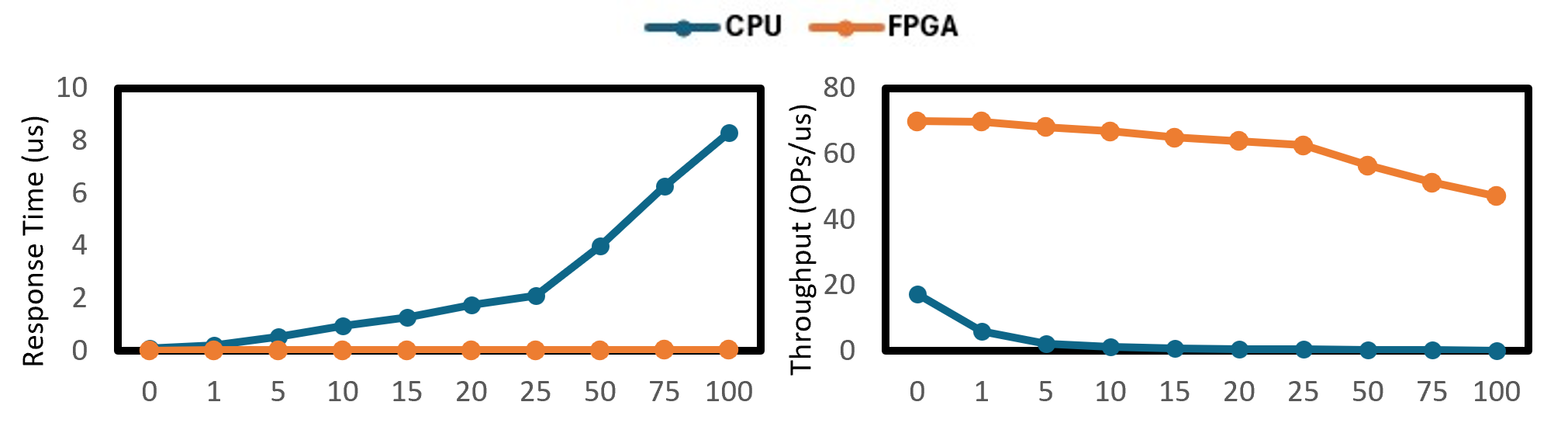}
\vspace*{-8mm}
\caption{Response time and throughput vs update percentage for SmallBank running SafarDB's CPU and FPGA-only configurations.} 
\label{fig:results:ablation:pcie}
\vspace{-5mm}
\end{figure}

\textcolor{black}{\textbf{PCIe Elimination and FPGA-resident Execution. (Design Principle 1)} To answer \ref{q:pci-elimination}, we conduct two experiments measuring the performance of SafarDB in two distinct configurations. In each experiment, we measure the response time and throughput of SafarDB running SmallBank operations in a CPU-only and FPGA-only configuration. Each configuration runs on two replicas and executes a total of 4 million SmallBank operations while varying the update percentage from 0\% to 100\%.
In the \textit{CPU-only} configuration, all operations are issued from the CPU-resident application and routed through PCIe to the FPGA network stack before transmission. The FPGA is present and handles all network operations, but the application logic and transaction path originate on the host CPU, crossing PCIe on every operation.
In the \textit{FPGA-only} configuration, all operations are issued directly from the FPGA-resident application via AXI Stream to the network stack. No PCIe crossing occurs on the critical path; the CPU is used only for initialization.
These two configurations are identical in hardware, network stack, replication protocol, and workload — the only difference is whether the application-to-NIC communication traverses PCIe or the on-chip AXI interconnect. The performance difference between them therefore directly quantifies the benefit of PCIe elimination supporting \textbf{Design Principle \# 1}, independent of all other design choices.
\autoref{fig:results:ablation:pcie} reports response time and throughput for SmallBank across update percentages from 0\% to 100\%, comparing CPU-only execution (all operations routed through PCIe) against FPGA-only execution (all operations routed through AXI). At 0\% updates, both configurations perform comparably, since read-only operations are 
served locally without triggering RDMA replication and the PCIe path is not on the critical path. As the update percentage increases, the CPU configuration degrades sharply — reaching $\sim$8.4\,$\mu$s response time and near-zero throughput at 100\% updates — while the FPGA configuration remains stable at $\sim$0.1\,$\mu$s response time and 
$\sim$47\,OPs/$\mu$s throughput. The divergence grows with update percentage because each write operation triggers RDMA: in the CPU configuration this crosses PCIe on every operation, whereas in 
the FPGA configuration the application and network stack communicate via on-chip AXI with no PCIe involvement. These results directly quantify the benefit of Design Principle~1: eliminating PCIe from the replication critical path while holding all other design choices constant reduces response time by up to $\sim$84$\times$ at write-heavy workloads.}

\textcolor{black}{\textbf{Memory Access Ablation (Design Principle~2).}} To answer \ref{q:memory-access}, we conduct five experiments in which we measure the performance of SafarDB. \textcolor{black}{These experiments serve as a direct ablation of Design Principle~2: the baseline configuration uses RDMA Write with memory polling, representing the system without any memory-access mitigation; the buffering configuration adds background polling to partially hide memory access latency; and the RPC configuration eliminates memory accesses entirely. The progression from baseline to buffering to RPC while holding all other design choices constant, therefore directly quantifies the individual contribution of each memory-access optimization.} In each experiment, we measure the response time and throughput of SafarDB when running CRDT or WRDT operations to see how memory accesses affect different transactions. The workloads run for 4 million operations while varying the number of nodes in the system and the percentage of operations that are updates versus reads. For example, in \autoref{fig:results:custom-verbs:reducible}a, the PN-Counter runs across 3--8 nodes, with 15\%, 20\%, or 25\% of the 4 million operations consisting of updates; the remaining operations are $\mathsf{query}$ transactions, a read-only operation that retrieves the application state, \eg returning the balance of a bank account.

The first pair of experiments (\autoref{fig:results:custom-verbs:reducible}) measures the impact of utilizing buffering and RDMA RPC to implement reducible transactions (\autoref{ssec:reducible-methods}) on the PN-Counter and Account. For PN-Counter, the buffered and RDMA RPC implementations exhibit $8\times$ lower response time and $7.8\times$ higher throughput than the baseline buffer-less design because each $\mathsf{query}$ and permissibility check does not incur memory latency. The buffered implementation's polling mechanism hides memory-access latency from the application's critical path, and RDMA RPC eliminates these accesses altogether. \textcolor{black}{This demonstrates that buffering captures a substantial fraction of the available gain, but RPC is needed to eliminate the residual memory accesses that background polling cannot hide.} Buffering and RPC perform comparably for Counter (\autoref{fig:results:custom-verbs:reducible}a and c).

For Account, buffering and RDMA RPC reduce response time by $6.3\times$ and $7\times$, while increasing throughput by $2\times$ and $2.4\times$, respectively (\autoref{fig:results:custom-verbs:reducible}b and d). WRDT conflicting transactions require an SMR protocol (\autoref{sec:wrdts}), which designates one replica as a leader (\autoref{sec:Mu}); the leader's execution time is much greater than that of the followers. We observe that buffering cannot fully hide memory accesses at the leader, but RDMA RPC eliminates them and reduces execution time. For WRDTs, throughput depends on the leader's execution time (noticeable impact), whereas response time is averaged across all replicas (marginal impact). \textcolor{black}{The gap between buffering and RPC is therefore larger for WRDTs than for CRDTs, because the leader's synchronous memory accesses on the conflicting transaction path cannot be hidden by background polling — only eliminated by RPC. This further isolates the contribution of memory-access elimination from that of background scheduling.}

The second pair of experiments (\autoref{fig:results:custom-verbs:irreducible}) measures the impact of utilizing RDMA RPC to implement irreducible transactions on the LWW-Register and Courseware. The results are similar to \autoref{fig:results:custom-verbs:reducible}. For the LWW register, all the memory accesses to the in-memory queues are hidden in the background, eliminating any possible advantage for RDMA RPC; this occurs because all replicas in a CRDT are peers (\autoref{fig:results:custom-verbs:irreducible} a and c). \textcolor{black}{This is an important control result: when background polling is sufficient to fully hide memory accesses, RPC provides no additional benefit, confirming that RPC's advantage is specifically attributable to eliminating accesses that polling cannot hide.}
% rather than to any other implementation difference

\begin{figure}[t]
\centering
\vspace*{-10mm}
\includegraphics[width=\linewidth]{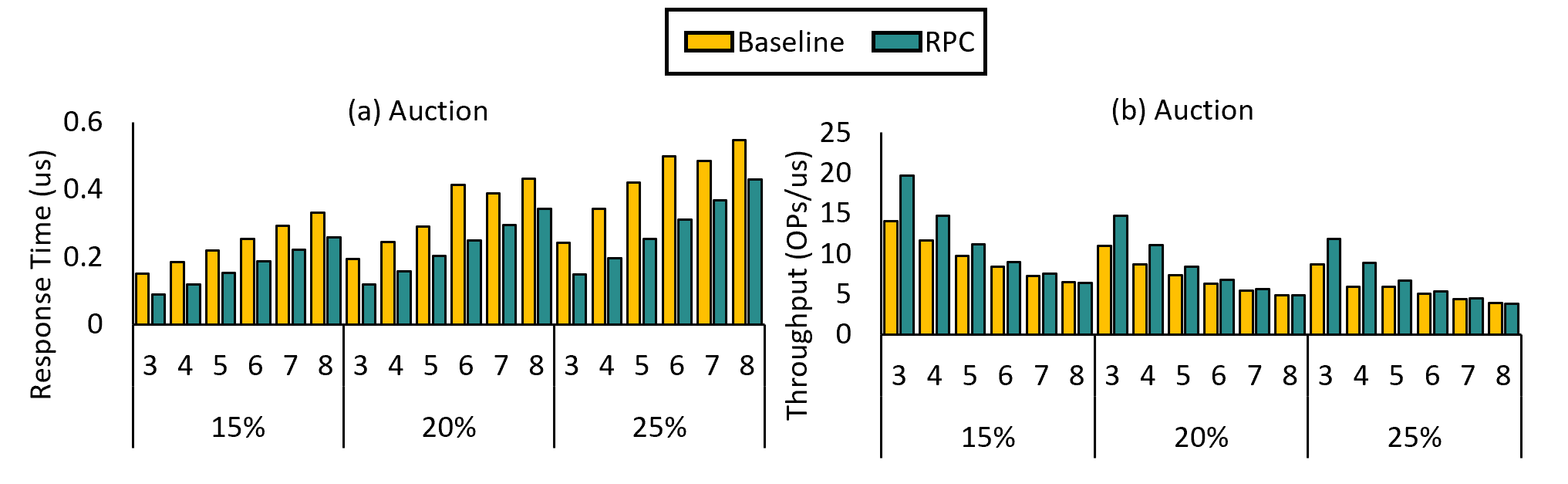}
\vspace*{-5mm}
\caption{Response time \& throughput for one WRDT Auction (\autoref{tab:wrdt-methods}) for two conflicting transaction implementations.} 
\label{fig:results:custom-verbs:conflicting}
\vspace*{-5mm}
\end{figure}

For Courseware, RPC achieves $1.04\times$ lower response times than the baseline, most visibly at 25\% write percentage and 6--8 nodes. While RPC achieves $1.14\times$ higher throughput than the baseline, the advantage narrows as the number of replicas increases. These trends occur because conflicting transactions (described in the next section) access memory more often than non-conflicting transactions at the leader, so the baseline has fewer opportunities to hide memory access latencies. With a fixed number of transactions to execute, each replica executes fewer transactions as the number of replicas increases, thereby suppressing RDMA RPC's throughput advantage at higher node counts.

The final experiment (\autoref{fig:results:custom-verbs:conflicting}) measures the performance impact of utilizing RDMA RPC Write-Through for conflicting transactions. When implemented with RDMA write, conflicting transactions must access memory for $\mathsf{query}$ transactions and permissibility checks before executing. Auction exacerbates these accesses because it has three synchronization groups and three replication logs (one per group). RDMA RPC Write-Through updates the WRDT state immediately, ensuring it is up to date before subsequent $\mathsf{query}$ transactions and permissibility checks, removing the need for them to access the log. RDMA RPC Write-Through on average lowers response time by $1.5\times$ and raises throughput by $1.1\times$ across all write percentages and numbers of nodes in the network, but only achieves higher throughput at low node counts. At lower node counts, followers benefit from fewer memory accesses, but at higher node counts, coordination (see \autoref{sec:Mu}) emerges as the bottleneck, eliminating the advantage over RDMA write. \textcolor{black}{Taken together, the five experiments in this section ablate Design Principle~2 across all three transaction categories: reducible, irreducible, and conflicting. In each case, the performance gap between baseline, buffering, and RPC is directly attributable to memory-access latency, since all other aspects of the system — hardware, network stack, replication protocol, and workload — are held constant across configurations.} These experiments answer \ref{q:memory-access} by showing that adding buffer/polling or RDMA RPC to the implementation of the different transactions improves SafarDB's overall performance.

\begin{figure}[t]
\centering
\vspace*{-10mm}
\includegraphics[width=\linewidth]{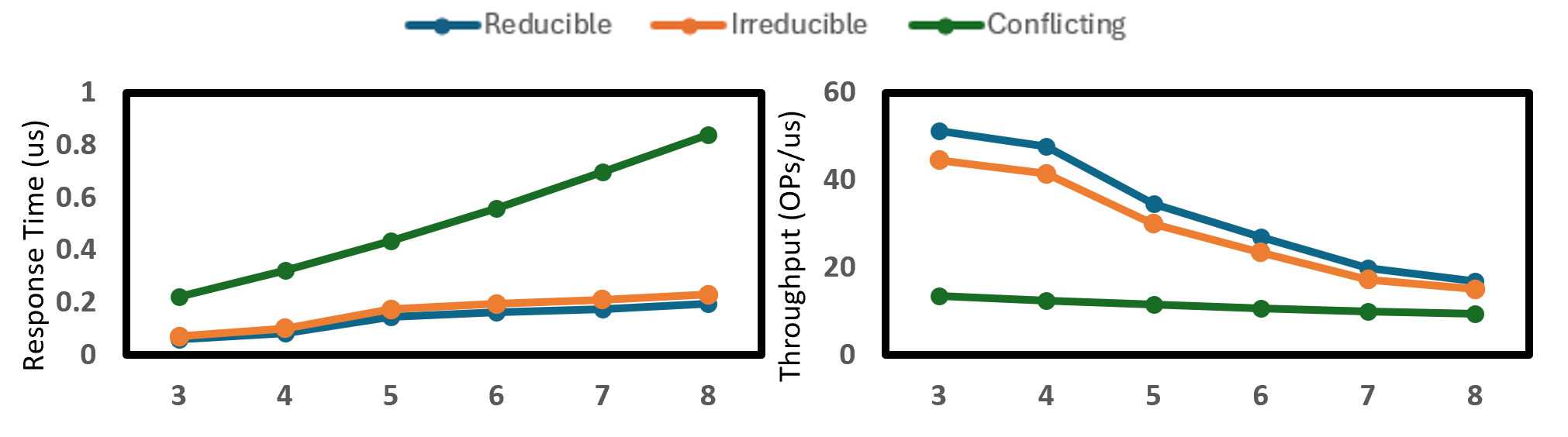}
\caption{\textcolor{black}{Response time \& throughput for SmallBank's deposit transaction implemented as different transaction types.}} 
\label{fig:results:ablation:deposit}
\vspace*{-5mm}
\end{figure}

\textcolor{black}{\textbf{RDT-Aware Consistency Selection Ablation} To isolate the benefit of RDT-aware consistency selection, we conduct a controlled ablation in which the same SmallBank deposit transaction is implemented under three different consistency configurations: reducible, irreducible, and conflicting. For each configuration, the three to eight-replica experiment executes a total of 4 million deposit operations (100\% updates). In the \textit{reducible} configuration,
deposit operations are treated as summarizable and conflict-free — multiple deposits are aggregated locally and propagated as a single RDMA RPC, bypassing SMR entirely. In the \textit{irreducible} configuration, deposit operations are treated as conflict-free but not summarizable — each deposit is propagated immediately via RDMA RPC without aggregation, still bypassing 
SMR. In the \textit{conflicting} configuration, deposit operations are forced through the SMR module for total ordering, regardless of whether ordering is semantically required.
\autoref{fig:results:ablation:deposit} reports the results. The reducible configuration achieves the lowest response time and highest throughput, since deposits are aggregated before propagation and SMR is never invoked. The irreducible configuration incurs higher response time than reducible due to the absence of aggregation, but remains substantially faster than conflicting since SMR is still bypassed. The conflicting configuration is the slowest, as every deposit must complete a full round of SMR consensus before committing. These results demonstrate that routing non-conflicting transactions away from SMR is one contributor to SafarDB's end-to-end performance.
}

\textcolor{black}{These ablations isolate individual sources of improvement, while Section~4.2 evaluates their combined effect in the complete system.}

% \autoref{fig:results:rpc:auction} reports response time and throughput for one representative WRDT (Auction), implementing conflicting method calls using RDMA Write and  RDMA RPC Write-Through. Subsection \ref{ssec:system-setup} describes the experimental setup: we vary the number of nodes in the network from 3-8 and consider workload write percentages of 15\%, 20\%, and 25\%. %These experiment indicate that  \textbf{RDMA\_RPC\_Write-Through} is the most efficient way to implement conflicting method calls.  
% \vspace{-6mm}
\subsection{Scalability}
To answer \ref{q:fpga-scaling} and \ref{q:update-percentage}, we evaluate SafarDB against Hamband and three FPGA/programmable-network replication baselines: Waverunner, Caribou, and P4xos.

\textcolor{black}{\textbf{Controlled same-cluster comparison.} We first compare SafarDB and Hamband on the same OCT hosts before evaluating scalability across clusters.}

\begin{figure}[t]
\centering
\vspace*{-10mm}
\includegraphics[width=\linewidth]{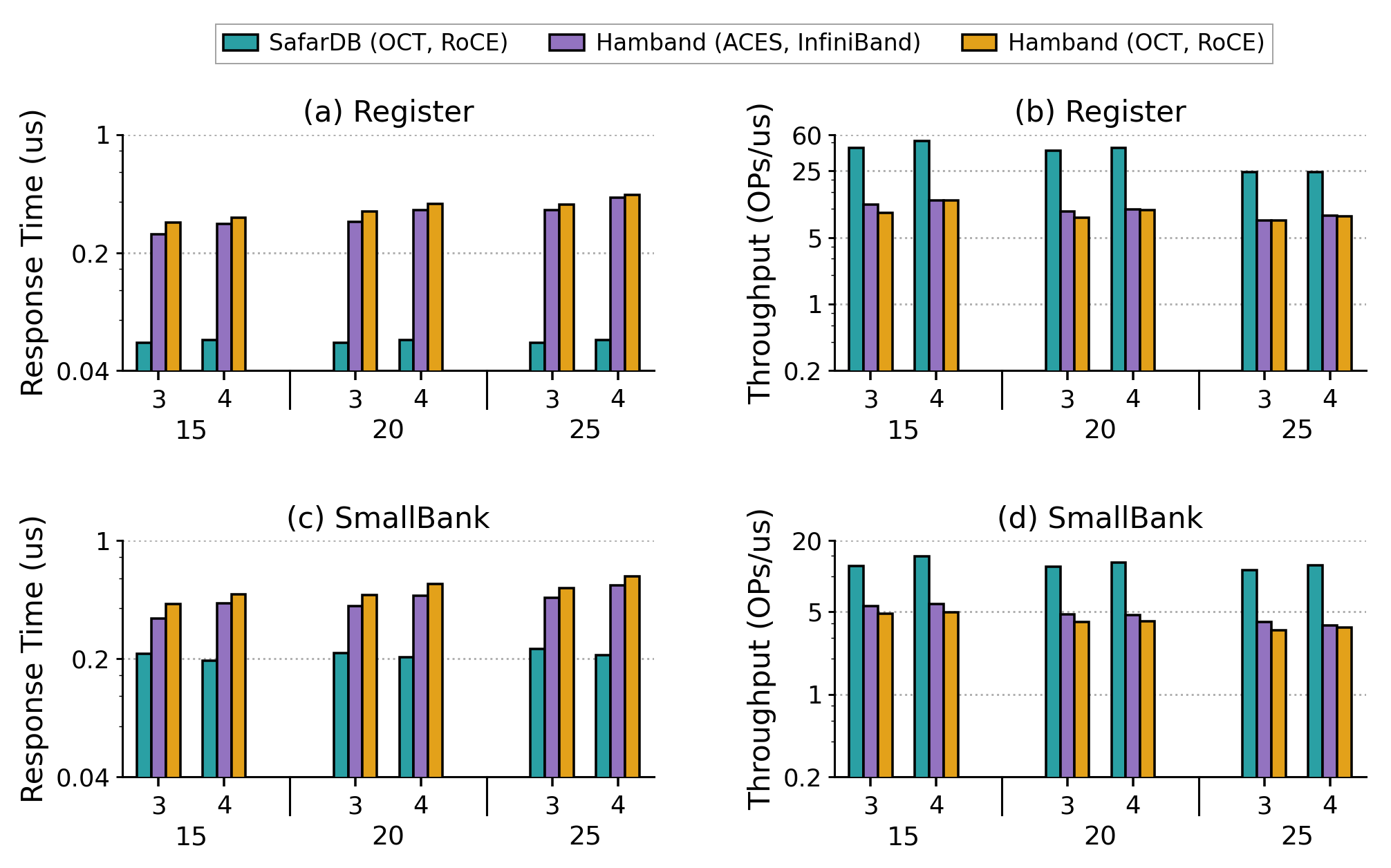}
\caption{\textcolor{black}{SafarDB on the OCT cluster with RoCE vs. Hamband on the ACES cluster with InfiniBand, and Hamband on the OCT cluster with RoCE (on 3 and 4 replicas at 15\%, 20\%, and 25\% updates). Panels (a) and (b) show Register response time and throughput; panels (c) and (d) show SmallBank response time and throughput. The y-axes use base-5 logarithmic scales; group labels show update percentages.}}
\label{fig:results:hybrid:roce}
\vspace*{-5mm}
\end{figure}

\textcolor{black}{The controlled OCT results in \autoref{fig:results:hybrid:roce} use SafarDB's FPGA-only configuration. To represent both consistency paths, we report Register as a conflict-free CRDT and SmallBank as a workload with conflicting transactions. Averaged across the matched OCT configurations at 15\%, 20\%, and 25\% updates, SafarDB provides $6.1\times$ lower response time and $4.1\times$ higher throughput than Hamband-RoCE for Register, and $2.4\times$ lower response time and $3.0\times$ higher throughput for SmallBank. Additional controlled comparison results are reported in \autoref{fig:app:crdts-roce} and \autoref{fig:app:wrdts-roce} in the appendix.}

\textcolor{black}{\textbf{Scaling beyond four replicas.} Because only four OCT hosts provide both the FPGA and RoCEv2 resources required for the controlled experiment, the same-cluster comparison cannot extend beyond four replicas. Averaged across the matched OCT configurations at 15\%, 20\%, and 25\% updates, Hamband's ACES InfiniBand deployment provides $1.1\times$ lower response time and $1.1\times$ higher throughput than its OCT RoCEv2 deployment for Register, and $1.2\times$ lower response time and $1.1\times$ higher throughput for SmallBank. We note that these differences compare the complete deployments rather than isolating InfiniBand and RoCE: the clusters differ in their CPUs, memory systems, RNICs, and network fabrics. We therefore use Hamband's ACES deployment to study scalability through eight replicas.}

\begin{figure*}[t]
\vspace{-7mm}
\centering
\includegraphics[width=\linewidth]{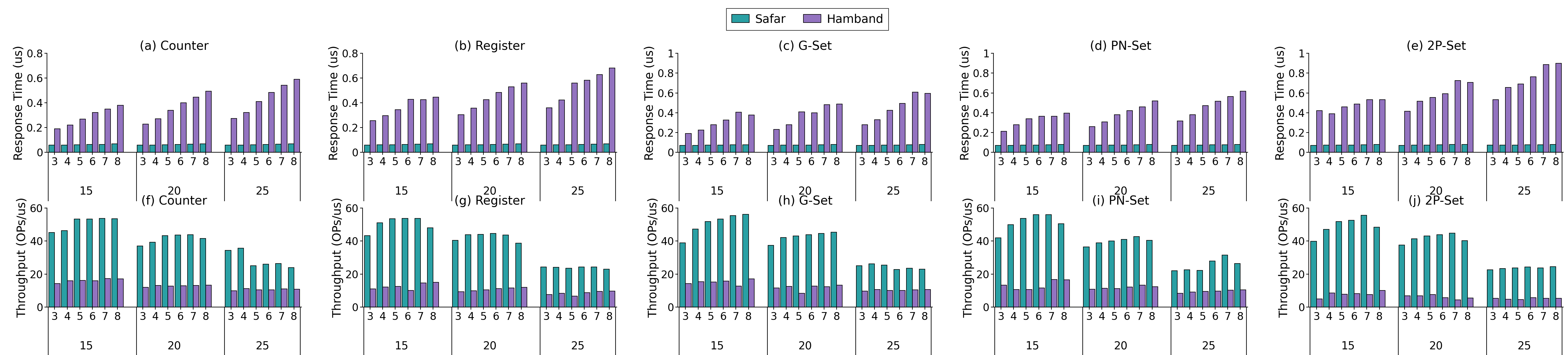}
\vspace*{-7mm}
\caption{Response time ($\mu$s) and throughput (OPs/$\mu$s) of CRDT use cases.}
\label{fig:results:crdts}
\Description[]{}
\end{figure*}

\begin{figure*}[t]
\vspace*{-1mm}
\centering
\includegraphics[width=\linewidth]{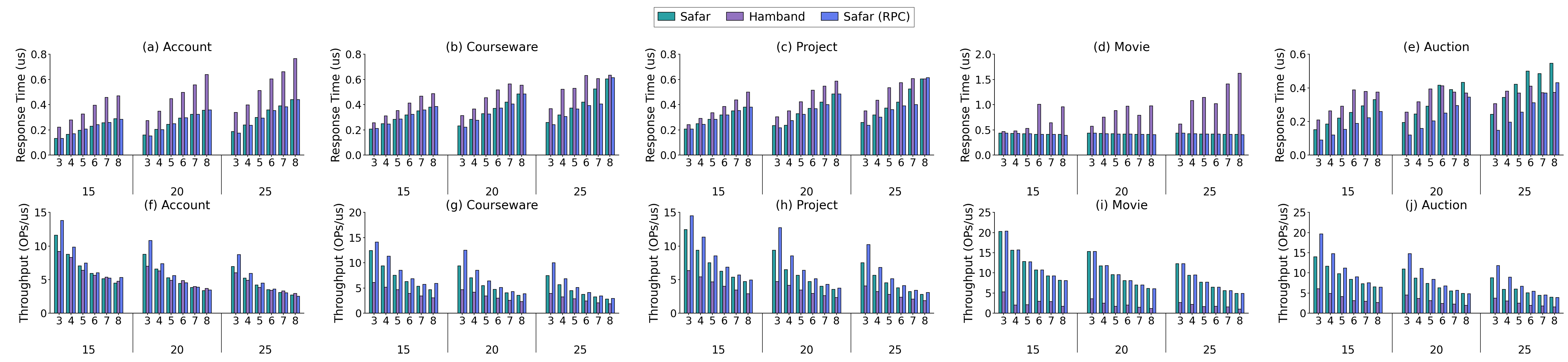}
\vspace*{-7mm}
\caption{Response time ($\mu$s) and throughput (OPs/$\mu$s) of WRDT use cases.}
\label{fig:results:wrdts}
\Description[]{}
\vspace*{-2mm}
\end{figure*}
\textbf{SafarDB vs. Hamband. \ }
% \subsubsection{SafarDB vs. Hamband}
Our first experiments compare SafarDB and Hamband on \emph{five CRDT microbenchmarks}; each run executes one CRDT use case, as shown in \autoref{fig:results:crdts}. Averaged across the 90 evaluated CRDT configurations, SafarDB provides $6.2\times$ lower response time and $3.8\times$ higher throughput than Hamband. Averaged across adjacent replica counts, response time increases by $2.8\%$ and $13.9\%$ per added replica for SafarDB and Hamband, respectively, while throughput increases by $2.0\%$ and $4.8\%$. Increasing the update percentage from $15\%$ to $25\%$ increases response time by $\sim\!1\%$ and $50\%$, and decreases throughput by $50\%$ and $31\%$ for SafarDB and Hamband, respectively. Increasing the number of replicas distributes operations across more replicas, which benefits throughput in both systems. However, Hamband's response time grows more quickly, and its throughput remains below SafarDB's. We believe a contributing factor is that Hamband must wait for completion queue ACKs before issuing the next RDMA verb or continuing with the application, whereas SafarDB's FPGA-resident applications can issue multiple RDMA verbs in sequence or interleave verbs with application logic through StRoM without waiting for ACKs. Across all evaluated CRDT configurations, SafarDB maintains lower response time and higher throughput than Hamband.

The next five experiments compare SafarDB and Hamband on \emph{five WRDT microbenchmarks}; each run executes one WRDT use case, as shown in \autoref{fig:results:wrdts}. Using the same per-configuration arithmetic mean, baseline SafarDB provides $1.5\times$ lower response time and $2.3\times$ higher throughput than Hamband, while SafarDB (RPC) provides $1.6\times$ lower response time and $2.5\times$ higher throughput. We believe that an important contributing factor is that the RDMA implementation on the network-attached FPGA eliminates sources of latency in a traditional CPU-based RDMA deployment (\autoref{fig:bg:cpu-rnic}). Increasing the number of replicas increases response time by $13.3\%$, $15.2\%$, and $14.8\%$ per replica for SafarDB, SafarDB (RPC), and Hamband, respectively, while decreasing throughput by $16.7\%$, $19.2\%$, and $14.1\%$ per replica. Increasing the update percentage increases the response time by $40.6\%$, $32.7\%$, and $47.9\%$, and decreases throughput by $40.2\%$, $39.8\%$, and $37.2\%$ for SafarDB, SafarDB (RPC), and Hamband, respectively, when increasing from $15\%$ to $25\%$ update percentage.

SafarDB (RPC) achieves comparable response times to the baseline SafarDB for most of the experiments, with some advantages at higher write percentages and node counts (e.g., Courseware at 25\% and 7--8 nodes). The one exception is Auction, where SafarDB (RPC) achieves lower response times than SafarDB (Baseline) across all runs, which we attribute to Auction having three synchronization groups, as well as two non-conflicting transactions. While Movie has two synchronization groups, it does not have a $\mathit{query}$ transaction or non-conflicting transactions, which eliminates the situations where SafarDB (RPC) can reduce response time in comparison to SafarDB. Similarly, SafarDB (RPC) achieves higher throughput than SafarDB at lower node counts, with its advantage reduced at higher node counts in most experiments. The one exception is Movie, where SafarDB (RPC) and SafarDB achieve comparable throughput, for the same reasons.

Ultimately, we see that both SafarDB and SafarDB (RPC) outperform Hamband on average; while the advantages of SafarDB (RPC) over SafarDB are somewhat limited, we see no instances in which SafarDB clearly outperforms SafarDB (RPC). For these reasons, we conclude that SafarDB (RPC) is a superior approach to implementing WRDTs for network-attached FPGAs; it can be used when system designers cannot augment the RDMA stack with FPGA-specific verbs.

\begin{figure}[t]
\vspace{-10mm}
\centering
\includegraphics[width=\linewidth]{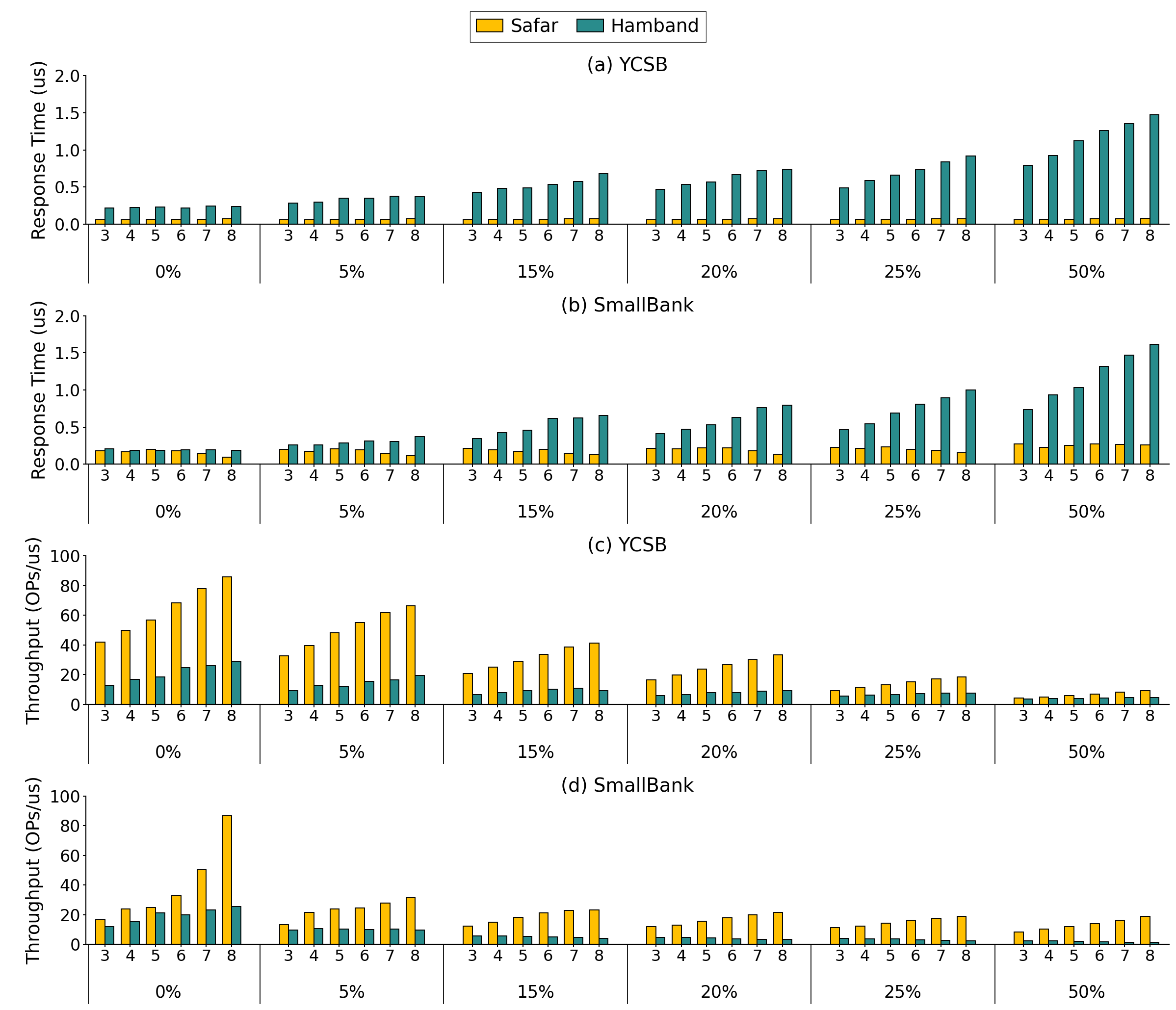}
\vspace{-5mm}
\caption{Response time \& throughput: YCSB/SmallBank} 
\label{fig:results:hybrid:scale}
\vspace{-5mm}
\end{figure}

Our last two experiments compare SafarDB and Hamband on the YCSB and SmallBank workloads (\autoref{fig:results:hybrid:scale}). SafarDB's throughput increases through eight replicas at all update percentages. Hamband's YCSB throughput also increases with the number of replicas, but its SmallBank throughput is nearly flat at 5\% updates and decreases at 15\%--50\% updates. Using the same per-configuration arithmetic mean, SafarDB provides $9.0\times$ lower response time and $2.8\times$ higher throughput for YCSB, and $3.0\times$ lower response time and $4.2\times$ higher throughput for SmallBank. The drop in SmallBank performance (Figure \ref{fig:results:hybrid:scale} b and d) from 0\% updates to 5\% updates for both SafarDB and Hamband is attributed to the need for SMR in the SmallBank workload.

\begin{figure}[t]
\vspace{-8mm}
\centering
\includegraphics[width=\linewidth]{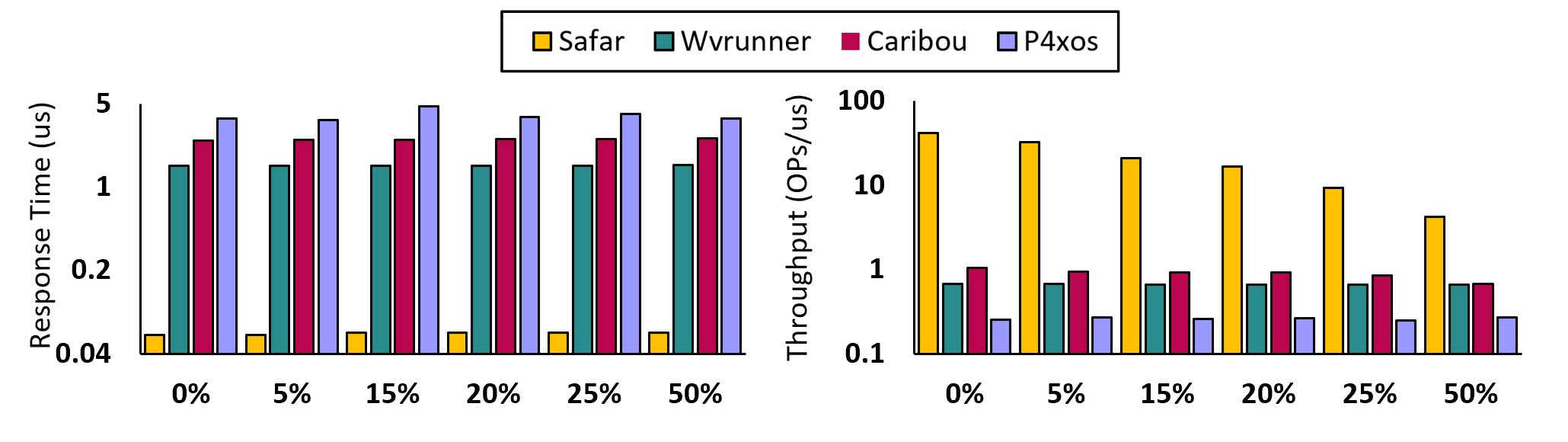}
\vspace{-6mm}
\caption{SafarDB vs. Waverunner, Caribou, and P4xos response time and throughput on three nodes.
% \todo{What are the numbers on the y axis. Do we need round them? Do we need horizontal lines?}
}
\label{fig:fpgaonly:waverunner}
\Description[]{}
\vspace{-5mm}
\end{figure}

% \subsubsection{SafarDB vs. FPGA replication baselines}
\textbf{SafarDB vs. FPGA replication baselines. \ }
The final FPGA-baseline experiment compares SafarDB to Waverunner running YCSB on three nodes (\autoref{fig:fpgaonly:waverunner}). We report results for different PUT/GET ratios. Waverunner is a state-of-the-art FPGA-based SmartNIC system that accelerates the consensus/SMR replication path in distributed systems \cite{waverunner}. Waverunner implements an accelerated Raft setup with three replicas (one leader and two followers). The leader is responsible for handling client requests (in the fast path these requests are processed as packets), while followers do not serve client requests directly. SafarDB provides $25.5\times$ lower response time and $31.3\times$ higher throughput. The main reason is that in SafarDB, the application is implemented inside the FPGA, so near-data processing reduces response time. In Waverunner, the application runs in software on the hosts, and the FPGA mainly accelerates replication. Due to hybrid consistency, SafarDB can handle client requests on all nodes, while in Waverunner, only the leader handles client requests; if a client contacts a follower, the follower rejects it, and the client must resend to the leader.

We also compare SafarDB with Caribou and P4xos on the same three-node OCT U280 request/response setup. Caribou is evaluated as a replicated key-value storage system with one leader and two followers. PUTs go to the leader because they must be ordered and replicated, while GETs can be served by any replica after it has received and applied the replicated update. P4xos is evaluated as a consensus-service baseline, matching its LevelDB-style model, with one proposer/coordinator FPGA and two acceptor/learner FPGAs. Client GET and PUT requests are submitted through the proposer/coordinator as Paxos values; the key-value state machine replies after a learner observes a quorum of acceptor votes and delivers the chosen request. Across 0\%--50\% writes, SafarDB provides $42.3\times$ lower response time and $22\times$ higher throughput than Caribou, and $67.08\times$ lower response time and $80.4\times$ higher throughput than P4xos. These results reflect SafarDB's combination of FPGA-resident YCSB operation handlers and RDT-aware consistency selection: read-only query transactions and other non-conflicting RDT operations can execute at any replica without consensus, while only conflicting operations use the strongly ordered SMR path. In contrast, Caribou exposes a TCP/IP key-value storage interface whose writes are replicated through leader-based atomic broadcast, and in the evaluated P4xos baseline, GET and PUT requests are submitted through the Paxos service before the key-value state machine responds.
% \todo{Resolved: filled SafarDB comparison multipliers above.}

% \begin{figure}[H]
% \centering
% \begin{subfigure}[b]{0.49\columnwidth}
%   \centering
%   \includegraphics[width=\linewidth]{figures/Waverunner/time.png}
%   \caption{Response Time}
%   %\label{fig:hybrid:assigned_time}
% \end{subfigure}\hfill
% \begin{subfigure}[b]{0.49\columnwidth}
%   \centering
%   \includegraphics[width=\linewidth]{figures/Waverunner/throughput.png}
%   \caption{Throughput}
%   %\label{fig:hybrid:assigned_throughput}
% \end{subfigure}
% \caption{Response time and throughput of SafarDB in comparison with Waverunner in three nodes.}
% \label{fig:fpgaonly:waverunner}
% \end{figure}

%\vspace*{-1mm}
%\subsection{Fault Tolerance} \todo{Should answer \autoref{q:crash-faults} and \autoref{q:permission-switch}}
\subsection{Fault Tolerance}

\begin{figure}[t]
\vspace{-7mm}
\centering
\includegraphics[width=\linewidth]{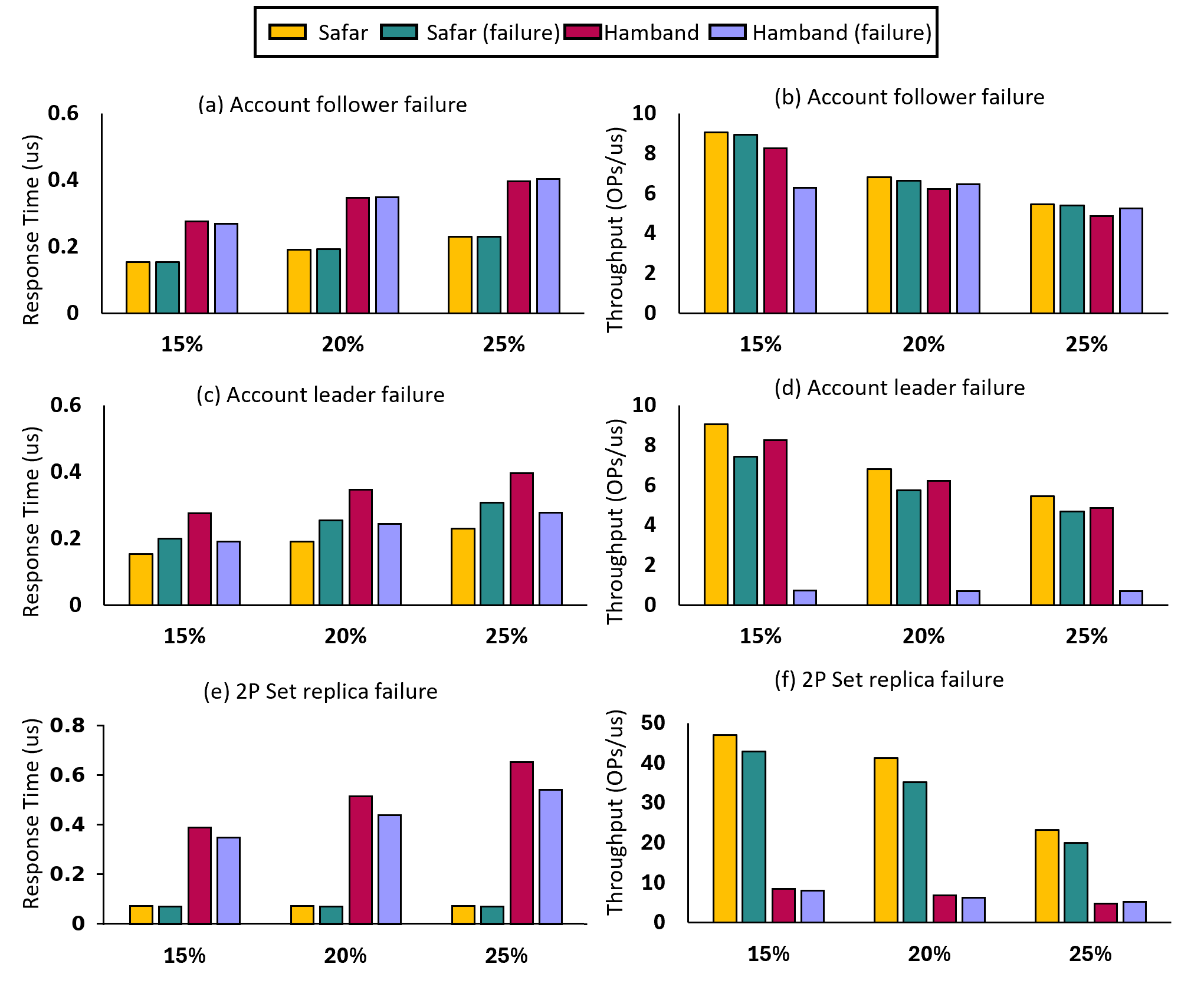}
\vspace*{-6mm}
\caption{Fault Tolerance experiments} 
\label{fig:results:failure}
\vspace{-7.1mm}
\end{figure}

To answer \ref{q:crash-faults} and \ref{q:permission-switch}, we evaluate (1) the latency/variability of permission switches and (2) the impact of single-node crash failures on both CRDT and WRDT executions. The detailed permission-switch histograms are provided in \autoref{fig:app:permission-switch} in the appendix. The measurements compare the round-trip time of changing write permissions in SafarDB's network-attached FPGA implementation and Hamband's traditional RDMA network \cite{hamband, mu}. The latter incurs overhead accruing to PCIe traffic, thread-switching, and RNIC caching, leading to high variability in round-trip time. In contrast, SafarDB's direct connection within the FPGA between the WRDT user-kernel accelerator and the Network kernel exhibits stable response times of either 17 or 24 $ns$ and is orders of magnitude faster than a traditional RDMA network. The impact of this reduced permission switch latency on failure recovery is elucidated in \autoref{fig:results:failure}. This directly benefits crash-fault recovery for WRDTs: after a leader failure, recovery requires revoking and re-granting write permissions at the newly elected leader, and SafarDB’s low and stable permission-switch latency reduces the recovery overhead on the critical path.

This subsection evaluates the impact of crash failures on one CRDT (2P-Set) and one WRDT (Bank Account). Experiments are performed with four nodes and varying write percentages. We simulate crash failures by stopping a preselected node during execution; the remaining operations are redistributed to the other replicas.

\textbf{Replica Failure:} In Figure \ref{fig:results:failure}e and \ref{fig:results:failure}f, the failure of one 2P-Set replica decreases SafarDB’s response time by $2\%$ and Hamband’s by $11$--$17\%$, and reduces SafarDB’s throughput by $12\%$. Hamband's throughput decreases by $6\%$ and $8\%$ at $15\%$ and $20\%$ updates, respectively, but increases by $8\%$ at $25\%$ updates. The response time decreases because there is one less remote node to replicate to. Throughput reflects two competing effects: losing one request-generating replica reduces parallelism, while having one fewer replication target reduces communication work. Even in the presence of a node failure, SafarDB outperforms Hamband by both measures, indicating that single-node failures do not compromise the benefits of FPGA acceleration.

\textbf{Follower Failure:} In \autoref{fig:results:failure} a and b, failure of one Account follower has no visible impact on SafarDB's response time and keeps Hamband's response time within $3\%$ of its no-failure baseline. It reduces SafarDB's throughput by around $2\%$. Hamband's throughput decreases by $24\%$ at $15\%$ updates, but increases by $4\%$ and $8\%$ at $20\%$ and $25\%$ updates, respectively. SafarDB's response time is not impacted because the other followers can immediately start executing the redistributed operations without waiting. For Hamband, losing one request generator reduces parallelism, while removing one replication target reduces communication work, producing the mixed throughput effect.

For WRDTs, stopping any replica prevents its heartbeat from incrementing. The leader maintains a list of known followers; when the leader detects that a follower fails, it removes the failed follower from its list. In Hamband, this update occurs in the foreground and impacts leader execution time; SafarDB updates the follower list in the background, which mostly hides the latency. 

\textbf{Leader Failure:} In \autoref{fig:results:failure} c and d, failure of Account's leader triggers a new leader election, with operations redistributed \textit{after} the election to the newly elected leader and two remaining followers. SafarDB's response time increases by around $25\%$, and its throughput is reduced by a modest $15\%$. Hamband's survivor-local response time decreases by around $30\%$, while its throughput decreases by $85$--$91\%$. The lower reported response time includes only calls processed by the surviving replicas and excludes the recovery pause; in contrast, throughput over the four-million-operation run includes failure detection, leader election, and the Mu permission handoff. SafarDB benefits from the faster permission switch (with detailed histograms provided in \autoref{fig:app:permission-switch} in the appendix), and maintains substantially higher throughput during leader failure.

These experiments answer \ref{q:crash-faults} by showing that crash faults affect throughput through reduced parallelism and, for leader failures, failure detection, election, and permission-handoff overheads. Response time can decrease when there is one less remote replica to update or when survivor-local measurements exclude the recovery pause. They also answer \ref{q:permission-switch} by showing that SafarDB’s fast and stable permission switch reduces the recovery overhead during leader failure, contributing to lower recovery cost than traditional RDMA deployments.

\subsection{Hybrid}
To answer \ref{q:hybrid-mode}, we evaluate the trade-offs of involving the host CPU in SafarDB’s hybrid mode using YCSB and SmallBank. We conduct three experiments: (1) varying the percentage of operations that target FPGA-resident data (Operations Assignment), (2) studying access skew under Zipfian workloads (Workload Distribution), and (3) applying summarization to batch remote updates (Summarization).

% In this section, we examine the flexibility of SafarDB when running in hybrid mode, and we evaluate performance using two applications: YCSB and SmallBank. We study the practical concern of what happens when the entire application does not fit inside the FPGA. We also examine the effect of skewed workloads and the impact of summarization on SafarDB’s performance.

% \subsubsection{Operations Assignment}
\textbf{Operations Assignment. \ }
With our first set of experiments (detailed results in \autoref{fig:app:hybrid:percent-ops} in the appendix), we study SafarDB’s flexibility and address the practical concern of when the whole application does not fit inside the FPGA. In hybrid mode, we implement part of YCSB and SmallBank on the host and part inside the FPGA. For YCSB, 100K keys are stored inside the FPGA, and the remaining 10M keys are in host memory. For SmallBank, 10M accounts are within the FPGA, and the remaining 90M accounts are stored within host memory. We then vary the fraction of the workload assigned to the FPGA versus the host—meaning we change the rate of client requests that target FPGA-resident keys/accounts versus the host-resident keys/accounts, and we report response time and throughput.

% With this design, we can place hot regions (keys/accounts that receive more client requests) inside the FPGA and keep cold keys on the host. 

As larger portions of operations are served by the FPGA, response time decreases and throughput increases approximately linearly as less PCIe communication occurs during execution (e.g., in YCSB at 50\% writes, response time decreases by $5.7\times$ and throughput increases by $4.7\times$ when we increase the fraction of operations assigned to the FPGA from 10\% to 90\%). By invoking the hybrid mode for these workloads, a small hit in performance is compensated for with an increase in application size while still delivering good response time and throughput for microsecond applications. This highlights a core trade-off in hybrid mode: host memory expands capacity beyond the FPGA, but performance improves as more requests are served by FPGA-resident hot keys/accounts.

% \subsubsection{Workload Distribution}
\textbf{Workload Distribution. \ }
Our second set of hybrid experiments (detailed results in \autoref{fig:app:hybrid:zipf} in the appendix) studies the effects of workload (uniform vs. Zipfian) distributions on SafarDB's hybrid mode.
%Now in this section we study the effect of access skew using a Zipfian distribution. Figure \ref{fig:hybrid:ycsbzipf} and \ref{fig:hybrid:smallbank:zipf} shows the effect of workload skew in data access patterns. 
We present the effect of skew on three different update ratios, 0\%, 5\%, and 50\%, on different hybrid configurations. 
In YCSB, the distribution applies to keys, while in SmallBank it applies to accounts. A value of $\theta = 0$ represents a uniform distribution, while $\theta = 2$ represents a highly skewed Zipfian distribution; \eg in SmallBank, $\theta = 2$ concentrates accesses on a small subset of hot accounts.

% For YCSB, client requests have different PUT/GET ratios, and for each ratio we vary the fraction of requests that target FPGA-resident keys. For SmallBank, the operations done on certain

Increasing the skew shows its effect when we have a higher fraction of reads, and when more requests target host-resident keys. For example, in YCSB at 0\% writes, when 20\% of operations are assigned to the FPGA, increasing the skewness from 0 to 1.2 decreases response time by $2.5\times$ and increases throughput by $2.3\times$; however, with the same write ratio, when 80\% of operations are assigned to the FPGA, we observe a $1.5\times$ decrease in response time and a $1.5\times$ increase in throughput. The reason is cache locality: with higher skew, the hot host keys are reused more and stay in the CPU caches, so host-side accesses become faster. Because of this, increasing skewness has little effect on the response time or throughput of (1) read operations that hit FPGA-resident keys, and (2) read/write operations where the key is inside the FPGA-resident set. Therefore, when the write ratio is higher, or when a larger fraction of requests are to FPGA-resident keys, the improvement from skew in response time and throughput becomes minimal. For example, in YCSB at 50\% writes, when 20\% of operations are assigned to the FPGA, increasing the skewness from 0 to 1.2 decreases response time by $1.3\times$ and increases throughput by $1.2\times$. Thus, SafarDB's hybrid mode exposes a second trade-off: skew can partially compensate for host accesses via CPU caching, but this benefit diminishes as writes increase or as more requests are served inside the FPGA.

\begin{figure}[t]
\vspace{-10mm}
\centering
\includegraphics[width=\linewidth]{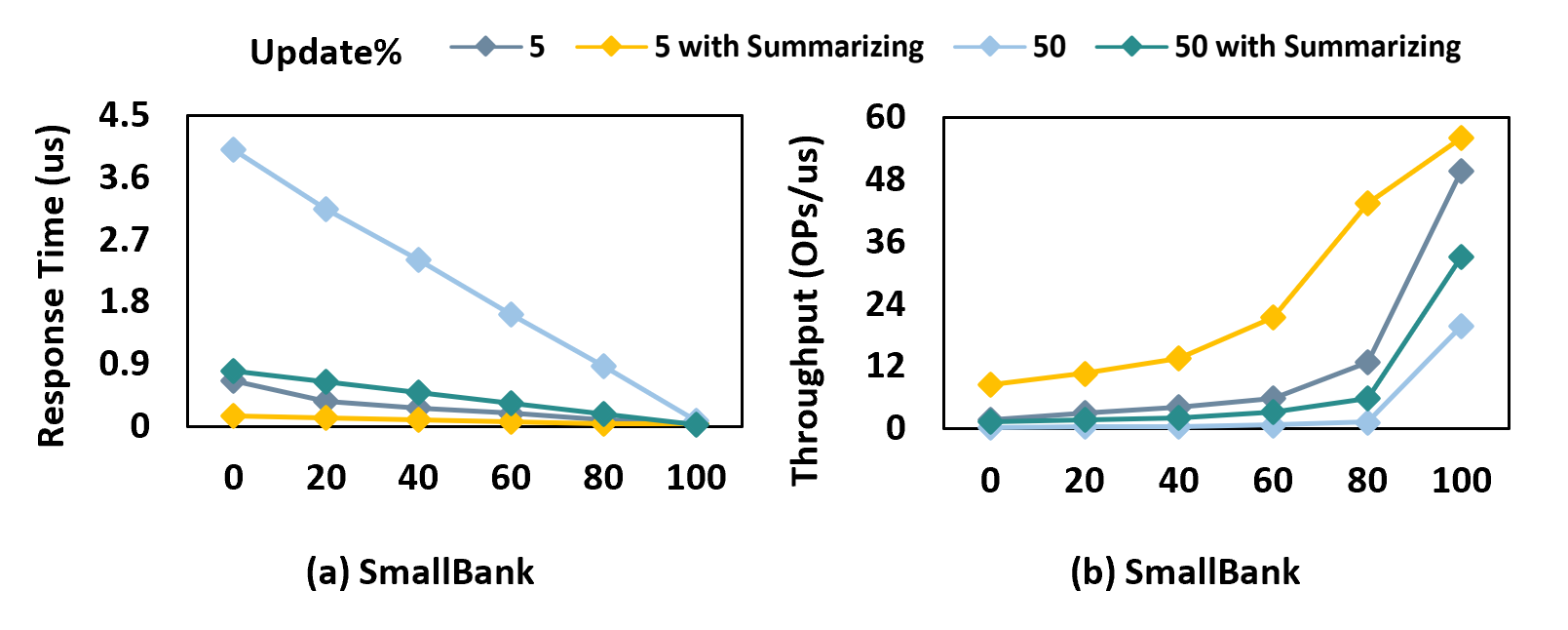}
\vspace*{-7mm}
\caption{Response time and throughput for SmallBank with a summarization size of 5 in different hybrid percentages.} 
\label{fig:results:hybrid:summ}
\vspace*{-6mm}
\end{figure}

\textbf{Summarization. \ }
% \subsubsection{Summarization}
% 
Our experiment studies the effect of summarization on SafarDB’s response time and throughput when we vary the percentage of operations assigned to the FPGA while running the SmallBank application. With summarization, instead of updating the remote replicas via RDMA or coordination after receiving client requests, we only update the local state, and whenever we reach the summarization threshold, we propagate the updates to the remote replicas. This helps avoid the cost of updating remote replicas for every transaction, since we update them only after processing a batch of transactions determined by the summarization size (in this experiment, the summarization size is 5). Figure~\ref{fig:results:hybrid:summ} shows the effect of summarization. As shown, summarization decreases response time and increases throughput across FPGA-assignment percentages; for example, in SmallBank under a write-heavy workload (50\% write), when 40\% of operations are assigned to the FPGA, summarization decreases response time by $4.9\times$ and increases throughput by $5\times$. This improves system performance, but it also increases staleness. Therefore, summarization introduces a third trade-off in hybrid mode: batching improves response time and throughput by reducing per-operation replication overhead, at the cost of higher staleness between synchronizations.

% \bigskip
These experiments answer \ref{q:hybrid-mode} by showing that SafarDB’s hybrid mode enables scaling to larger datasets by placing cold keys on the host while keeping hot keys on the FPGA; performance improves as more operations hit FPGA-resident data, host-side skew can mitigate CPU accesses via caching, and summarization improves performance by batching updates at the cost of increased staleness.
 % New
% \input{05_experiments} % Old

\vspace{-2mm}
\section{Related Work}
\label{sec:related_work}

\noindent\textbf{Positioning and RDT foundations.}
SafarDB is not the first FPGA-based replicated storage system, RDMA-based replication system, or hardware consensus system. It builds on StRoM's FPGA RoCE/RDMA stack~\cite{StRoM}, Mu's RDMA-based SMR protocol~\cite{mu}, and Hamsaz/Hamband's RDT and WRDT semantics~\cite{hamsaz,hamband}. CRDT/RDT and coordination-avoidance work provides the semantic basis for avoiding unnecessary ordering~\cite{shapiro11,BurckhardtGotsman14,BailisFekete14,indigo,ipa,ecros}; SafarDB changes the execution substrate by placing the RDT/WRDT handlers, RDMA-capable endpoint, and strong-ordering support inside the network-attached FPGA.

\noindent\textbf{SMR and FPGA/network replication.}
SafarDB uses strongly ordered replication only for conflicting WRDT operations, building on SMR protocols for message-passing and RDMA networks~\cite{OkiLiskov88,Lamport98,CorbettDean13,OngaroOusterhout14,HuntKonar10,zab_protocol,dare,apus,tailwind,sift,rethinking,mu,odyssey,hermes,acuerdo,ubft,whale}. Hardware systems such as HovercRaft~\cite{kogias2020hovercraft}, Waverunner~\cite{waverunner}, and CBox~\cite{cinbox} accelerate Raft/ZAB-style ordering or replication control paths. SafarDB instead combines FPGA-resident RDT execution with a relaxed path for non-conflicting operations and an FPGA-resident SMR path for conflicting operations.

\noindent\textbf{FPGA-based storage/replication and programmable-network consensus.}
Caribou~\cite{caribou} places replicated key-value storage on network-attached FPGAs using TCP/IP, FPGA-attached memory, near-data processing, and ZAB-based replication. SafarDB differs by exposing an RDMA/RoCE-style RDT/WRDT execution endpoint, supporting key-value workloads such as YCSB through application-level handlers, allowing non-conflicting operations to bypass strong ordering, and supporting hybrid FPGA/host placement. Waverunner~\cite{waverunner} accelerates Raft with an FPGA SmartNIC while keeping the replicated application on the host; SafarDB places the RDT/WRDT handlers and RDMA-capable endpoint inside the FPGA. P4xos and NetPaxos~\cite{p4xos,netpaxos} accelerate Paxos-style agreement in programmable network hardware; SafarDB keeps replicated state and RDT execution at the FPGA-attached storage endpoint, so NetPaxos is treated analytically rather than as a directly comparable U280 storage/RDT baseline.

\noindent\textbf{RDMA/RPC and SmartNIC systems.}
Prior work extends RDMA/RPC and programmable-NIC interfaces to reduce host overheads or support remote computation, including Prism, StRoM, RMC, KV-Direct, and nanoPU~\cite{prism,StRoM,rmc,kv-direct,nanopu}. SmartNIC and in-network systems similarly offload networking or computation to programmable NICs and accelerators~\cite{catapult,configurable_cloud,vfp,azure_smartnic,configurable_cloud_dnn,ipipe,xenic,off_path_characterization,spin}. \autoref{app:extended-related-work} provides the extended related-work discussion.

Therefore, SafarDB should be viewed as a network-attached FPGA replication engine rather than as a conventional CPU-hosted database with an offloaded NIC function. The relevant distinction is that the RDMA-capable endpoint, replication control path, and RDT operators are co-designed inside the same FPGA, which lets SafarDB remove PCIe and host-memory interactions from the common FPGA-resident execution path.

% \sadoghi{I suggest dividing the related work into a few categories. Having better organization, for example, we can survey the use of RDTs/RMDA/FPGA in DB, programming languages, and systems.}

% \vspace{-5mm}
\section{Conclusion}
SafarDB demonstrates that replicated data types can be efficiently implemented directly on network-attached FPGAs, significantly outperforming state-of-the-art RDMA-based and FPGA-accelerated designs. 
% Compared to Hamband \cite{hamband}, SafarDB improves response time by $7.0\times$–$12\times$ and throughput by $5.3\times$–$6.8\times$.
% while simultaneously supporting both CRDTs and WRDTs. 
These gains stem from tightly integrating RDT accelerators with an RDMA-capable NIC on a single FPGA, enabling low-latency communication among FPGA-resident components, direct access to on-chip storage, and specialized RDMA verbs that eliminate software and host CPU overheads. 
Importantly, experiments show that these benefits extend beyond microbenchmarks to replicated transactional workloads and database replication (e.g., YCSB and SmallBank).
% Importantly, experiments show that these benefits extend beyond microbenchmarks to full distributed applications. 
% When running YCSB \cite{ycsb} and SmallBank \cite{oltp}, SafarDB achieves a $8\times$ reduction in response time and a $5.2\times$ increase in throughput over Hamband, and improves response time by $25.5\times$ and throughput by $31.3\times$ compared to Waverunner \cite{waverunner}. 
Together, these results suggest that co-designing NICs and consistency primitives on programmable network-attached hardware is a promising direction to build low-latency, high-throughput replicated transactional systems and database replication.
%distributed systems.

% \sadoghi{Once we have reworked the introduction, then we can adjust the conclusions accordingly.}

% There is reason to believe that many distributed applications could benefit from a similar approach; while the RDMA verbs proposed in this paper represent general use cases that are not limited to RDTs, other distributed applications could benefit from application-specific verbs that combine data communication with computation at a sub-RPC granularity. 

% The ability of FPGA-resident accelerators to directly access the NIC state, which significantly sped up leader elections may be a narrow use-case; we posit that system services that pertain to memory protection and failure recovery beyond leader elections may also benefit if they can be integrated with FPGA-resident distributed application accelerators.

\bibliographystyle{ACM-Reference-Format}
\providecommand{\safardbbibliographyfiles}{sample-base}
\bibliography{\safardbbibliographyfiles}

\clearpage
\begin{appendices}
% \newpage

\section{Replicated Data Types (RDTs)}
\label{sec:rdts}
\label{sec:crdts}
\label{sec:wrdts}

A \textit{Replicated Data Type (RDT)} is a class of distributed objects that comprises a \textit{data} type and a set of \textit{transactions} (or method calls) \cite{BurckhardtGotsman14}. Clients of an object can request transactions at any replica, and replicas coordinate them. Transactions may not necessarily execute in the same order across replicas. Thus, the states of replicas may \textit{diverge} if replicas execute certain transactions in different orders. A \textit{replicated execution} is \textit{convergent} if all replicas reach the same state after all transactions are propagated and executed at all replicas. 
An RDT may include an \textit{integrity or consistency invariant} (the C in ACID) \cite{BailisFekete14, bailis2015feral, whittaker2021interactive}, a predicate on its data that must be maintained between transactions. Examples of integrity properties include requiring that a bank account's balance is always non-negative, or referential integrity in relational schemas, \textit{e.g}., only a student registered at a university may enroll in a course.

% \sadoghi{We need a statement here that we reformulate transactions using RDTs in the same way that people in the past have used CRDTs for transaction processing, which we can refer to as an example.}

% A call is \textit{permissible} if it preserves the integrity property in the post-state.
% A \textbf{permissibility check} determines if the post-state will preserve the integrity property;
% Every call is checked locally at its issuing replica to be permissible.
Every transaction can be checked locally to be \textit{permissible}, \textit{i.e.}, preserve integrity at its issuing replica. We refer to this local invariant/precondition validation performed before executing a transaction as a \emph{permissibility check}. 
% \sadoghi{What does integrity mean here? Do we mean a database integrity constraint? We need to align our language with the DB transaction language here as well. You can use ChatGPT (advanced model) to help with the alignment, not necessarily just using its output as is, but it can be helpful for doing quick research.} See previous paragraph for relational schema
% Calls that are not \textit{locally impermissible} are aborted and may be retried later. 
% Remote replicas do not perform permissibility checks when they receive calls.
%
However, when transactions are issued concurrently at different replicas and then propagated,
they can become impermissible, \textit{i.e.}, violate integrity at the target replicas.
% \sadoghi{DB community language is ACID and consistency models, we can use this premissiblity, but we need to translate/align with DB concepts as well. For example, they have a language as a permissible schedule/run, which is what we mean here.} It is defined above. It is a term we are adopting from multiple previous work of ours.
% 
% they can become impermissible in the target replicas.
% We say that they \emph{invariant-conflict}.
% with calls that execute concurrently at other replicas.
% When they propagate to other replicas, they violate integrity.
% \sadoghi{we need to use transaction, not process, not request.} We replaced calls with transactions
For example, suppose that two processes execute locally permissible withdrawal transactions from a bank account that ensures a positive balance; when propagated, 
% the calls may no longer be permissible if 
the sum of the two withdrawals can overdraft the account.
We say that a pair of transactions has a \textit{convergence conflict} or an \textit{integrity conflict} if reordering them violates convergence or integrity, respectively.
If either holds, we say that the pair conflicts.
A transaction is conflict-free for convergence or integrity
if there is no convergence conflict or integrity conflict with any other transaction, respectively.
If both hold, we say that the transaction is conflict-free.
% An object is conflict-free for convergence or integrity
% if it has no pairs of calls that convergence or integrity-conflict respectively.
% If both hold, we say that the object is conflict-free.
% \sadoghi{We need transaction abstraction or at least align it.} Used transaction
% These notions are conservatively lifted from method calls to methods.
An object is conflict-free if all its transactions are conflict-free.
% 
% A \textit{conflict} is either a \textit{state-conflict} or an \textit{invariant-conflict}.
% 
In order to preserve integrity,
a transaction may be \textit{dependent} on preceding transactions in the originating replica.
For example, a bank account withdrawal may depend on a preceding deposit to ensure a positive account balance.
In order to preserve integrity, 
a remote replica must execute transactions only after transactions it depends on have already been executed.
% to preserve integrity.
A transaction is dependence-free if it does not depend on any other transaction.

RDTs are implemented using a \textit{consistency model}. %which varies in terms of expressiveness and convergence, and integrity guarantees. 
\textit{Strong consistency} 
(also known as State Machine Replication (SMR))
% \sadoghi{SMR may also rely on weaker consistency models. I am not sure if this is what was intended.} Yes, we get to weak too. 
ensures convergence and integrity by imposing a linear ordering of transactions across replicas; this, however, limits responsiveness and availability \cite{Vogels08,park2019exploiting,lakshman2010cassandra,cooper2008pnuts,MemcachedWeb, Abadi12,brewer2012cap,brewer2000towards}, which inhibits practical deployments \cite{CorbettDean13,Burrows06,HuntKonar10}. 
\textit{Relaxed consistency} models \cite{shapiro2016consistency,nair2020proving} provide better performance than strong consistency \cite{pedone2002handling,lamport2004generalized}, but guarantee convergence or integrity properties for a limited subset of RDTs.
 For example,
\textit{Convergent Replicated Data Types (CRDTs)} \cite{shapiro11, calm}
are convergence conflict-free RDTs,
and can be implemented using relaxed consistency.
CRDTs have been successfully used as building blocks for distributed applications, including distributed databases \cite{calm}, key-value stores \cite{dynamo}, online data stores \cite{flighttracker}, and collaborative editing \cite{peritext}. 
%\js{distributed databases, collaborative text editing and high-scale backend logic\cite{ShapiroPreguica11,shapiro11, calm}.}. 
% \footnote{
We consider \textit{operation-based} CRDTs.
% } 
\autoref{tab:crdt-methods} lists six CRDTs, including a register and different types of counters and sets, which serve as our benchmarks; the Supplement describes each CRDT in greater detail.
For example, increment and decrement on the counter type are conflict-free for convergence.
% \sadoghi{Better to refer to as appendix, and we need proper linkage. Also, it would be good describe the type of operation in a sentence, for example, increment, decrement, or a common property for all of these operations, which is, for example, modification is not required reading first.} Added example of counter.
% The Supplement differentiates operation-based from state-based CRDTs.}
We note that RDTs include a \textit{query()} transaction, which reads values from the local object state; for brevity, we omit discussion of \textit{query()}, except when needed.
  
\textit{Well-coordinated Replicated Data Types (WRDTs)} \cite{hamsaz, hampa, hamband}
preserve both convergence and integrity,
and are implemented using
a hybrid consistency model \cite{balegas12ipa, LiPorto12,LiLeit14,LiLeit15, BalegasDuarte15,BalegasDuarte15_2, GotsmanYang16, SivaramakrishnanKaki15, lewchenko2019sequential, de2021ecros}:
% 
% \textit{Hybrid Consistency} models offer a middle ground:
% the methods that preserve convergence and integrity 
conflict-free transactions are implemented using relaxed consistency,
and conflicting transactions are implemented using strong consistency.
We will explain how WRDTs operate below.
% implement a .
% that necessitates strong consistency for a third category of \textit{conflicting} methods.
We note that WRDTs ensure convergence and integrity properties, whereas CRDTs only provide the former; CRDTs are a special case of WRDTs whose integrity predicate is the trivial assertion $true$. 

WRDTs divide transactions into three mutually exclusive categories with increasing coordination cost:
\textit{reducible},
\textit{irreducible},
and
\textit{conflicting}
transactions.
The first two are conflict-free.
% 
% As we described for CRDTs,
% the former two are conflict-free
% \footnote{WRDTs spell out irreducible as \textit{irreducible conflict-free}}.
%
Reducible transactions are conflict-free, dependence-free, and summarizable. 
A transaction is \textit{summarizable} if a sequence of invocations of that transaction can be aggregated into a single transaction.
Summarizable transactions can be implemented efficiently: 
they can be aggregated at the local replica, and then a single transaction can be invoked at each remote replica to propagate the aggregated transaction. 
% A method is reducible if it is summarizable and if it has no dependencies on other methods.
For example, the G-Counter contains a single non-negative value, $\mathit{cnt}$, initialized to zero. Each transaction $\mathit{Increment}(x)$ adds $x \geq 0$ to $\mathit{cnt}$; repeated transactions can be summed locally, and then the aggregate $\mathit{Increment}(\mathit{cnt})$ is propagated once to each remote replica. 
Irreducible conflict-free transactions are conflict-free but are either not dependence-free or not summarizable.
Conflicting transactions require strong consistency to ensure the same order of execution across all replicas.
Transactions that conflict with one another must be ordered consistently; together, they form a synchronization group.
Table \ref{tab:wrdt-methods} (SG) lists the synchronization groups for each WRDT.
As an example, consider the \textit{Movie} WRDT for a movie theater system.
Movie provides the transactions
$\mathit{addMovie}$, $\mathit{deleteMovie}$, 
$\mathit{addCustomer}$, and $\mathit{deleteCustomer}$, 
which add and delete movies from a theater's database, 
and reserve customer seats for movie showings. 
(The transactions $\mathit{deleteMovie}$ and $\mathit{addCustomer}$ succeed only if the given movie exists.)
Since add and remove operations on a set have a convergence conflict,
the first two transactions are a synchronization group, 
and the other two transactions are another synchronization group.
% The two methods $\mathit{addMovie}$ and $\mathit{deleteMovie}$ are a synchronization group, and the two methods $\mathit{addCustomer}$ and $\mathit{deleteCustomer}$ are another synchronization group.
% 
We will describe the State Machine Replication (SMR) implementation in \autoref{sec:Mu}.
Transactions of a synchronization group use the same instance of SMR \cite{schneider1990implementing}.
% in more detail
% subsequently 
% consensus
% \todo{We need to know the mechanism that the leader has a buffer and remove replicas have a buffer}
% Each replica keeps a replication log for the calls of each synchronization group.
% The leader decides the order of the calls, and propagates them in order to other replicas.
% writes 
% the replication logs of 
% remote 

\autoref{tab:wrdt-methods} lists five WRDTs 
that we adopted \cite{de2021ecros, hamband} to
serve as benchmarks in our experiments.
% the Supplement describes each WRDT in greater detail.

% \sadoghi{It is important to have a formal definition of what we mean by these types of conflicts. We could have new preliminary sections that define these concepts formally. Formalism is valued at the DB conference.} They were formalized before, and we abstracted over all of that in favor of readability. However the text above does have precision.

% Table \ref{tab:wrdt-methods} lists five WRDTs which serve as benchmarks in our experiments; the Supplement \todo{Cite}\js{\cite{de2021ecros, hamband}} describes each WRDT in greater detail. %Variants of these benchmarks have been used in prior research on design and synthesis of RDTs \todo{\cite{}}; 

\begin{figure}[t]
%\vspace{-7mm}
\begin{subfigure}{.47\textwidth}
  \centering
  % include first image
  \includegraphics[width=0.9\linewidth]{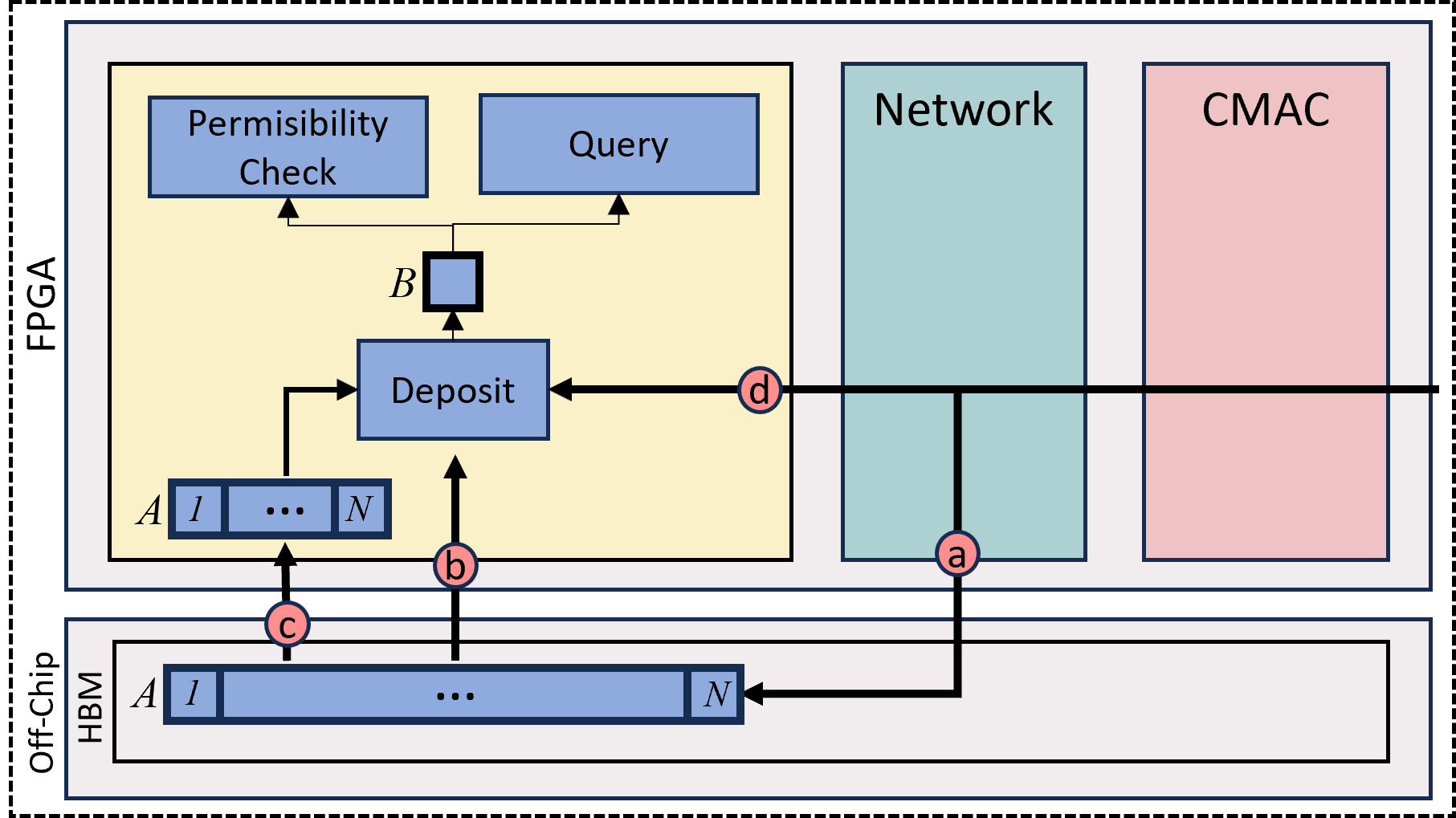}
  %\vspace*{-2mm}
  \caption{Reducible Transactions}
  \label{fig:arch:reducible}
\end{subfigure}
\begin{subfigure}{.47\textwidth}
  \centering
  \includegraphics[width=0.9\linewidth]{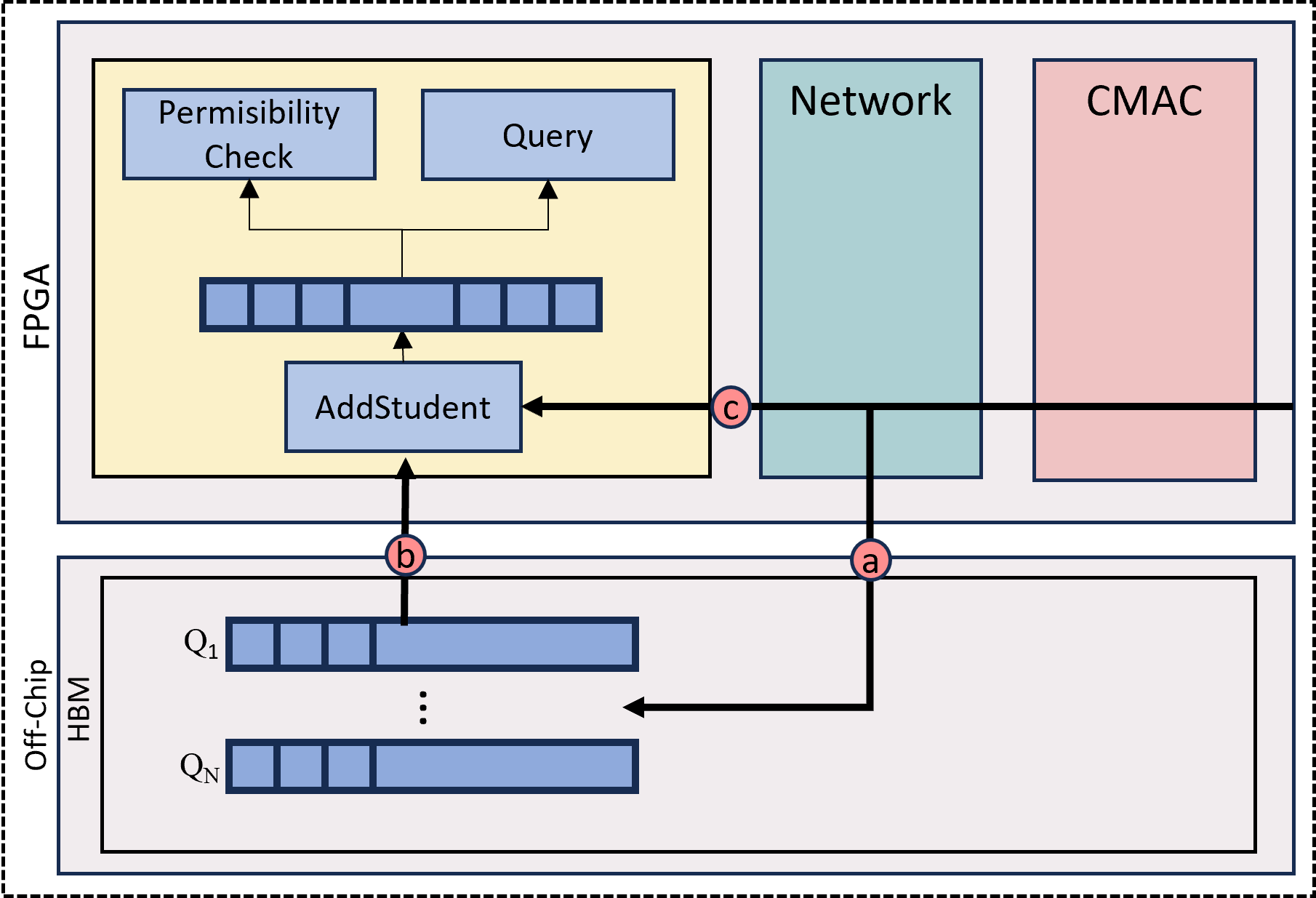}
  %\vspace*{-2mm}
  \caption{Irreducible Transactions}
  \label{fig:arch:irreducible}
\end{subfigure}
\begin{subfigure}{.47\textwidth}
  \centering
  % include second image
  \includegraphics[width=0.9\linewidth]{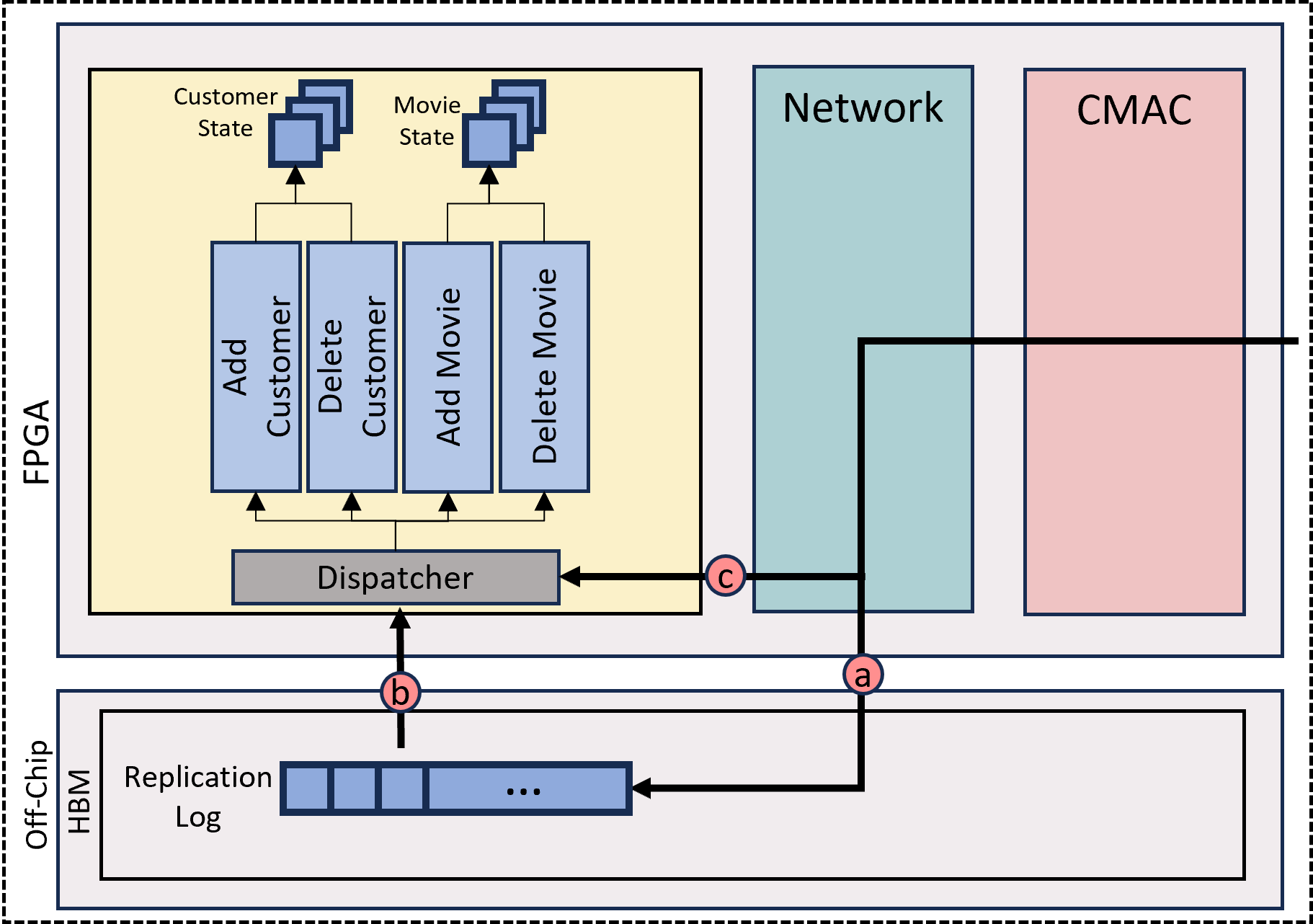}
  %\vspace*{-2mm}
  \caption{Conflicting Transactions}
  \label{fig:arch:conflicting}
\end{subfigure}
%\vspace*{-4mm}
\caption{The implementation of the three transaction categories.}
%\vspace*{-4mm}
\label{fig:arch:methods}

\end{figure}

\textbf{Example.} In \autoref{fig:arch:reducible}, assume the system comprises $N$ replicas implementing a Bank Account with balance $B$ and a reducible transaction $deposit(x)$, which increments $B$ by $x$. Each replica allocates an $N$-element array, $A$, where $A_i$ stores the deposits of replica $i$. When replica $i$ executes $deposit(x)$, it first increments its local $B$ by $x$ and then writes $B+x$ into the remote replicas' memory at $A_i$ (through $a$), overwriting the previous values. In configuration (1), when a remote replica needs to read the total balance $B$, it reads from memory (through $b$) and combines all replica balances. In configuration (2), a background module pulls the remote balances from memory (through $c$), so when the balance is required, only fast on-chip BRAM is read. In configuration (3), the remote replicas' on-chip BRAM is directly updated from the network (through $d$), removing the need to poll memory.

\textbf{Example.} In \autoref{fig:arch:irreducible}, assume the system comprises $N$ replicas implementing a Course management system with a set of students $S$ and an irreducible transaction $addStudent(x)$, which appends student $x$ to $S$. Each replica allocates $N$ queues, where $Q_i$ stores the transactions performed by replica $i$. When replica $i$ executes $addStudent(x)$, it first appends $x$ locally to $S$ and then appends the transaction to $Q_i$ at the remote replicas (through $a$). In configuration (1), a background module polls the remote queue to apply the outstanding $addStudent(x)$ transaction (through $b$) to the on-chip BRAM set $S$. In configuration (2), the remote replicas' on-chip set $S$ is directly updated from the network (through $c$), removing the need to poll memory.

\textbf{Example.} In \autoref{fig:arch:conflicting}, assume the system comprises $N$ replicas implementing a Movie ticketing system with a set of movies $M$ and conflicting transactions $addMovie(x)$ and $deleteMovie(x)$, which add and remove movie $x$ from $M$, respectively. $addMovie(x)$ and $deleteMovie(x)$ depend on each other and therefore form a synchronization group (see \autoref{sec:rdts}). Each replica allocates a replication log in memory for each synchronization group (see \autoref{sec:Mu}). When a replica executes $addMovie(x)$, it first triggers its SMR module, which performs a round of consensus, commits the transaction locally, and then replicates it to other replicas by appending it to the replication log (through $a$). In configuration (1), the replica polls the replication log for committed transactions (through $b$) and also polls the log when $M$ is accessed to ensure the most up-to-date data. In configuration (2), the SMR module uses the custom RPC both to append the replication log (through $a$) and to directly update the FPGA-resident data (through $c$).

\section{Conflict-free Replicated Data Types}
\label{sec:CRDTs}

A CRDT is a data type (or data structure) that can be replicated across a distributed computer system in a manner that provides several properties. Updates can be performed independently and concurrently at any replica; algorithms within the CRDT that are transparent to the user/programmer propagate updates to all other replicas, resolving any inconsistencies that may occur; and convergence under the strong eventual consistency model is guaranteed. It is important to note that strong eventual consistency \textit{does not} guarantee any timeline by which convergence will be achieved.

% \subsection{Operation vs. State-based CRDTs}

Two different CRDT design and implementation strategies have been proposed: operation-based~\cite{Letia10, Baquero14} and state-based \cite{Baquero99, Almeida15}. Both operation-based and state-based CRDTs provide strong eventual consistency, and they are theoretically equivalent, meaning that any operation-based CRDT can emulate its state-based counterpart, and vice-versa \cite{shapiro2011conflict}. State-based CRDTs require periodic transmission of each replica's state to other replicas, requiring a non-trivial state merge operation, depending on the level of sophistication of the CRDT. In contrast, operation-based CRDTs require each replica to broadcast each method invocation to the other replicas, which increases network communication. The CRDT Wikipedia page\footnote{\url{https://en.wikipedia.org/wiki/Conflict-free_replicated_data_type}} provides further details.

\subsection{CRDT Benchmark Summary}

\begin{table}[t]
\caption{\textbf{CRDTs}}
\label{tab:crdt-methods}
\centering
%\small
\footnotesize
\begin{tabular}{|l||l|l|l|}
 \hline
 \textbf{CRDT} & \textbf{State} & \textbf{Reducible}  & \textbf{Irreducible} \\
 \hline
G-Counter & Scalar \textit{C} & \begin{tabular}[c]{@{}l@{}}\textit{increment}\end{tabular} &  \\
 \hline
PN-Counter & Scalar \textit{C} & \begin{tabular}[c]{@{}l@{}}\textit{increment}\\ \textit{decrement}\end{tabular} &  \\
 \hline
LWW-Register & \begin{tabular}[c]{@{}l@{}}Scalar \textit{R}\\ Scalar \textit{T}\\ \end{tabular} & \textit{assign} &  \\
 \hline
G-Set & Set \textit{S} &  \textit{insert} &  \\
 \hline
PN-Set & Array  \textit{C} &  & \begin{tabular}[c]{@{}l@{}}\textit{insert}\\\textit{remove}\\ \end{tabular} \\
 \hline
2P-Set & \begin{tabular}[c]{@{}l@{}}Set \textit{A}\\ Set \textit{R}\\ \end{tabular} &  & \begin{tabular}[c]{@{}l@{}}\textit{insert}\\\textit{remove}\\ \end{tabular} \\
 \hline
\end{tabular}

\end{table}

The CRDTs used for experimental evaluation are described as follows. Our evaluation is limited to scalar instances of each CRDT. 

\bigskip
\noindent
\textbf{G-Counter}\footnote{We do not include G-Counter as a benchmark; it is a building block for the PN-Counter.}\textbf{:} A grow-only counter, which can only be incremented.

\bigskip
\noindent
\textbf{PN-Counter:} A positive-negative counter which supports increment and decrement operations. A PN-Counter comprises two G-Counters: one for increments, and one for decrements. 

\bigskip
\noindent
\textbf{LWW-Register:} A last-writer-wins register, which supports assignment of values to the register. Unique timestamps are associated with each assignment, which ensure a total order of operations. The register always retains the most recently written value.

\bigskip
\noindent
\textbf{G-Set:} A grow-only set which supports insertion, but not removal, of set elements. 

\bigskip
\noindent
\textbf{PN-Set:} A counter is associated with each element of a set: insertion increments the counter; removal decrements the counter. An element is present in the set if its counter is positive. A PN-Set may represent a non-traditional multiset, in which a positive counter value represents the number of occurrences of an element in the set; however, when the counter becomes negative, multiple insertion operations are needed to reintroduce the element to the multiset.

\bigskip
\noindent
\textbf{2P-Set:} A two-phase set that supports insertion and removal of set elements. Once removed, an element cannot be reinserted into the set. A 2P-Set is typically implemented using two G-Sets. 

\section{Well-Coordinated Replicated Data Types}

WRDTs generalize CRDTs with the addition of conflicting methods that require strong consistency \cite{hamsaz, hampa, hamband}. Strong consistency necessitates an SMR protocol to synchronize conflicting calls. Like CRDTs, WRDTs provide strong eventual consistency for non-conflicting method categories. WRDTs thereby guarantee convergence for both conflicting and non-conflicting method categories. WRDTs also provide invariants that are satisfied through permissibility checks, thereby guaranteeing integrity.

\subsection{WRDT Benchmark Summary}

The WRDTs used for experimental evaluation are described as follows. 

\begin{table*}[t]
\centering
\caption{\textbf{WRDT Methods and Invariants}}
%\small
\footnotesize
\begin{tabular}{|l||l|l|l|l|l|l|l|}
 \hline
 \textbf{RDT} & \textbf{State} & \textbf{Reducible}  & \textbf{Irreducible} & \textbf{Conflicting} & \textbf{Sync Group} \\
 \hline
Bank Account & Scalar \textit{B} & $\mathit{deposit}(d)$ & & $\mathit{withdraw}(w)$ where $B - w \geq 0$ & 1  \\
 \hline
Courseware & \begin{tabular}[c]{@{}l@{}}Set \textit{S}\\ Set \textit{C}\\ Set \textit{E}\end{tabular} &  & $\mathit{addStudent}(s)$ where $s \not\in S$& \begin{tabular}[c]{@{}l@{}}$\mathit{addCourse}(c)$ where $c \not\in C$ \\ $\mathit{deleteCourse}(c)$ where $c \in C$\\ $\mathit{enroll}(s, c)$ where $s \in S, c \in C, \langle s,c \rangle \not\in E$\end{tabular} & 1 \\
 \hline
Project & \begin{tabular}[c]{@{}l@{}}Set \textit{E}\\ Set \textit{P}\\ Set \textit{A}\end{tabular} &  & $\mathit{addEmployee}(e)$ where $e \not\in E$ & \begin{tabular}[c]{@{}l@{}} $\mathit{addProject}(p)$ where $p \not\in P$\\  $\mathit{deleteProject}(p)$ where $p \in P$\\  $\mathit{assign}(e, p)$ where $e \in E, p \in P, \langle e,p \rangle \not\in A$\end{tabular} & 1\\
 \hline
Movie & \begin{tabular}[c]{@{}l@{}}Set \textit{C}\\ Set \textit{M}\end{tabular} &  &  & \begin{tabular}[c]{@{}l@{}} $\mathit{addCustomer}(c)$ where $c \not\in C$\\  $\mathit{deleteCustomer}(c)$ where $c \in C$\\ \hline $\mathit{addMovie}(m)$ where $m \not\in M$ \\ $\mathit{deleteMovie}(m)$ where $m \in M$\end{tabular} & 2\\
 \hline
Auction & \begin{tabular}[c]{@{}l@{}}Set \textit{U}\\ Set \textit{A}\\ Set \textit{I} \\ Array $S[]$\end{tabular} & $\mathit{sellItem}(i, u)$ & $\mathit{openAuction}(a)$ where $a \not\in A$ & \begin{tabular}[c]{@{}l@{}}$\mathit{registerUser}(u)$ where $u \not\in U$\\ \hline $\mathit{buyItem}(i, u)$ where $i \in I, S[i] \geq 1, u \in U$\\ \hline $\mathit{placeBid}(a, bid, u)$ where $a \in A, u \in U$ \\  $\mathit{closeAuction}(a)$ where $a \in A$\end{tabular} & 3 \\
 \hline
\end{tabular}
\label{tab:wrdt-methods}
\end{table*}

\bigskip
\noindent
\textbf{Bank Account:} A distributed bank account that maintains a scalar balance; $deposit()$ and $withdraw()$ methods respectively increase and reduce the balance. Withdrawals that cause overdrafts (funds less than 0) are impermissible. 

\bigskip
\noindent
\textbf{Courseware:} A university registrar that manages courses to be taught ($addCourse()$, $deleteCourse()$), enrolls students in the university ($addStudent()$), and enrolls students in courses ($enroll()$).

\bigskip
\noindent
\textbf{Project:} Business software that manages projects ($addProject()$, $deleteProject()$), enrolls personnel ($addEmployee()$), and assigns personnel to work on projects ($assign()$).

\bigskip
\noindent
\textbf{Movie:} A database of movies ($addMovie()$, $deleteMovie()$) and customers ($addCustomer()$, $deleteCustomer()$).

\bigskip
\noindent
\textbf{Auction:} An e-commerce site that manages users ($registerUser()$), lists items for direct sale ($sellItem()$), allows users to purchase items ($buyItem()$), and facilitates item sales via auction ($openAuction()$, $placeBid()$, $closeAuction()$). 

\bigskip
Bank Account's state comprises a single scalar, i.e., one account with one balance. The other WRDTs' states comprise two or three sets, and in the case of Auction, three sets and one array.

\section{RDMA}

\subsection{Remote Direct Memory Access (RDMA)}
\label{sec:rdma}

\begin{table}[t]
%\vspace*{-3mm}
    \centering
    \footnotesize
    \caption{Average latency to issue an RDMA verb on a traditional PCIe-attached RNIC versus a network-attached FPGA using AXI Stream. Latency measures the local submission overhead between the application and the network stack; network transmission time is excluded in both cases.}
    %\vspace*{-3mm}
    \begin{tabular}{|l|l|l|}
    \hline
        & Read Latency & Write Latency\\
    \hline
        %CPU / RNIC / PCIe & 1.8235$\mu$s &  1.9582$\mu$s \\
        Traditional RDMA Network & 1.8 $\mu$s & 2.0 $\mu$s \\
    \hline
        %FPGA (Alveo U280 / HBM)  & 0.009023 $\mu$s & 0.008945 $\mu$s \\
        Network-attached FPGA & 0.0090 $\mu$s & 0.0089 $\mu$s \\
    \hline
    \end{tabular}
    \label{tab:rdma-ops-read-write}
    %\vspace*{-5mm}
\end{table}

\textit{Remote Direct Memory Access (RDMA)} allows user-space programs to access the memory of another node in a network without interrupting the remote CPU. State-of-the-art RDTs \cite{hamband} and SMR protocols \cite{mu} are specialized for RDMA networks. While the RDMA standard specifies a wide variety of RDMA \textit{verbs}, RDTs are implemented using $\mathsf{Read}$ and $\mathsf{Write}$ exclusively. 

\textbf{Traditional RDMA. \ }
% \subsection{Traditional RDMA}
\label{sec:cpu-rdma}
%\autoref{fig:bg:cpu-rnic} depicts a node in a CPU-based RDMA network. 
The CPU, memory, and \textit{RDMA-enabled NIC (RNIC)} are connected via PCIe. Local and remotely-accessible memory regions are allocated in user space. The \textit{Queue Pair Context (QPC)} maintains a set of permissions which govern local and remote access to memory; applications may dynamically adjust permissions at runtime \cite{mu}. The CPU accesses a \textit{doorbell register} via memory-mapped I/O to inform the RNIC when a verb is ready to transmit.

\textbf{RDMA for Network-Attached FPGAs. \ }
% \subsection{RDMA for Network-Attached FPGAs}
\label{sec:rdma-fpga}
% \ml{If this is our contribution, we should explicitly say it like: we introduce ...} \py{It's not our contribution just an explanation of how StRoM works, our contributions are in the FPGA-Specific RDMA Verbs paragraph.}
%\autoref{fig:bg:fpga-rnic} depicts a network-attached FPGA. 
The FPGA card connects to a host CPU via PCIe, and features onboard \textit{high-bandwidth memory (HBM)}. The host CPU programs and initializes the FPGA, initializes the internal state, and starts the FPGA execution. The \textit{Ethernet Medium Access Control Block (CMAC)} and \textit{Network} kernels implement a \textit{`soft'} RNIC, which supports RoCEv2, and communicate with the \textit{application} kernel. 

The application kernel and soft RNIC are connected via \textit{AXI (Advanced eXtensible Interface)} Streams (AXI\_S), and are connected to HBM via \textit{Memory-Mapped AXI (MM-AXI)}. The RNIC QPs do not conform to the RDMA standard \cite{StRoM}. 
The send queue SQ is an AXI stream that is written by the application and is read by the Network kernel; the receive queue RQ resides in the Network kernel, and there is no completion queue CQ. Application kernels communicate with the QPC to modify permissions. The supplement details the execution of FPGA-based one-sided RDMA verbs.

% \begin{figure}[t]
% \centering
% \includegraphics[width=\linewidth]{figures/arch-new-colors/fpga-rnic.png}
% \caption{RDMA architecture for a network-attached FPGA featuring a soft RNIC \cite{StRoM}.
% } 
% \label{fig:bg:fpga-rnic}
% \Description[]{}
% \end{figure}

%The User kernel issues an RDMA verb to the RNIC by pushing a WQE to the SQ AXI stream; The network kernel dequeues the WQE and transmits the RDMA verb to the remote node, while simultaneously placing a WQE in the RQ. The Network kernel removes the WQE from the RQ when an ACK is received from the remote node, and, if data is delivered (e.g., the result of an \textbf{RDMA\_Read} verb), it is written into AXI memory where the user kernel can directly access it. This obviates the need to allocate a CQE and for the User kernel to poll the CQ to detect data. At the remote node, the User kernel polls AXI memory to detect new values written by \textbf{RDMA\_Write} operations. 
%\todo{For ``vanilla'' RDMA read operations, is the data delivered directly to the network kernel, or is it written to HBM?}. 

AXI-based communication within the FPGA is faster than PCIe. To quantify this, we generated 1 million random $\mathsf{Read}$ and $\mathsf{Write}$ requests on a traditional RDMA network and a network-attached FPGA cluster. \autoref{tab:rdma-ops-read-write} reports the average latency to issue an RDMA verb under two communication substrates: AXI Stream (network-attached FPGA) and PCIe (traditional CPU-based RDMA). The reported latency measures the time from verb submission by the issuing process to acknowledgment by the network stack — specifically, the AXI Stream handshake latency on the FPGA and the PCIe doorbell round-trip on the CPU. Network transmission time is excluded from both measurements; the table therefore isolates the \emph{local} communication overhead between the application and the NIC. Verbs implemented on the network-attached FPGA are two orders of magnitude faster than the traditional RDMA network. 

%There is an important difference between \autoref{fig:bg:cpu-rnic} and \autoref{fig:bg:fpga-rnic} in terms of scalability. A server-class CPU runs orders of magnitude more threads compared to the number of user kernels that can fit onto a modern high-capacity FPGA. The former necessitates orders of magnitude more QPs than the latter, and, given limited RNIC cache capacities, it makes sense to allocate them in pinned memory. Future work may evolve the soft RNIC to encompass larger-scale applications that are partitioned between FPGA-resident accelerators and software threads that run on the host CPU. 

\begin{figure}[t]
%\vspace*{-7mm}
\centering
\includegraphics[width=\linewidth]{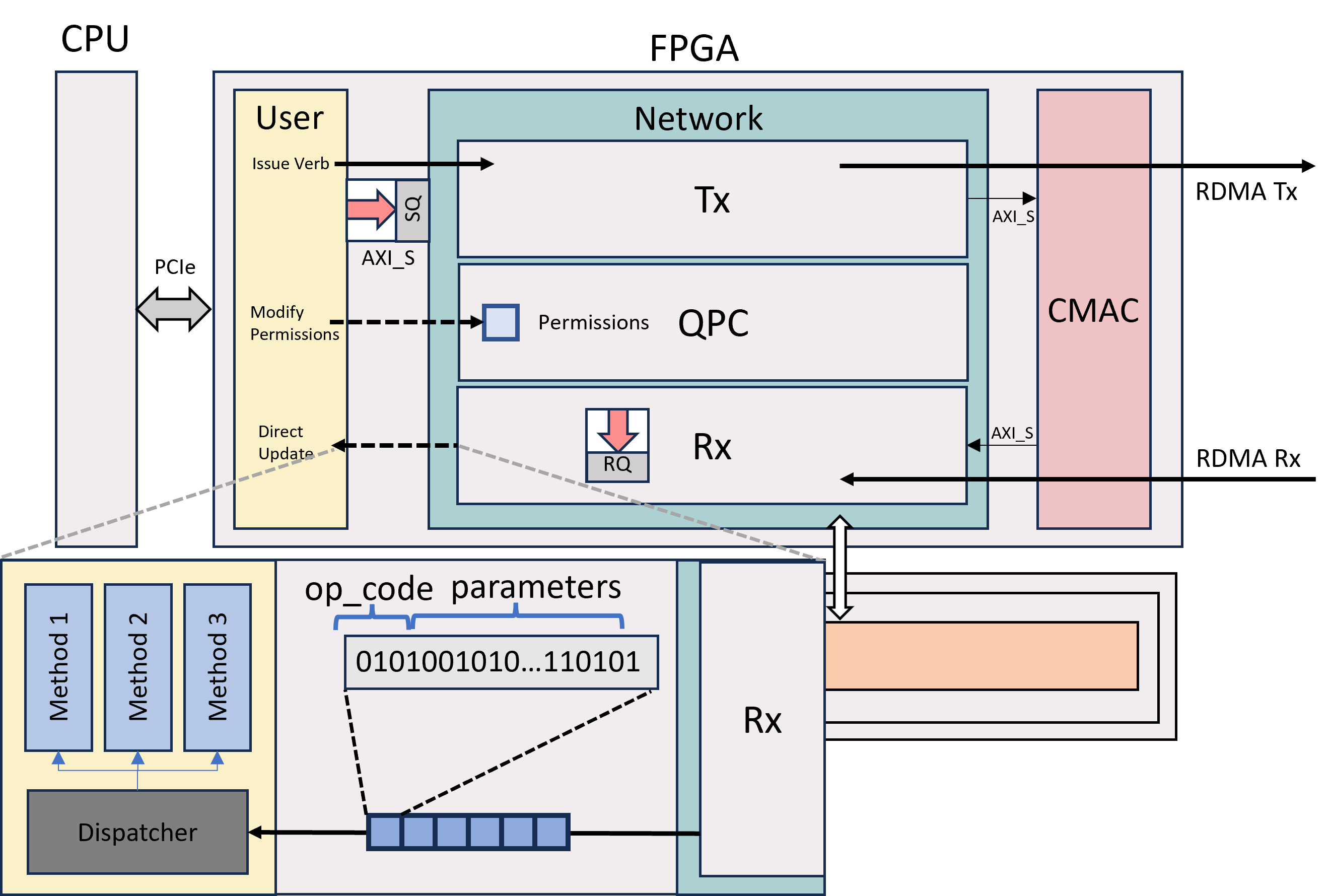}
%\vspace*{-6mm}
\caption{FPGA-specific RDMA\_Write verbs enable RPCs.} 
\label{fig:bg:rpc}
\Description[]{}
\end{figure}

% \subsection{RDMA vs. RPC, Redux}
\textbf{RDMA vs. RPC \ }
% \todo{Redux?}
\label{sec:rdma-rpc}
% 
% Hamband \cite{hamband} is a high-performance RDMA-based implementation of CRDTs and WRDTs.
% Hamband
We observed that in a replicated execution,
most of the data that remote replicas $\mathsf{Write}$ consists of transaction IDs and parameters, indirectly implementing \textit{Remote Procedure Calls (RPCs)} via RDMA. RPCs exhibit orders of magnitude greater latency than RDMA verbs \cite{attack}, and there have been successful proposals to reduce RPC overhead \cite{eRPC}. We show that network-attached FPGAs eliminate this disparity.
%We acknowledge, but eschew detailed discussion of, a larger debate around RDMA vs. RPCs \todo{\cite{}} in data center networks. Hamband might be able to further improve performance by replacing some \textbf{RDMA\_Writes} with efficient RPC calls, e.g., using RPC \todo{\cite{}}. 
In \autoref{fig:bg:rpc}, FPGA-specific $\mathsf{Write}$ verbs transparently enable RPCs that invoke FPGA-resident accelerators: the payload comprises a transaction ID (\textit{opcode}) and parameters; a \textit{Dispatcher} selects the appropriate accelerator, forwards the parameters, and invokes the accelerator. 
%This $\mathsf{RPC}$ does not need to return the accelerator's output when used for replicated calls.
% when limited to replicated calls that implement RDT convergence. 

\textbf{Two-sided RDMA alternative.} Two-sided RDMA SEND/RECV is a natural point of comparison because it delivers requests to the receiver without requiring the sender to name a remote address in the receiver's memory region. However, the receiver still must pre-post registered receive buffers through Receive Queue Entries (RQEs) before SEND operations can complete. On a CPU-based system, a software thread can replenish the receive queue and poll completion queues. On a network-attached FPGA without a general-purpose CPU in the datapath, this model would either reintroduce host CPU involvement over PCIe or require additional FPGA queue-management logic. SafarDB's custom RPC verb instead encodes the transaction opcode and parameters directly in the RDMA Write payload. The receiving FPGA's Dispatcher kernel reads the opcode from the AXI Stream interface and routes the request to the appropriate accelerator without pre-posted receive buffers. This design is better matched to SafarDB's FPGA-resident execution path because it requires no receive-queue management, avoids completion-queue polling for request delivery, and allows the Write-Through variant to update both the replication log in HBM and the application state in on-chip BRAM in a single network operation.

RDMA was originally proposed for high-performance computing (HPC) clusters through the InfiniBand protocol, and was later adapted for data center networks as RDMA over Converged Ethernet (RoCE), a link-layer protocol; RoCEv2 is implemented on top of UDP/IPv4 or UDP/IPv6 as an internet-layer protocol, which supports packet routing. Further details can be found on the RoCE Wikipedia page\footnote{\url{https://en.wikipedia.org/wiki/RDMA_over_Converged_Ethernet}}.

\subsection{RDMA vs. Socket-based Networking}

\begin{figure}[t]
\begin{subfigure}{.475\textwidth}
  \centering
  \includegraphics[width=\linewidth]{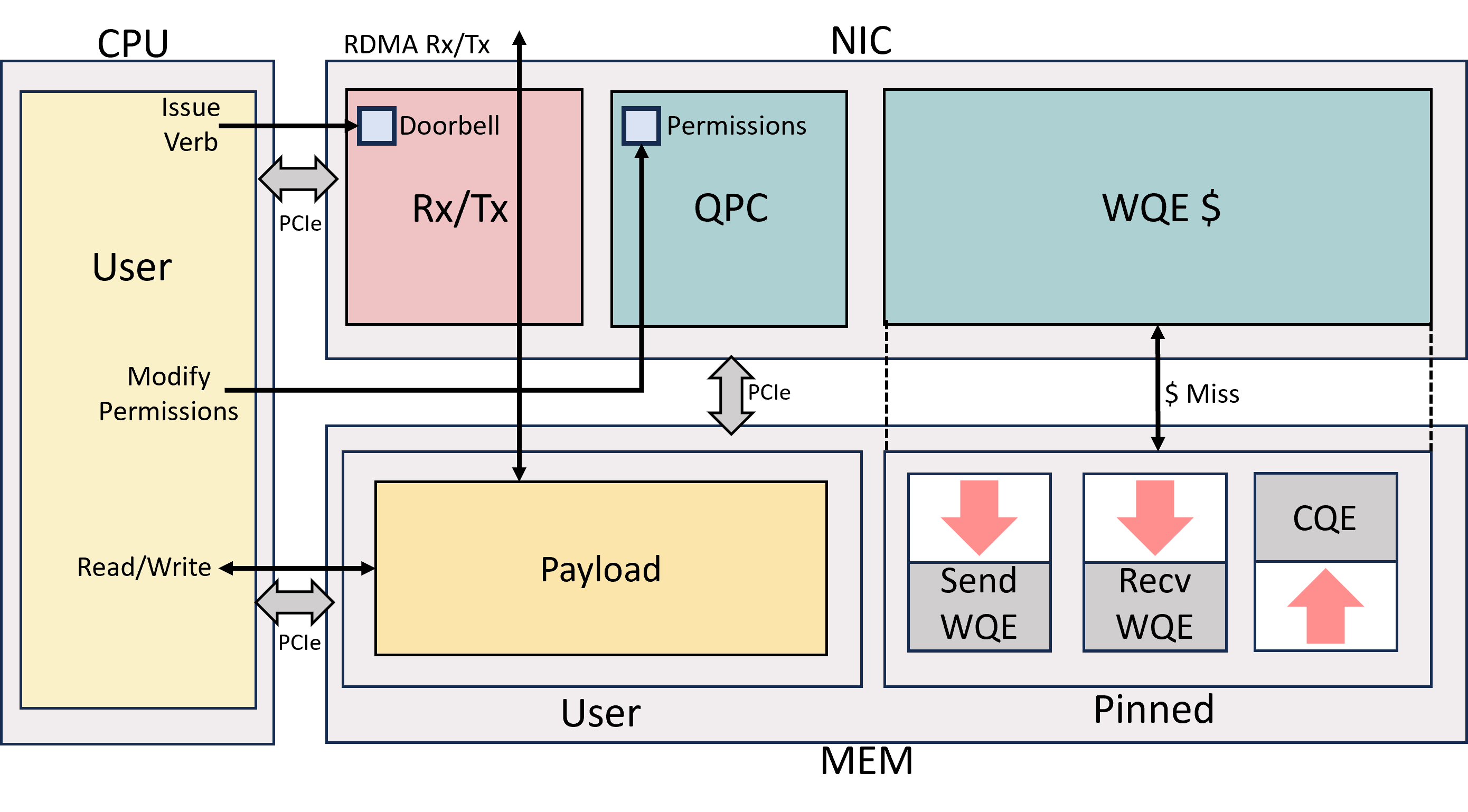}  
  %\vspace*{-8mm}
  \caption{Architectural components that implement RDMA for a traditional CPU-based network.}
  \label{fig:bg:cpu-rnic-s}
\end{subfigure}
\begin{subfigure}{.475\textwidth}
  \centering
  \includegraphics[width=\linewidth]{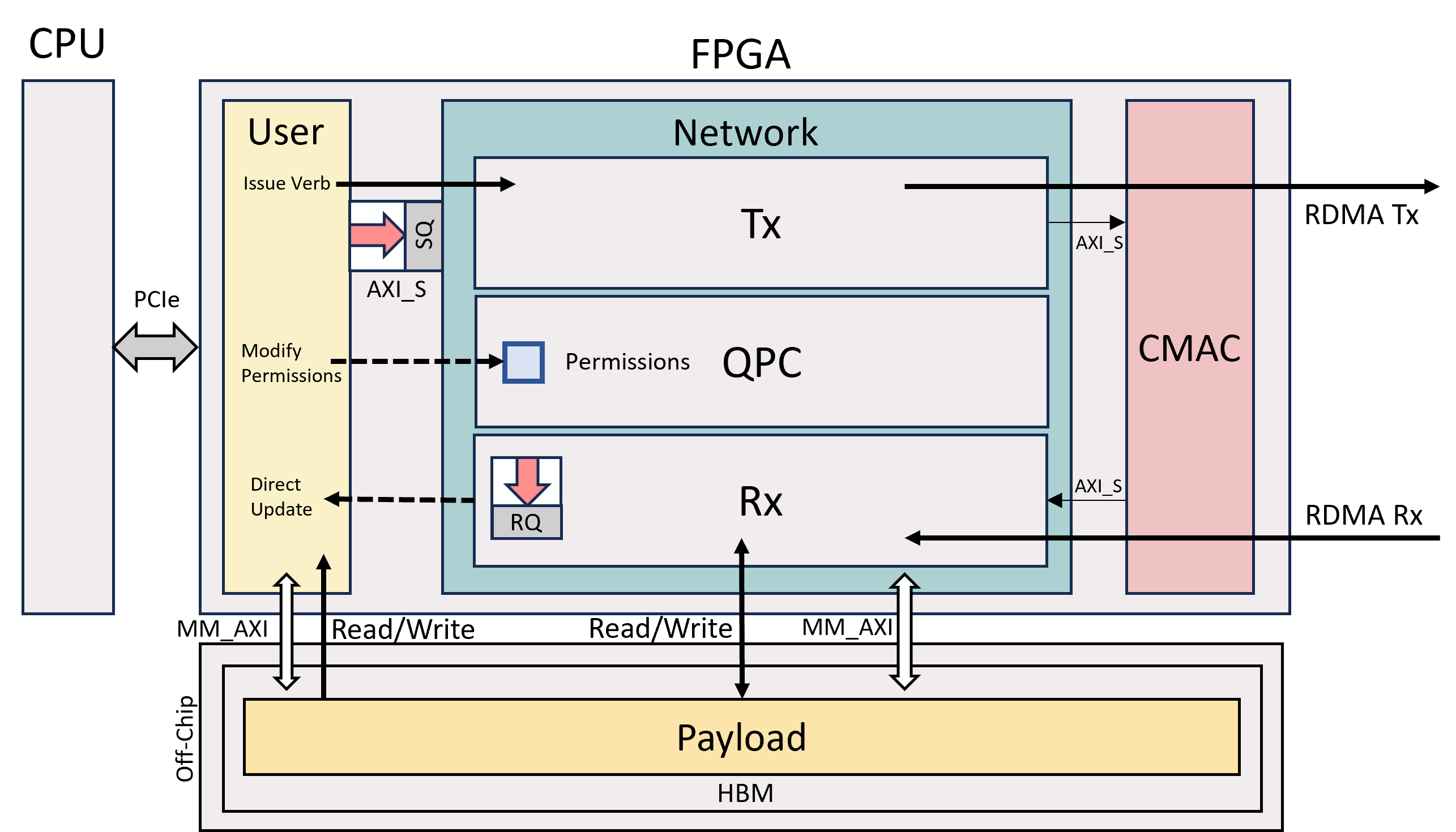} 
  
  \caption{RDMA architecture for a network-attached FPGA featuring a soft RNIC \cite{StRoM}.}
  \label{fig:bg:fpga-rnic-s}
\end{subfigure}
%\vspace*{-7mm}
\caption{Architectural components that implement RDMA for both a CPU and FPGA.}
\label{fig:bg:socket-nic}
%\vspace*{-5mm}
\end{figure}

% \begin{figure}[t]
% \centering
% \includegraphics[width=\linewidth]{appendix/figures/rdma/socket-nic.png}
% \caption{A node in a traditional socket-based network.} 
% \label{fig:bg:socket-nic}
% \Description[]{}
% \end{figure}

The two subfigures in Figure \ref{fig:bg:socket-nic} respectively depict the components (CPU, memory, and NIC) that implement socket-based and RDMA networks; both subfigures assume that all three components are connected via PCIe.

In \autoref{fig:bg:socket-nic}, the operating system (OS), but not the user-space application, communicates with the NIC. User-space applications must incur the overhead of system calls to transmit and receive data. All data received by the NIC is written to memory managed by the OS kernel; once received, the OS copies data to user-space memory, before the user-space application can access the data.

In \autoref{fig:bg:socket-nic}, the user-space application communicates directly with an RDMA-enabled NIC (RNIC), bypassing the OS kernel. As the RNIC provides remote access to user-space memory, the RNIC's Queue Pair Context (QPC) assumes responsibility for managing memory access permissions, a function typically performed by the OS in other contexts. The three queues that comprise an RDMA queue-pair (QP) for each RDMA connection are allocated in a pinned memory segment. The RNIC contains a cache that provides fast access to QPs. Data is remotely written to user-space memory, eliminating the need to copy data from kernel-managed memory into user-space memory in \autoref{fig:bg:socket-nic}. User-space applications that expect to receive data poll their memory segments to access data written by remotely-initiated \textbf{RDMA\_Write}s, and their QPs to detect completion of locally-initiated \textbf{RDMA\_Read}s.

To refresh terminology, the QP on each side of a connection comprises three queues: a Send Queue (SQ), a Receive Queue (RQ), and a Completion Queue (CQ). The user-space application issues a work request by posting a Work Queue Entry (WQE) to the QP: a Send Queue Entry (SQE) is posted to the SQ, a Receive Queue Entry (RQE) is posted to the RQ, and a Completion Queue Entry (CQE) is posted to the CQ. 

% \begin{figure}[t]
% \centering
% \includegraphics[width=\linewidth]{appendix/figures/rdma/cpu-rnic.png}
% \caption{A node in a traditional CPU-based RDMA network.} 
% \label{fig:rdma:cpu-rnic}
% \Description[]{}
% \end{figure}

\subsection{RDMA\_READ}

\begin{figure}[t]
\centering
\includegraphics[width=0.9\linewidth]{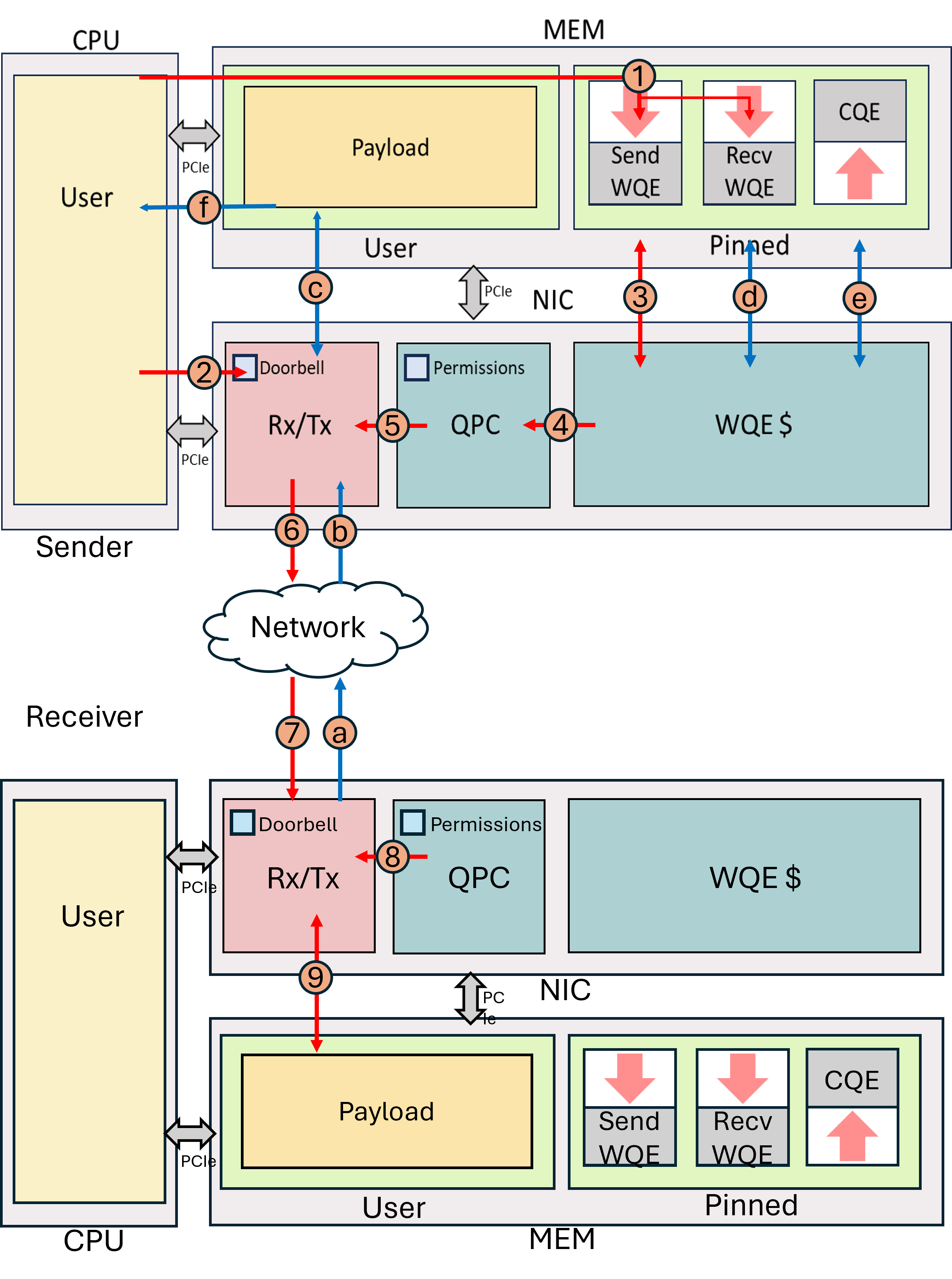}
\caption{CPU NIC read path.}
\label{fig:rdma:cpu-rnic-read-path}
\Description[]{}
\end{figure}

\autoref{fig:rdma:cpu-rnic-read-path} depicts the \textbf{RDMA\_Read} path for a CPU-based RDMA network. The user application (1) posts an \textit{SQE} and \textit{RQE} to the \textit{SQ} and \textit{RQ} and (2) triggers the RNIC by writing to the doorbell register; both of these steps entail PCIe transactions and the first step accesses memory. The RNIC (3) fetches the \textit{SQE} from memory, (4) checks the QP state, (5) verifies permissions, and (6) packages the \textbf{RDMA\_Read} operation for transmission over the network; step (4) incurs PCIe transaction and memory-access latencies. On the receiving side, the RNIC (7) retrieves and unpacks that \textbf{RDMA\_Read} operation, (8) verifies permissions, and (9) reads the payload from memory over PCIe at the address specified in the operation. The receiver's RNIC (a, b) returns the payload over the network to the sender, whose RNIC (c) writes the received payload to memory, (d) fetches the \textit{RQE} from the \textit{RQ}, and (e) writes a \textit{CQE} to the \textit{CQ}, signaling completion of the operation; steps (c), (d), and (e) incur PCIe transaction and memory access latencies. The sender's user application can (f) access memory at any time to retrieve the remotely read payload.

%Once an RDMA operations is received, the QPC is checked to verify read and write permissions of the incoming operation (8). The requested memory address is read over PCIe (9), then transmitted over the network (a) and received by the original sender at (b) and written into a local memory address (c). On receiving the read reply, the \textit{RQE} is popped from \textit{RQ} (e) and placed into \textit{CQ} (e) signaling the completion of the operation. The user-application can then poll the change at the locally written memory address (f). The CPU-based RNIC is not cache coherent causing the user-application to incur a cache miss and memory access latency.

\subsection{RDMA\_WRITE}

\begin{figure}[t]
\centering
\includegraphics[width=0.9\linewidth]{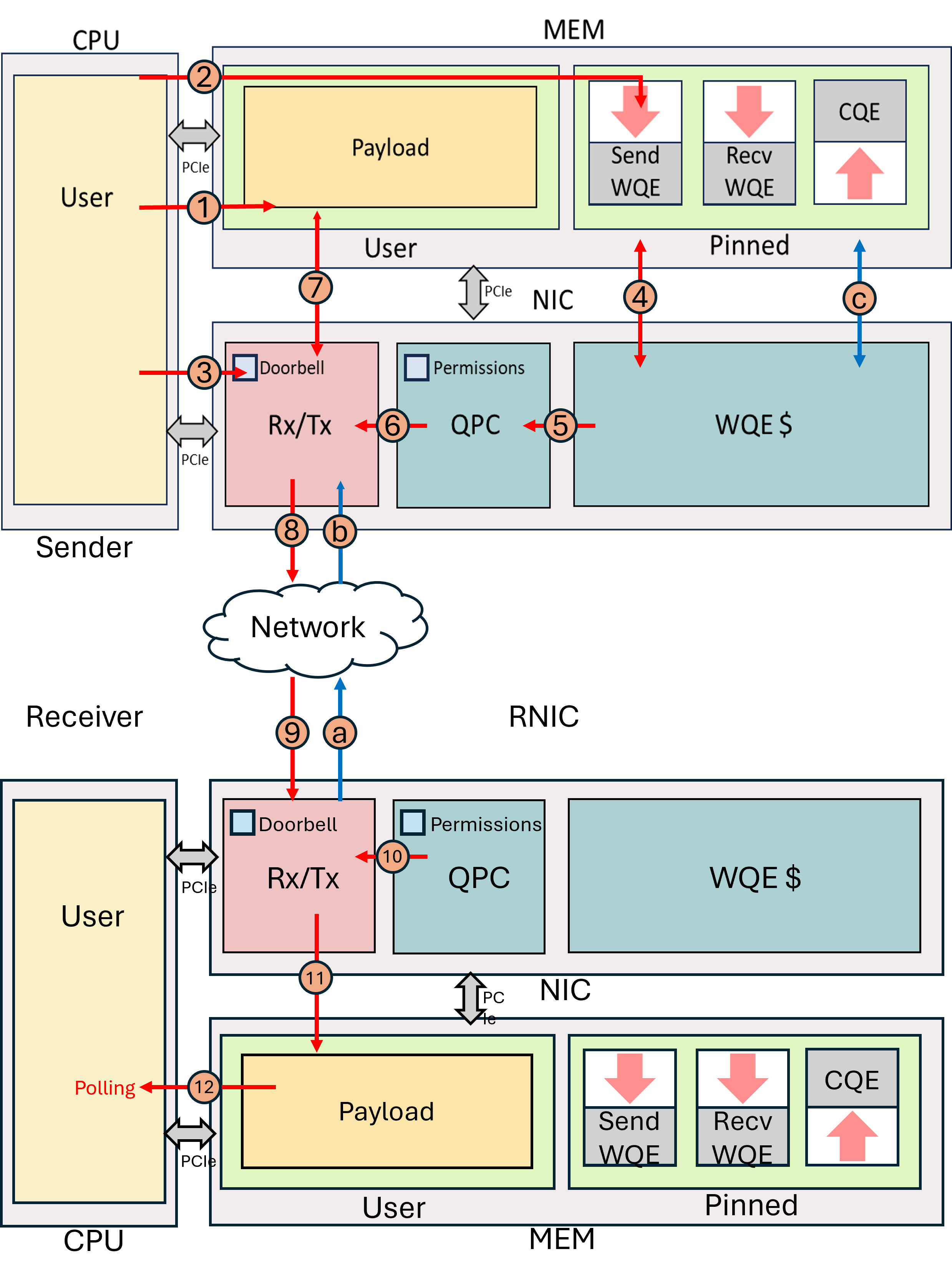}
\caption{CPU NIC write path.}
\label{fig:rdma:cpu-rnic-write-path}
\Description[]{}
\end{figure}

\autoref{fig:rdma:cpu-rnic-write-path} depicts the \textbf{RDMA\_Write} path for a CPU-based RDMA network. The user application (1) writes the payload intended for transmission into memory; (2) posts an \textit{SQE} to the \textit{SQ}; and (3) triggers the RNIC by writing to a doorbell register; all three of these steps entail PCIe transactions, and the first two access memory. The RNIC (4) fetches the \textit{SQE} from memory, (5) checks the QP state, (6) verifies permissions and packet header information, (7) retrieves the payload from memory, and (8) packages the \textbf{RDMA\_Write} operation for transmission across the network; steps (4) and (7) incur PCIe transactions and memory access latencies. On the receiving side, the RNIC (9) retrieves and unpacks the \textbf{RDMA\_Write} operation, (10) verifies permissions, and (11) writes the payload to memory over PCIe. The receiver's RNIC (a, b) returns an acknowledgment (ACK) to the sender, whose RNIC completes the \textbf{RDMA\_Write} operation by posting a CQE to the CQ. The receiver's user application can (12) access memory at any time to retrieve the remotely written payload.

\subsection{RDMA\_READ (FPGA Implementation)}

\begin{figure*}[!tbp]
\centering
\includegraphics[width=\linewidth]{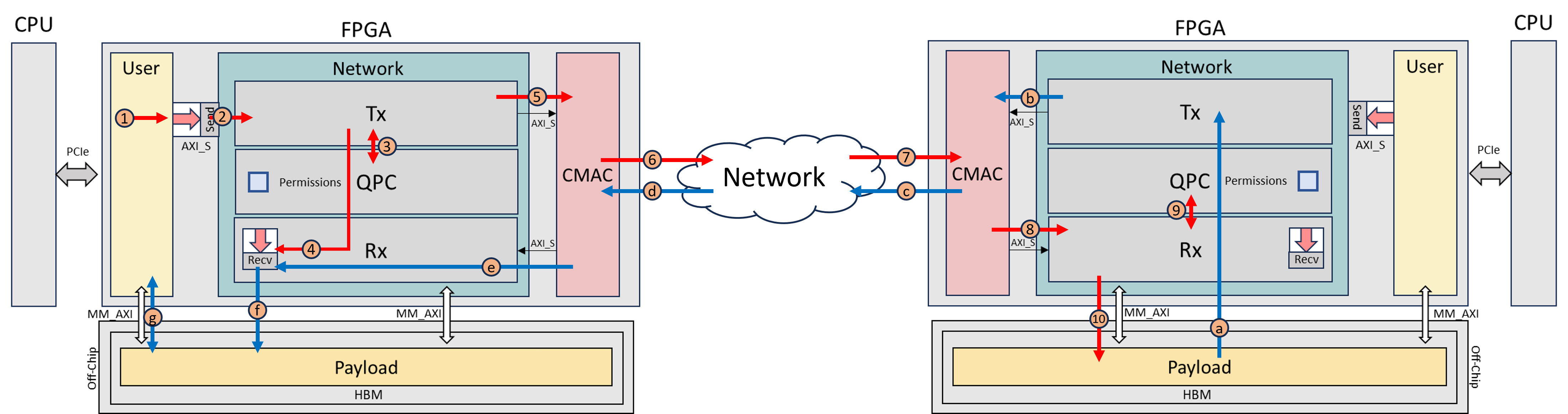}
\caption{FPGA RDMA read path} 
\label{fig:rdma:fpga-rnic-read-path}
\Description[]{}
\end{figure*}

\autoref{fig:rdma:fpga-rnic-read-path} depicts the \textbf{RDMA\_Read} path for a pair of network-attached FPGAs. 

The user kernel (1) pushes the \textbf{RDMA\_Read} operation to the \textit{SQ}, implemented as an AXI Stream interface. (2) The Network kernel pops the verb from the \textit{SQ} and (3) checks the \textit{QPC} for permissions and connection information. Since the operation is \textbf{RDMA\_Read}, (4) an \textit{RQE} is posted to the \textit{RQ}, which resides in on-chip BRAM. The RDMA packet is (5) packaged and transferred to the CMAC through an AXI Stream. The CMAC takes the RDMA packet, adds Ethernet headers, and (6, 7) transmits the packet over the network. At the receiving node, the CMAC strips the Ethernet header and (8) passes the RDMA packet to the Network kernel. The remote Network kernel (9) checks permissions, (10) issues the read operation to memory through a memory-mapped AXI interface to obtain the payload, and (a) passes the payload to the Network kernel for transmission back to the node that initiated the verb. The RDMA packet is (b, c, d) returned and (e) received by the Network kernel. The Network kernel (f) pops the \textit{RQE} from the \textit{RQ} and writes the payload to the requested address in memory. The User kernel (g) accesses the payload through a memory-mapped AXI interface.

%In a traditional RDMA network (\autoref{fig:rdma:cpu-rnic-read-path}), the application, which executes on the CPU, must initiate a PCIe transaction to retrieve the data from memory. In contrast, the FPGA-based RNIC allows the user to kernel to access a memory-mapped AXI buffer, which has a much lower access time than is otherwise achievable in a traditional RDMA network. 

\subsection{RDMA\_WRITE (FPGA Implementation)}

\begin{figure*}[t]
\centering
\includegraphics[width=\linewidth]{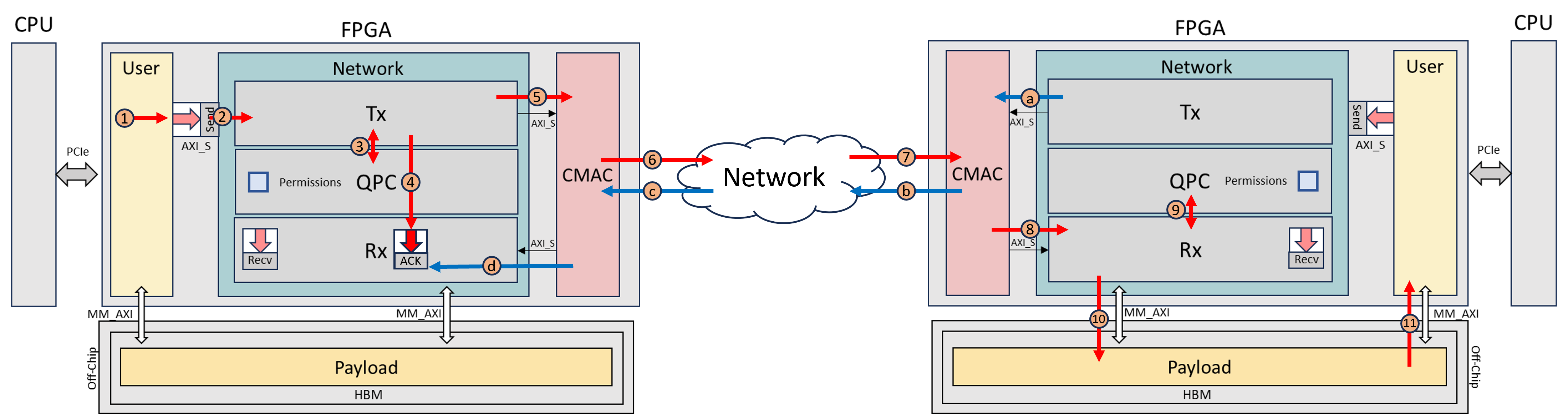}
\caption{FPGA RDMA write path} 
\label{fig:rdma:fpga-rnic-write-path}
\Description[]{}
\end{figure*}

\autoref{fig:rdma:fpga-rnic-write-path} depicts the \textbf{RDMA\_Write} path for a pair of network-attached FPGAs. 

The user kernel (1) pushes the \textbf{RDMA\_Write} verb to the \textit{SQ}, implemented as an AXI Stream interface. (2) The Network kernel pops the verb from the \textit{SQ} and (3) checks the \textit{QPC} for permissions and connection information. The Network kernel maintains a queue of expected acknowledgments (ACKs) for retransmission purposes; before transmitting an RDMA packet, the Network kernel posts an expected ACK into the ACK queue. The RDMA packet is (5) packaged and transferred to the CMAC through an AXI Stream. The CMAC takes the RDMA packet, adds Ethernet headers, and (6, 7) transmits the packet over the network. At the receiving node, the CMAC strips the Ethernet header and (8) passes the RDMA packet to the Network kernel. The remote Network kernel (9) checks permissions and (10) writes the payload to memory through a memory-mapped AXI interface. The User kernel (11) can then access the payload through its own memory-mapped AXI interface. The receiving node generates an ACK and (a) transmits it to the CMAC and (b, c) across the network. The sender receives the ACK, notifying it that the \textbf{RDMA\_Write} verb successfully completed. The Network kernel (d) removes the expected ACK from the ACK queue, thereby completing the operation.

\subsection{FPGA-Specific RDMA Verbs. \ }
% \subsection{FPGA-Specific RDMA Verbs}
\label{sec:fpga-specific-verbs}

\begin{figure}[t]
\begin{subfigure}{.475\textwidth}
  \centering
  \includegraphics[width=\linewidth]{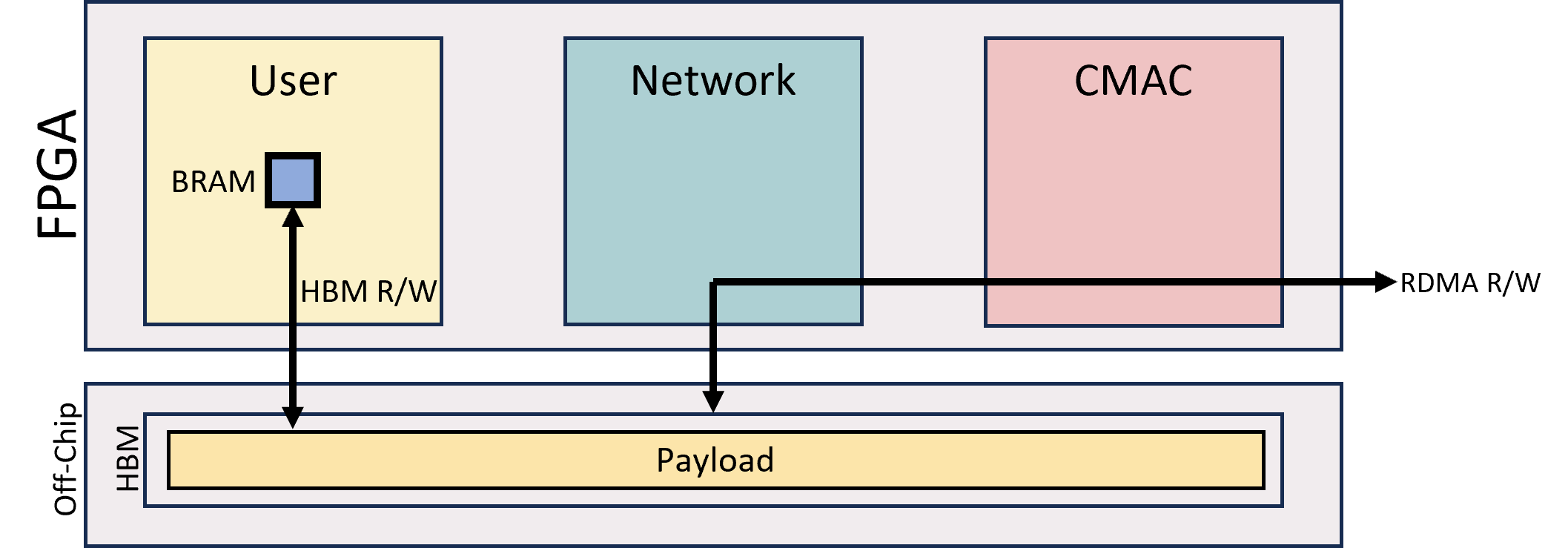}  
  % \vspace*{-8mm}
  \caption{FPGA-based RDMA\_Read and RDMA\_Write verbs.}
  \label{fig:bg:cpu-rnic}
\end{subfigure}
\begin{subfigure}{.475\textwidth}
  \centering
  \includegraphics[width=\linewidth]{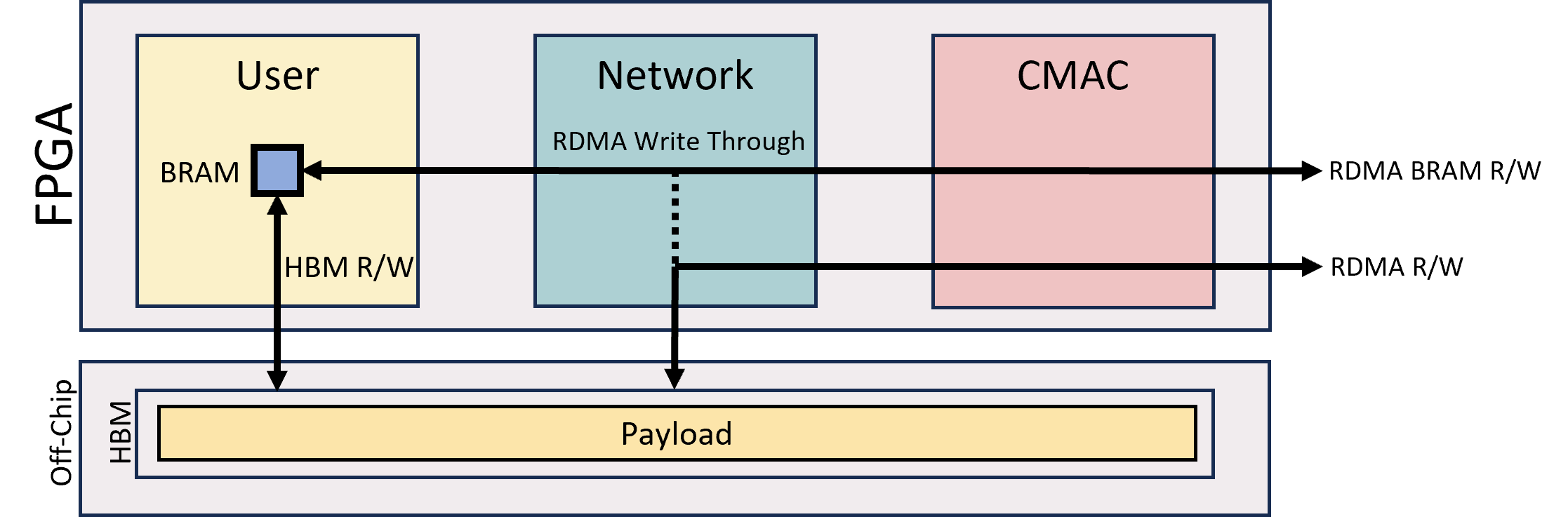} 
  
  \caption{FPGA-specific RDMA verbs that access integrated storage.}
  \label{fig:bg:fpga-rnic}
\end{subfigure}
% \vspace*{-7mm}
\caption{RDMA verbs for network-attached FPGAs.}
\label{fig:rdmafpgaverbs}
% \vspace*{-5mm}
\end{figure}
$\mathsf{Read}$ and $\mathsf{Write}$ verbs communicate with remote CPUs via host memory \footnote{RDMA does not support direct access to the CPU register file or cache hierarchy.}. An analogous implementation of these verbs on a network-attached FPGA utilizes HBM as an intermediary between the local User kernel and the remote node \cite{StRoM} (\ref{fig:rdmafpgaverbs}).

In \ref{fig:rdmafpgaverbs}, we propose RDMA verbs specialized for network-attached FPGAs. Modern FPGAs provide integrated storage. For example, the AMD Ultrascale FPGA in the Alveo U280 accelerator card integrates 2,607,000 registers, 2,016 \textit{Block RAMs (BRAMs)} (36 Kb each) and 960 \textit{UltraRAMs} (288 Kb each); plus any portion of the programmable fabric can be configured as storage. Our proposed FPGA-specific RDMA verbs can read or write to these resources, along with $\mathsf{Write\text{-}Through}$ verbs which concurrently write to integrated storage and HBM.

\autoref{tab:rdma-ops-custom} reports the latencies of some $\mathsf{Write}$ and $\mathsf{Write\text{-}Through}$ verbs that write to FPGA integrated storage resources. The reported latencies include network transmission, RDMA stack, and HBM, BRAM, or register access latencies. In the case of the $\mathsf{Write\text{-}Through}$ verbs, the times reported are for BRAM and register writes.
%\todo{whichever? access latency for the three? It is not clear. Revise for clarity.}
\autoref{tab:rdma-ops-custom} \textit{does not} report the time to generate and receive ACKs. We identify one RDT use-case for a write-through verb (\autoref{sec:conflicting}).
%; the lowest-latency \textbf{RDMA\_Register\_Write-Through} suffices. 

\begin{table}[t]
    \centering
    \footnotesize
    \caption{Average latencies measured for several RDMA verbs that write to FPGA integrated storage.}
    % \vspace*{-3mm}
    \begin{tabular}{|l|l|}
    \hline
     Operation & Latency \\
    \hline
        $\mathsf{Write}$ & 413 $n$s\\
    \hline
        $\mathsf{BRAM\_Write}$ & 309 $n$s\\ 
    \hline
        $\mathsf{BRAM\_Write\text{\_}Through}$ & 309 $n$s\\
    \hline
        $\mathsf{Register\_Write}$ & 285 $n$s\\
    \hline
        $\mathsf{Register\_Write\text{\_}Through}$  & 285 $n$s\\
    \hline
    \end{tabular}
    % \vspace*{-5mm}
    \label{tab:rdma-ops-custom}
\end{table}

\section{Extended Related Work}
\label{app:extended-related-work}
{

{
\noindent\textbf{Positioning and scope.}
We use this claim narrowly: SafarDB is not the first FPGA-based replicated storage system, nor the first RDMA-based replication system. SafarDB builds on four existing lines of work: StRoM's FPGA RoCE/RDMA stack \cite{StRoM}, Mu's RDMA-based SMR protocol \cite{mu}, Hamsaz/Hamband's RDT and WRDT semantics \cite{hamsaz,hamband}, and FPGA or programmable-network replication systems such as CBox, Caribou, Waverunner, P4xos, and NetPaxos \cite{cinbox,caribou,waverunner,p4xos,netpaxos}. SafarDB's contribution is to co-locate RDMA communication, RDT execution, and the strong-consistency control path inside a network-attached FPGA, and to add the custom RDMA/RPC mechanism, direct permission switching through RNIC state, an FPGA implementation of the adopted hybrid relaxed-consistency/strong-consistency RDT model, and hybrid FPGA/host memory mode.
}

{
\noindent\textbf{RDTs and coordination avoidance.}
Prior work on CRDTs, RDTs, and coordination avoidance provides the semantic foundation for SafarDB. CRDTs provide convergence for commutative operations \cite{shapiro11,BurckhardtGotsman14}, while systems and analyses for highly available or weakly consistent databases identify when coordination is required to preserve application invariants \cite{BailisFekete14,indigo,ipa,ecros}. Hamsaz and Hamband specialize this direction for WRDTs and RDMA-based replicated execution \cite{hamsaz,hamband}. SafarDB differs from these systems by changing the execution substrate: instead of executing the RDT logic on CPUs over RDMA, it moves the RDT operation handlers, RDMA endpoint, and strong-ordering support onto the FPGA.
}

\noindent\textbf{Strongly ordered replication and SMR.}
SafarDB targets a different point in this design space: FPGA-resident execution of RDT/WRDT operations with both relaxed and strongly ordered replication paths. Consensus/SMR work relates to SafarDB's support for strongly ordered operations. SMR protocols, which are needed for WRDT conflicting transactions, have been designed for message-passing \cite{OkiLiskov88,Lamport98,CorbettDean13,OngaroOusterhout14,HuntKonar10,zab_protocol} and RDMA \cite{dare,apus,tailwind,sift,rethinking,mu,odyssey,hermes,acuerdo,ubft,whale} networks. In terms of hardware acceleration, HovercRaft \cite{kogias2020hovercraft} and Waverunner \cite{waverunner} accelerate Raft \cite{OngaroOusterhout14} using an in-network programmable P4 ASIC and an FPGA-based Soft NIC, respectively. Consensus-in-a-Box (CBox) \cite{cinbox} implements the full Zookeeper Atomic Broadcast (ZAB) protocol \cite{HuntKonar10,zab_protocol}, including leader election, failure recovery, and application code on the FPGA while storing the replication log in onboard DRAM. We opted to implement Mu \cite{mu} on the FPGA as our RDT implementations were RDMA-based; non-RDMA-based RDTs could be ported to network-attached FPGAs using either Waverunner \cite{waverunner} or CBox \cite{cinbox} to provide consensus.

{
\noindent\textbf{FPGA storage and network-accelerated replication.}
Caribou \cite{caribou} is particularly relevant to SafarDB's FPGA-only configuration because it also places replicated storage on network-attached FPGAs. Caribou exposes a replicated key-value storage interface over regular TCP/IP sockets, stores key-value data in FPGA-attached memory, performs near-data filtering/scans in the FPGA, and replicates writes through a ZAB-based protocol over the same network. SafarDB differs in both the communication and replication abstractions: it uses an RDMA/RoCE-style network endpoint rather than a TCP/IP key-value interface; it executes application-level RDT/WRDT handlers, including key-value workloads such as YCSB, rather than exposing only a key-value storage API; it allows non-conflicting RDT operations to bypass strong ordering; and it accelerates the strong ordering path using FPGA-resident SMR and direct RNIC-state access.

Waverunner \cite{waverunner} is also closely related because it accelerates state-machine replication using an FPGA-based SmartNIC and evaluates replicated services, including key-value workloads. SafarDB differs from Waverunner in where application execution and consistency selection occur. Waverunner keeps the replicated application on the host and uses the FPGA primarily to accelerate the Raft replication path. As a result, Waverunner accelerates the common strongly ordered Raft replication path, while leaving application logic in software and not providing FPGA-resident RDT execution or RDT-aware consistency selection. SafarDB places RDT/WRDT operation handlers and the RDMA-capable endpoint inside the network-attached FPGA, allowing non-conflicting operations to be executed on the relaxed path and conflicting operations to be strongly ordered and executed on FPGA-resident SMR.

P4xos and NetPaxos also accelerate consensus using programmable network hardware \cite{p4xos,netpaxos}. They focus on moving Paxos roles or ordering logic into switches or programmable forwarding devices, whereas SafarDB keeps the replicated application state and RDT execution at the FPGA-attached storage endpoint. Thus, those systems are complementary consensus accelerators; they do not provide RDMA-based RDT execution, hybrid relaxed-consistency/strong-consistency replication paths, or FPGA/host memory placement for replicated data. We therefore treat NetPaxos as a programmable-network consensus accelerator rather than as an experimental storage/RDT baseline.
}

\noindent\textbf{RDMA/RPC communication primitives.}
SafarDB relates to prior work that extends RDMA/RPC and programmable-NIC interfaces. For example, Prism \cite{prism} introduces new primitives to support common remote access patterns, StRoM \cite{StRoM} and RMC \cite{rmc} provide RDMA or RDMA-like extensions for remote computation and RPC-style execution, and KV-Direct \cite{kv-direct} uses programmable NICs to extend RDMA-style access for direct key-value operations. The nanoPU provides an RDMA-inspired fast path between the NIC and CPU register file to support $\mu$s-scale RPCs \cite{nanopu}. Either Hamband's or SafarDB's RDTs may further benefit from Prism's primitives or nanoPU acceleration. SafarDB's SMR could similarly complement KV-Direct-style programmable-NIC key-value access in a replicated key-value store.

{
This line of work also highlights a design trade-off between RDMA and RPC abstractions. One-sided RDMA avoids remote CPU involvement, but it exposes memory operations rather than application-level execution. Two-sided RDMA SEND/RECV provides request delivery, but still requires receive queues and dispatch logic at the target. RPC-style NIC extensions instead allow a network request to directly invoke application-specific handlers. SafarDB follows this last direction, but specializes it for FPGA-resident RDT execution: incoming operations can both move data and trigger the appropriate RDT or SMR logic without routing the common case through host memory.
}

{
\noindent\textbf{FPGA database query processing.}
Prior FPGA database systems accelerate event processing, multi-query stream processing, and flexible query execution on reconfigurable hardware \cite{sadoghi2011fpga_topss,sadoghi2012multiquery,najafi2013fqp,najafi2015fqpvision}. These systems show how database query-processing operators can be optimized on FPGAs. SafarDB targets a different layer of the data-management stack: replicated transactional processing, where the main challenge is coordinating RDT/WRDT operations across replicas using relaxed and strongly ordered replication paths.
}

\noindent\textbf{SmartNIC architectures.}
SmartNICs \cite{catapult,configurable_cloud,vfp,azure_smartnic,configurable_cloud_dnn} integrate programmable accelerators, allowing servers to offload tasks relating to networking to the network interface. \textit{On-path} (``bump in the wire'') SmartNICs place programmable elements on the packet processing path, which adds latency when they are not utilized, while \textit{off-path} SmartNICs \cite{ipipe,xenic,off_path_characterization} integrate a network switch, which removes the programmable elements from the packet processing path, but adds a small amount of routing latency. Prior work has noted a lack of versatility in SmartNIC functionality, as SmartNICs are often limited to specific tasks such as RDMA \cite{spin}. Unlike SmartNICs, SafarDB's network stack is implemented within the FPGA; while less efficient than using an ASIC, SafarDB eliminates inter-chip communication overheads that are inherent to SmartNICs while enabling application developers to design and deploy application- or domain-specific on-path, off-path, or hybrid communication architectures.

{
Therefore, SafarDB should be viewed as a network-attached FPGA replication engine rather than as a conventional CPU-hosted database with an offloaded NIC function. The relevant distinction is that the RDMA-capable endpoint, replication control path, and RDT operators are co-designed inside the same FPGA, which lets SafarDB remove PCIe and host-memory interactions from the common FPGA-resident execution path.
}
}

\section{Additional Experimental Results}

\begingroup
\begin{table*}[!t]
\color{black}
\centering
\small
\captionsetup{labelfont={bf,color=black},textfont={bf,color=black}}
\caption{\textcolor{black}{Application-State and Dataset Configurations. CRDT and WRDT rows report the total configured logical application state stored by each replica in FPGA-only experiments, rather than per-object or per-element sizes. YCSB and SmallBank rows report the configured FPGA/host dataset partitions.}}
\label{tab:application-state-datasets}
\begin{tabular}{llp{9.0cm}p{3.0cm}}
\toprule
\textbf{Category} & \textbf{Use case} & \textbf{Total state per replica or dataset partition} & \textbf{Placement} \\
\midrule
CRDT & Counter & One 64-bit counter (8~B) & FPGA-only \\
CRDT & Register & One 16-bit value and one 16-bit timestamp (4~B) & FPGA-only \\
CRDT & G-Set & 100K 32-bit elements (400~KB) & FPGA-only \\
CRDT & PN-Set & 100K pairs of 32-bit elements and counters (800~KB) & FPGA-only \\
CRDT & 2P-Set & Two 60K-element sets of 32-bit values (480~KB) & FPGA-only \\
\midrule
WRDT & Account & Three 32-bit balance components (approximately 12~B) & FPGA-only \\
WRDT & Courseware & 2K students, 2K courses, and 250K 32-bit enrollments (approximately 1~MB) & FPGA-only \\
WRDT & Project & 2K employees, 2K projects, and 250K 32-bit assignments (approximately 1~MB) & FPGA-only \\
WRDT & Movie & 2K movie and 2K customer Boolean flags (approximately 500~B logical state) & FPGA-only \\
WRDT & Auction & 200 auction records with four 32-bit fields, 200 32-bit balance/direct-buy entries, and 200 user flags (approximately 4~KB logical state) & FPGA-only \\
\midrule
Benchmark & YCSB & 100K FPGA-resident keys and 10M host-resident keys & Hybrid FPGA/host \\
Benchmark & SmallBank & 10M FPGA-resident accounts and 90M host-resident accounts & Hybrid FPGA/host \\
\bottomrule
\end{tabular}
\end{table*}
\endgroup

\subsection{Hardware Comparison for Cross-Cluster Evaluation}

\begingroup

\begin{table}[!b]
\centering
\captionsetup{font=footnotesize}
\setlength{\abovecaptionskip}{2pt}
\setlength{\belowcaptionskip}{2pt}
\tiny
\caption{\textcolor{black}{Hardware comparison between OCT and ACES. OCT provides the FPGA accelerator and both SafarDB's FPGA soft-RNIC path and Hamband's 100 Gbps hardware-RoCEv2 path. ACES provides a newer CPU, greater memory and cache capacity, and higher network bandwidth, while OCT has slightly higher CPU base and turbo frequencies. Intel's specification sheets report memory speed using different units for the two CPUs.}}
\label{tab:cluster-comparison}
\resizebox{\columnwidth}{!}{%
\begin{tabular}{lll}
\toprule
\textbf{Dimension} & \textbf{OCT (SafarDB/Hamband-RoCE)} & \textbf{ACES (Hamband-InfiniBand)} \\
\midrule
CPU model         & Xeon Gold 6226R \cite{intel_xeon_gold_6226r}        & Xeon Platinum 8468 \cite{intel_xeon_platinum_8468}     \\
CPU generation    & Cascade Lake (2020)    & Sapphire Rapids (2023)  \\
CPU allocation / replica & \shortstack[l]{SafarDB: initialization only;\\Hamband-RoCE: 4 physical cores} & Hamband-InfiniBand: 4 physical cores \\
Base / turbo MHz  & 2900 / 3900            & 2100 / 3800             \\
Memory type       & 192GB DDR4             & 512GB DDR5-4800         \\
L3 cache / CPU    & 22 MB                  & 105 MB                  \\
Network bandwidth & 100 Gbps (GbE)         & 200 Gbps (NDR IB)       \\
Switch fabric     & 100GbE Ethernet        & NDR400 InfiniBand       \\
RDMA NIC          & \textcolor{black}{\shortstack[l]{FPGA soft RNIC (SafarDB);\\100 Gbps hardware RoCEv2 RNIC (Hamband)}} & ConnectX-7 (hardened)   \\
FPGA accelerator  & Alveo U280 (8\,GB HBM) & None                    \\
PCIe Generation   & PCIe 3.0               & PCIe 5.0                \\
Memory speed      & 2933 MHz               & 4800 MT/s               \\
\bottomrule
\end{tabular}
}
\end{table}
\endgroup

% \begin{figure}[t]
% \centering
% \includegraphics[width=0.7\linewidth]{appendix/figures/results/read-only-cluster-padded.png}
% \caption{Performance of Hamband running SmallBank on OCT vs ACES.}
% \label{fig:results:read-only-cluster}
% \end{figure}

\begin{figure}[!t]
\centering
\includegraphics[width=\linewidth]{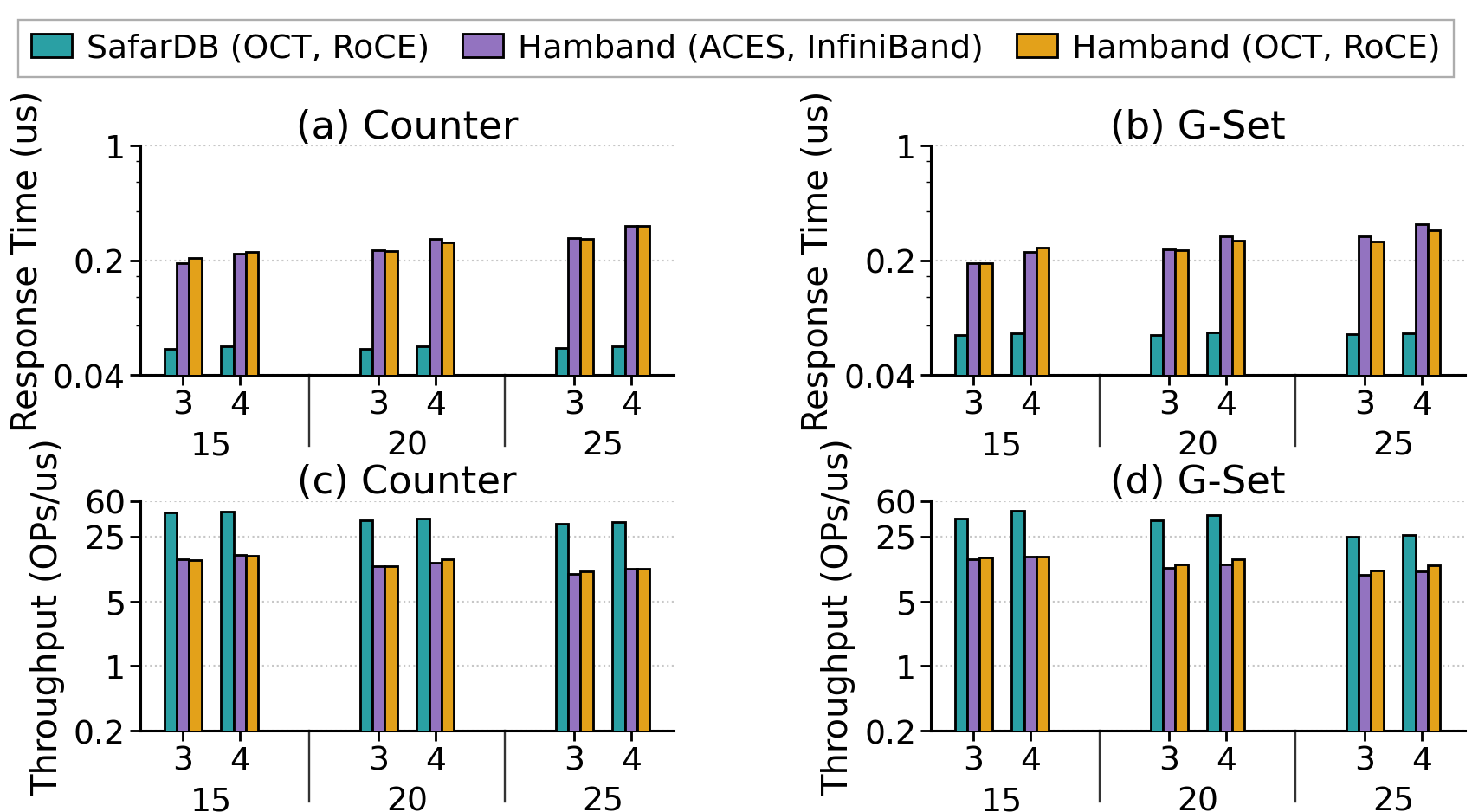}
\caption{\textcolor{black}{SafarDB on the OCT cluster with RoCE, Hamband on the ACES cluster with InfiniBand, and Hamband on the OCT cluster with RoCE. 
Response time and throughput for Counter and G-Set with three and four replicas.
Results use base-5 logarithmic y-axes, and numbers below each replica group show the update percentage.}}
\label{fig:app:crdts-roce}
\end{figure}

\begin{figure*}[!t]
\centering
\includegraphics[width=\textwidth]{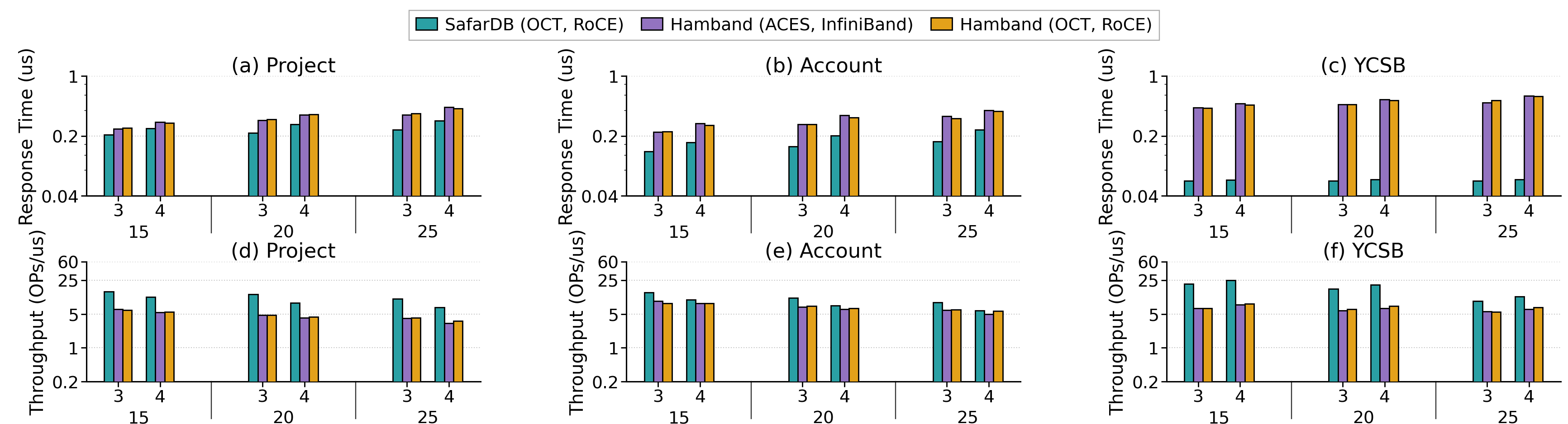}
\caption{\textcolor{black}{Project, Account, and YCSB response time and throughput with three and four replicas. SafarDB runs on OCT/RoCE (RPC for Project and Account; FPGA-only for YCSB), and Hamband runs on ACES/InfiniBand and OCT/RoCE. The y-axes use base-5 logarithmic scales; group labels show update percentages.}}
\label{fig:app:wrdts-roce}
\end{figure*}

\subsection{Latency breakdown}

\begin{figure}[!b]
\centering
\includegraphics[width=0.7\linewidth]{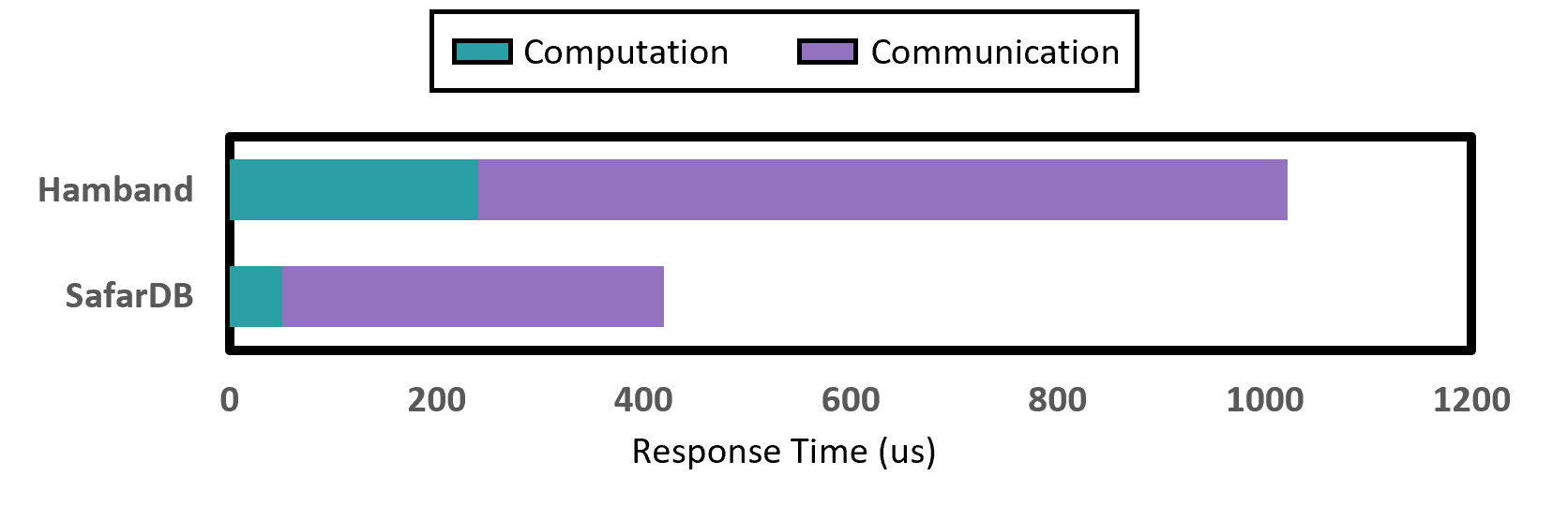}
\caption{Latency breakdown of Hamband and SafarDB.} 
\label{fig:results:comp-vs-comm}
\end{figure}

\begin{figure}[!t]
\centering
\includegraphics[width=0.7\linewidth]{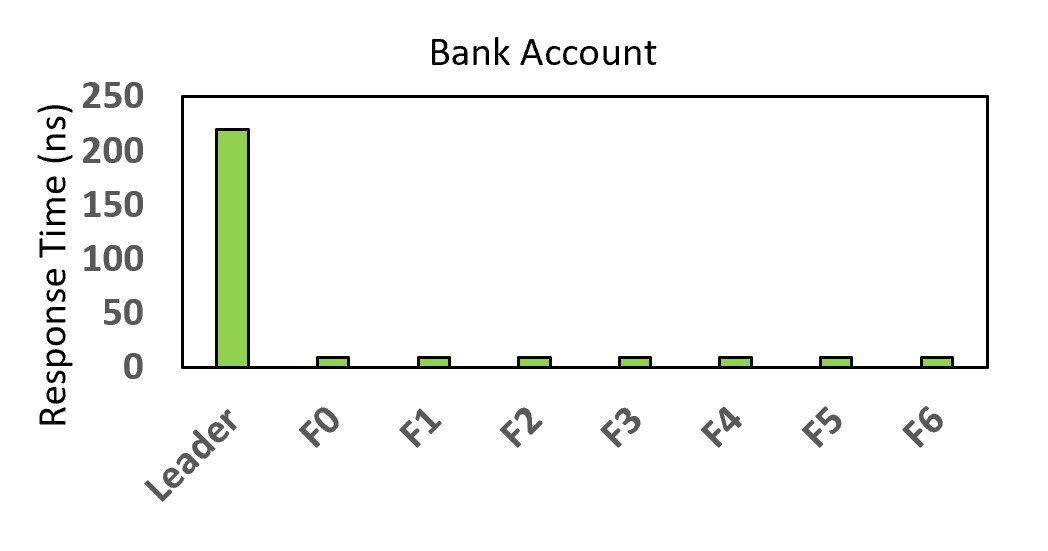}
\caption{Execution time of all replicas for the Bank Account WRDT in an 8-node configuration with 15\% writes.}
\label{fig:results:splits}
\end{figure}

\autoref{fig:results:comp-vs-comm} reports the latency breakdown of computation versus communication for conflicting transactions. In this experiment, computation refers to the latency of modifying local data for the given workload, while communication refers to the latency of coordinating conflicting transactions. Most of the time in both systems is spent coordinating the transaction, with computation contributing only minimal overhead. SafarDB's computation is 4x faster than Hamband's, while its communication is 2x faster. SafarDB leverages FPGA hardware acceleration to perform both computation and coordination; the FPGA's RDMA implementation allows coordination to be completed efficiently.

\subsection{Leader and Follower Execution Breakdown}
\autoref{fig:results:splits} reports the execution time of the Bank Account WRDT with 8 replicas and a write percentage of 15\%. The leader's execution time is more than twice as large as the execution times of the seven follower replicas F0-F6; the follower execution times are approximately equal. Throughput is defined as the total number of completed operations divided by the total execution time. The total execution time is constrained by the longest-running replica, which is the leader in \autoref{fig:results:splits}. Presuming that Bank Account is representative of all WRDTs, \autoref{fig:results:splits} indicates that the most promising way to improve throughput is to improve the performance of the leader, which means focusing on how conflicting transactions and coordination are implemented.

%Throughput is measured by taking the total executed operations divided by the total execution time of the system. Throughput includes time taken to execute operations on the remote replica as well, meaning throughput is the slowest replica plus communication and remote replica execution time. The slowest node in WRDT workloads is usually a leader; so we can say that throughput is bounded by the leader replicas execution time. 

% \begin{figure}[]
% \centering
% \includegraphics[width=0.8\linewidth]{figures/results/leader.png}
% \caption{Execution time of Leader node in Courseware WRDT} 
% \label{fig:results:leader}
% \end{figure}

% \begin{figure}[]
% \centering
% \includegraphics[width=0.8\linewidth]{figures/results/followers.png}
% \caption{Average execution time of all followers in Courseware WRDT} 
% \label{fig:results:follower}
% \end{figure}

\begin{figure}[t]
\centering
\includegraphics[width=0.7\linewidth]{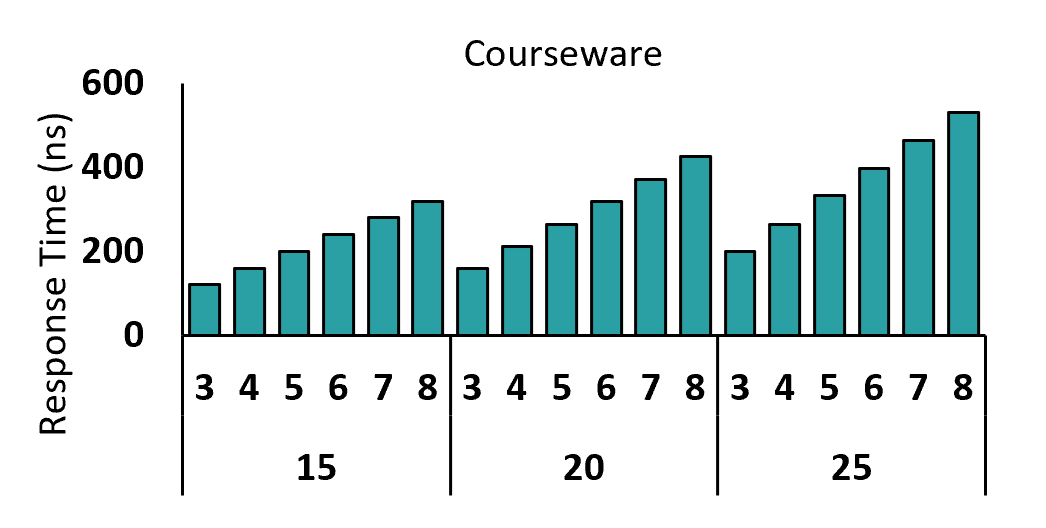}
\caption{Execution time of leaders in Courseware WRDT} 
\label{fig:results:leader}
\end{figure}

\begin{figure}[t]
\centering
\includegraphics[width=0.7\linewidth]{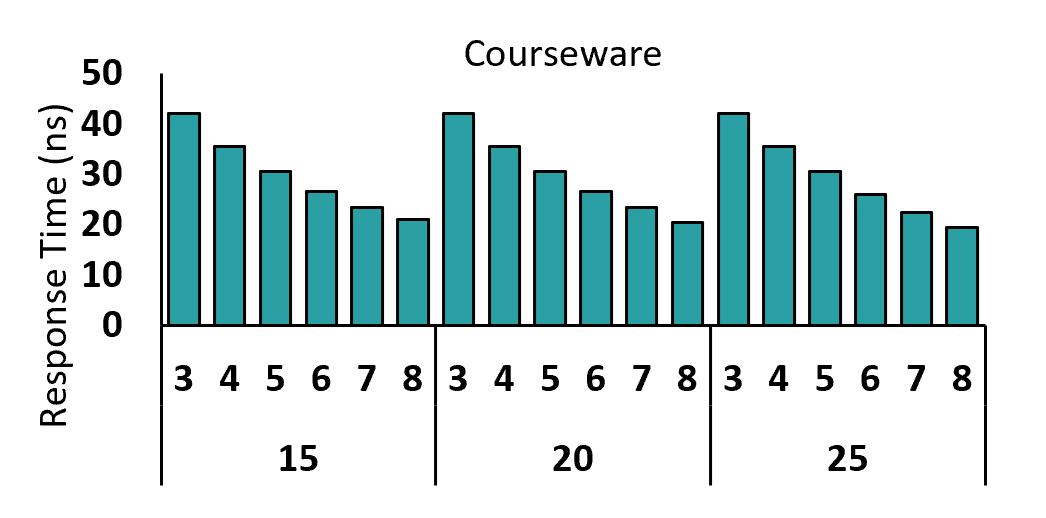}
\caption{Execution time of followers in Courseware WRDT} 
\label{fig:results:follower}
\end{figure}

\begin{figure}[t]
\centering
\includegraphics[width=0.5\textwidth]{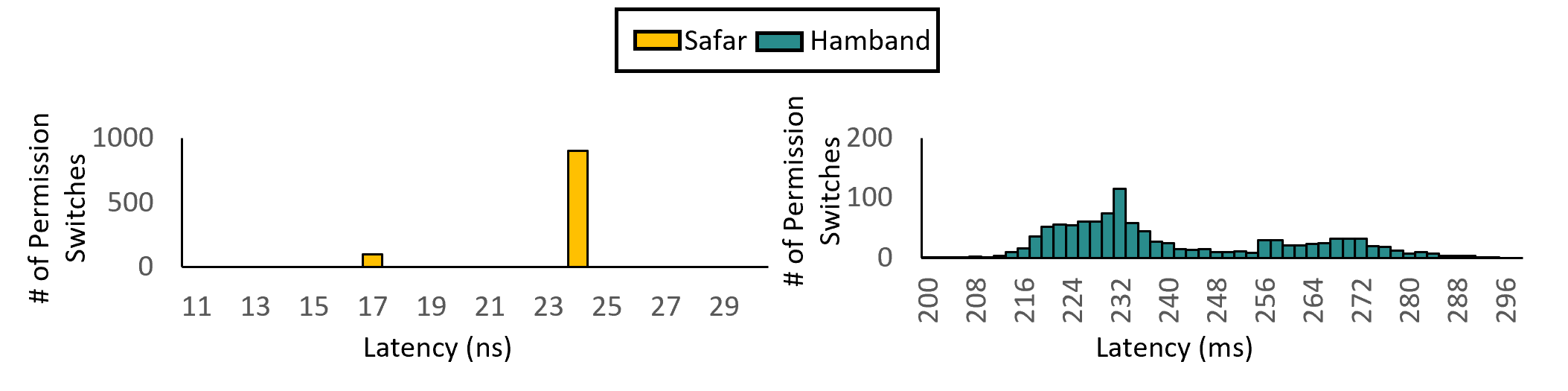}
\caption{Permission Switch histograms}
\label{fig:app:permission-switch}
\Description[]{}
\end{figure}

\autoref{fig:results:leader} reports the leader execution times across 3--8 replicas and write percentages of 15\%, 20\%, and 25\%, for the Courseware WRDT. Increasing the write percentage tends to increase leader execution time, since there are more conflicting transactions. To recall, the write percentage is the fraction of methods that write state, which includes reducible, irreducible conflict-free, and conflicting methods, whereas all remaining calls invoke the $query()$ method. Increasing the number of replicas tends to increase leader execution time due to the increased cost of coordinating a larger number of followers.

Conversely, \autoref{fig:results:follower} reports the average execution time of all followers across the same range of replicas and write percentages. As the number of replicas increases, each follower invokes fewer calls, which reduces execution time. Increasing the write percentage marginally increases execution time by adding more conflict-free transactions to execute; as conflict-free methods lack coordination, the slowdown is minimal. Conflicting calls do not impact follower execution time to the same extent that they impact leader execution time.

\subsection{Permission Switch Histograms}

\subsection{Power Consumption}

FPGA power consumption was measured following AMD Xilinx Documentation \cite{xilinx-docs}. CPU and I/O power consumption was measured with a combination of PowerTop \cite{powertop} and Scaphandre \cite{scaphandre}.

\begin{figure}[t]
\centering
\includegraphics[width=0.384\textwidth]{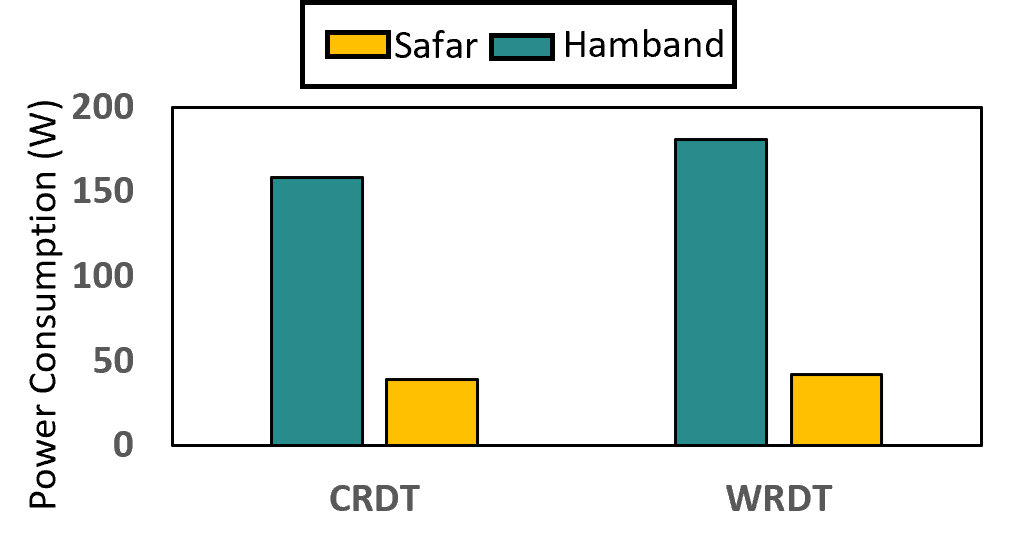}
\caption{Power Consumption}
\label{fig:app:power-consumption}
\Description[]{}
\end{figure}

\autoref{fig:app:power-consumption} reports the peak power consumption of CRDTs and WRDTs averaged across use cases and write percentages. SafarDB consumes about $35$ W, while Hamband consumes about $160$ W, so Hamband consumes approximately $4.5\times$ as much power as SafarDB while SafarDB provides improved performance. SafarDB's power consumption includes the entire Alveo U280 card and HBM, while Hamband's includes the full system with CPU, RNIC, memory, and I/O. For Hamband, approximately two-thirds of the power consumption comes from the CPU, and one-third comes from I/O, including memory, the RDMA NIC, and the PCIe interface. This disparity is expected for FPGAs because they operate at lower clock frequencies, avoid CPU-style instruction fetch/decode overheads, and reduce data movement through the cache hierarchy \cite{OsFPGA}.

\begin{figure}[!t]
\centering
\includegraphics[width=0.5\textwidth]{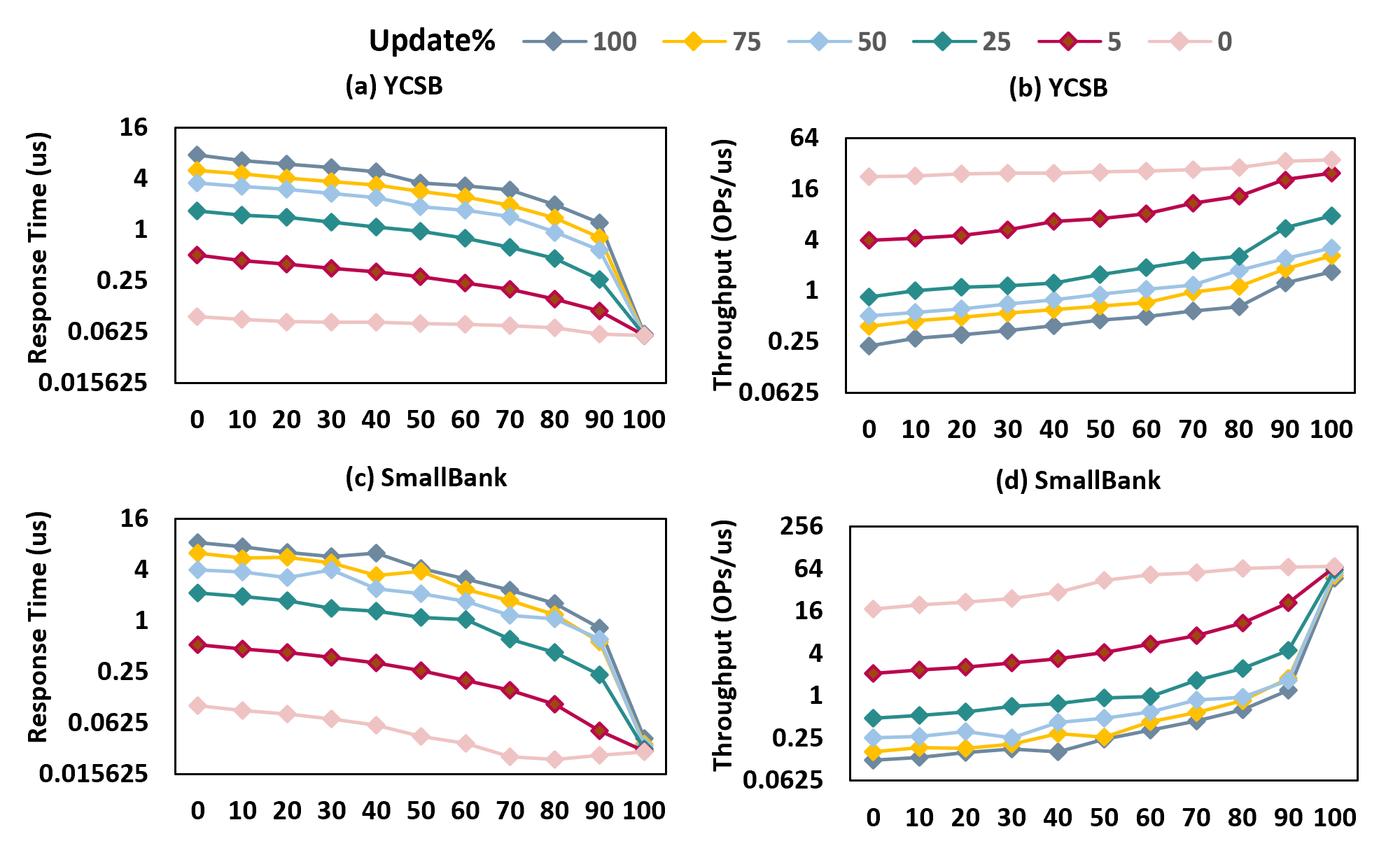}
\caption{Response time and throughput for YCSB/SmallBank with different percentages of OPs assigned to the FPGA.}
\label{fig:app:hybrid:percent-ops}
\Description[]{}
\end{figure}

\begin{figure*}[!t]
\centering
\includegraphics[width=\textwidth, keepaspectratio]{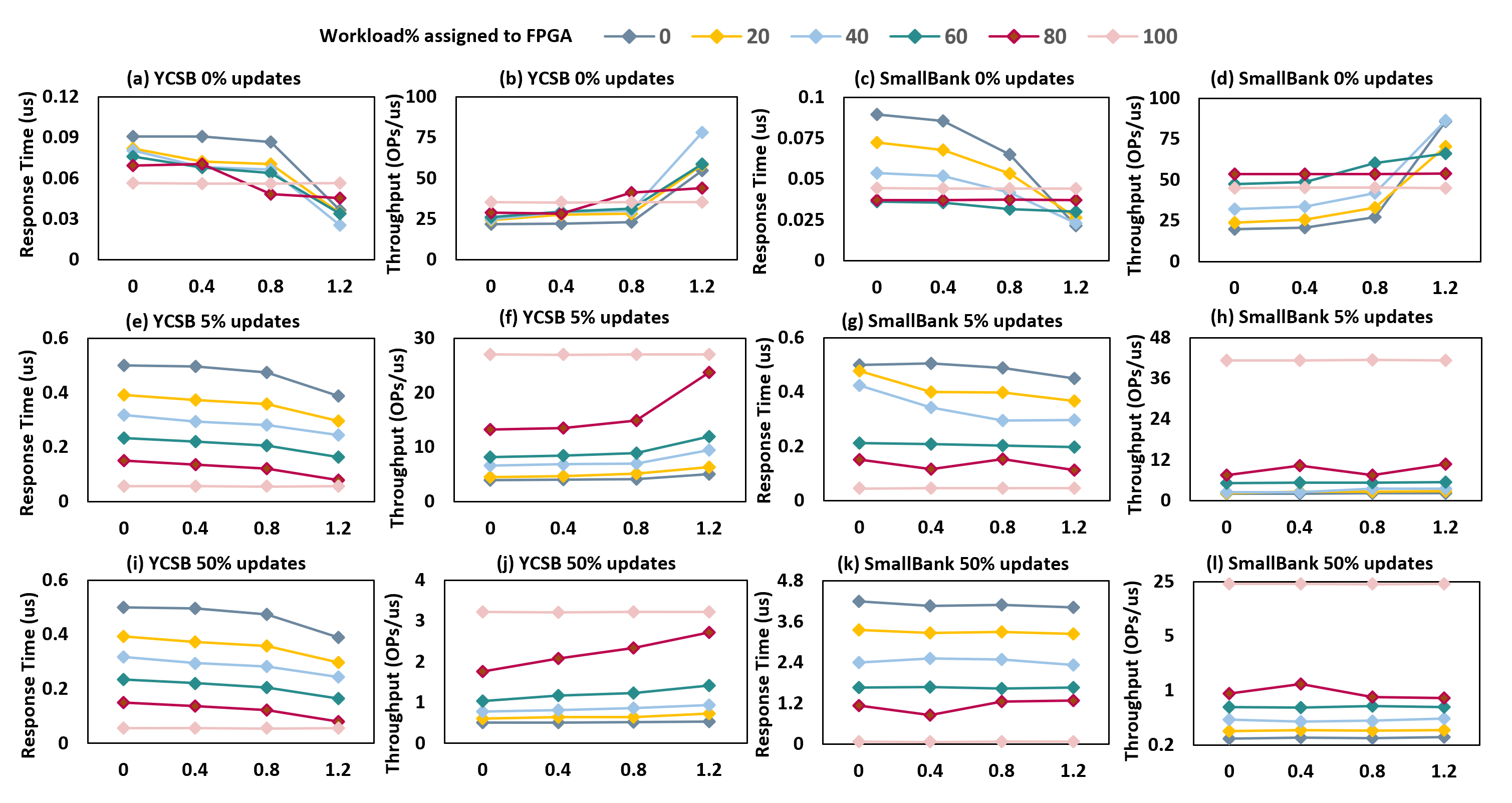}
\caption{Response time and throughput for YCSB and SmallBank with different $\theta$ values for Zipfian distributions.}
\label{fig:app:hybrid:zipf}
\Description[]{}
\end{figure*}

\autoref{fig:app:hybrid:percent-ops} and \autoref{fig:app:hybrid:zipf} provide the detailed hybrid-mode results summarized in the Evaluation section. \autoref{fig:app:hybrid:percent-ops} varies the fraction of YCSB and SmallBank requests that target FPGA-resident keys/accounts versus host-resident keys/accounts. These results show the capacity/performance trade-off of hybrid mode: host memory expands the dataset beyond FPGA memory, while response time decreases and throughput increases as more requests are served by FPGA-resident state. \autoref{fig:app:hybrid:zipf} varies the Zipfian skew and shows that skew can improve host-side accesses through CPU cache locality, but this benefit becomes smaller as the update ratio increases or as more requests already target FPGA-resident state.
\FloatBarrier
\subsection{Hybrid-Mode Detailed Results}
The percentage-assignment experiment separates capacity from placement. The workload remains the same, but the fraction of keys/accounts served from FPGA memory increases along the x-axis. The 100\% point corresponds to the FPGA-only case, while lower percentages model deployments where only part of the replicated state is placed in FPGA memory and the remaining state stays in host memory. This explains why response time improves most consistently as the FPGA-resident fraction grows: fewer requests cross the FPGA-host boundary, and more operations complete in the FPGA-resident RDT handlers.

The Zipfian experiment exercises a different effect. Higher $\theta$ concentrates requests on fewer records, which can improve CPU-cache locality for host-resident state. This benefit is most visible when a substantial fraction of requests still access host memory. When most requests already target FPGA-resident state, or when the workload has more updates, the benefit from skew is smaller because replication, synchronization, and FPGA-side execution costs dominate. Together, these detailed plots support the main-text summary that hybrid mode is both a capacity mechanism and a placement knob for keeping frequently accessed state in FPGA memory while preserving one replication interface for both storage locations.

\end{appendices}

\end{document}